 \documentclass{aa}  
\usepackage{amsmath}
\usepackage{txfonts}
\usepackage{graphicx}
\usepackage{epstopdf}
\usepackage{morefloats}
\usepackage{subfigure}

\usepackage{natbib}
\bibpunct{(}{)}{;}{a}{}{,} % to follow the A&A style

\def\cdo{\hbox{$\rm ^{13}CO$}}
\def\cao{\hbox{$\rm C^{18}O$}}
\def\cdot{\hbox{$\rm ^{13}CO$(2--1)}}
\def\caot{\hbox{$\rm C^{18}O$(2--1)}}
\def\kkms{\hbox{$\,\mathrm{K\,km\,s^{-1}}$}}
\def\kms{\hbox{$\,\mathrm{km\,s^{-1}}$}}
\def\taud{\hbox{$\tau_{13}$}}
\def\taua{\hbox{$\tau_{18}$}}
\def\tex{\hbox{$T_\mathrm{ex}$}}

\def\hzw{\hbox{$\mathrm{H_2}$}}
\def\vlsr{\hbox{$v_\mathrm{LSR}$}}

\def\modot{\hbox{$\,\mathrm{M_\odot}$}}
\def\planck{{\it Planck}}
\def\herschel{{\it Herschel}}
\def\lupus{\hbox{Lupus~I}}

\newcommand{\nhz}[2]{\hbox{$#1\times10^{#2}\,\mathrm{cm^{-2}}$}}

\begin{document}
    \title{Squeezed between shells? On the origin of the Lupus~{\rm I} molecular cloud}
    
    \subtitle{II. APEX\thanks{This publication is based on data acquired with the Atacama Pathfinder Experiment (APEX). APEX is a collaboration between the Max-Planck-Institut fur Radioastronomie, the European Southern Observatory, and the Onsala Space Observatory.} CO and GASS \ion{H}{I} observations}

%    \author{B.~Gaczkowski\inst{1} \and T.~Preibisch\inst{1} \and T.~Stanke\inst{2} \and M.~Krause\inst{1,3,4} \and et al.}
   \author{B.~Gaczkowski\inst{1}  \and V.~Roccatagliata\inst{2,1} \and S.~Flaischlen\inst{1} \and D.~Kr{\"o}ll\inst{1,3} \and M.G.H.~Krause\inst{5,1,3,4} 
\and A.~Burkert\inst{1,3} \and R.~Diehl\inst{3,4} \and K.~Fierlinger\inst{1,4} \and J.~Ngoumou\inst{1}  \and T.~Preibisch\inst{1}   }

   \institute{
   Universit\"ats-Sternwarte M\"unchen, 
        Ludwig-Maximilians-Universit\"at,
          Scheinerstr.~1, 81679 M\"unchen, Germany;
	  \email{bengac@usm.uni-muenchen.de};  \email{roccatagliata@arcetri.astro.it}
\and
INAF-Osservatorio Astrofisico di Arcetri, L.go E. Fermi 5, 50125 Firenze, Italy
\and
Max-Planck-Institut f\"ur extraterrestrische Physik, Postfach 1312, 85741 Garching, Germany
\and
Excellence Cluster Universe, Boltzmannstrasse 2, 85748 Garching, Germany
\and
School of Mathematics \& Physics, Private Bag 37, University of Tasmania, Hobart 7001, Australia
}

 \titlerunning{Lupus~{\rm I} - APEX  CO and GASS \ion{H}{I} observations}

   \date{Received 14.03.2016;  Accepted 13.10.2017}

\abstract{The Lupus~{\rm I} cloud is found between the Upper-Scorpius (USco) and the Upper-Centaurus-Lupus (UCL) sub-groups of the
Scorpius-Centaurus OB-association, where the expanding USco H\,{\rm I} shell appears to interact  
with a bubble currently driven by the winds of the remaining B-stars of UCL.}
{
We investigate if the Lupus I molecular could have formed in a colliding flow, and in particular, how the kinematics of the cloud might have been influenced by the larger scale gas dynamics.
%We study how collisions
%of large-scale interstellar gas flows form and influence new dense clouds in the ISM.
% We study how large-scale interstellar gas flows collide with ambient gas and how this may relate to dense and star forming clouds in the ISM.
}
% {We study the origin of the Lupus~{\rm I} cloud in light of the interaction of the feedback driven shells in USco and UCL.}
% 
% 
{We performed APEX \cdot{} and \caot{} line observations of three distinct parts of Lupus~{\rm I} that provide kinematic information 
on the cloud at high angular and spectral resolution. We compare those results to the atomic hydrogen data
from the GASS \ion{H}{I} survey and our dust emission results presented in the previous paper. 
Based on the velocity information, we present a geometric model for the interaction zone between the USco shell and the UCL wind bubble.}
{We present evidence that the molecular gas of \lupus{} is tightly linked
to the atomic material of the USco shell. The CO emission in \lupus{} is found mainly at velocities between $\vlsr=3$--$6\kms$ which
 is in the same range as the \ion{H}{I} velocities. 
Thus, the molecular cloud is co-moving with the expanding USco atomic \ion{H}{I} shell.
The gas in the cloud shows a complex kinematic structure with several line-of-sight components that overlay each other.
%This could be a sign that the filament is composed of different velocity fibers.
The non-thermal velocity dispersion is in the transonic regime in all parts of the cloud and could be injected
by external compression. 
Our observations and the derived geometric model agree with a scenario where \lupus{} is located in the interaction zone between the USco shell and the UCL wind bubble. 
% Our geometric model shows that the observations agree with the assumption
% that \lupus{} is located between the shell and the wind bubble.
}
{%We conclude that the \lupus{} molecular cloud was formed through the collision process of the 
%USco shell with the UCL wind-bubble and is now expanding with the shell.
%This collision was the factor that enhanced local inhomogeneities
%in the USco shell at a position that favored the formation of \lupus.
 The kinematics observations are consistent with a scenario where the \lupus{} cloud formed via shell instabilities. The particular location of Lupus I between USco and UCL suggests that counter-pressure from the UCL wind bubble and pre-existing density enhancements, perhaps left over from the gas stream that formed the stellar subgroups, may have played a role in its formation.}

   \keywords{
		    radio lines: ISM --
		    ISM: clouds -- 
		    ISM: kinematics and dynamics --
		    ISM: molecules -- 
		    ISM: structure
% 		    Stars: formation --
%              Stars: protostars --
		    ISM: bubbles --
% 	     ISM: clouds --
% 	     ISM: dust, extinction --
	     individual objects: \object{Lupus~{\rm I}}
               } 
 
   \maketitle
%
%________________________________________________________________

\section{Introduction}

% In the current picture of the dynamic interstellar medium (ISM), 
Molecular cloud
formation is attributed to collisions of large-scale flows in the ISM 
% \citep[see review by][
% on 'Formation of Molecular Clouds and Global Conditions for Star Formation'
% and references within]{mc-formation-ppvi-review-dobbs-2014}. 
% Such flows can be driven
that can be driven by stellar feedback processes such as UV-radiation, winds, and supernovae \citep[see review by][]{mc-formation-ppvi-review-dobbs-2014}.
At the interface of the colliding flows, compression, cooling, and fragmentation of the diffuse atomic medium produces
cold sheets and filaments that later may become molecular and self-gravitating.
Realizations of this general concept include expanding (super-) shells, 
where an expanding hot bubble collides with the ambient medium 
\citep[e.g.][]{superbubbles-vishniac-krause-2013}
and direct collisions of stellar winds 
\citep[e.g.][]{clump-formation-winds-calderon-2016}.

%  and dominate the
% appearance of the ISM as observed today \citep[see review by][]{filaments-sf-ppvi-review-andre-2014}.
% In this picture 
% also provides an explanation for 
% the fast formation (and dispersion) of molecular clouds and the often simultaneous onset 
% of star formation within \citep[see][]{rapid-formation-mcs-hartmann-2001,mc-formation-stars-vazquez-semadeni-2007,
% clumps-mhd-mc-formation-banerjee-2009,filaments-in-simulations-mc-formation-gomez-vazquez-semadeni-2014} appears plausible.
% 
Such a  process might take place
% large-scale flow is an expanding shell or super-shell around e.g. an OB-association or in general
% driven by multiple stellar feedback of a star cluster or association \citep[see review by][]{supershells-mc-formation-review-dawson-2013}.
% Molecular clouds may then either form inside the wall of such a shell \citep{mcs-in-supershells-dawson-nov-2011} or at the interface
% region when two such shells collide with each other. The latter has been recently investigated by \cite{gmc-formation-colliding-shells-dawson-2015}
% for a young giant molecular cloud (GMC) at the interface of two colliding super-shells. From the comparison
% of CO observations with high-resolution 3D hydrodynamical simulations they found that the GMC assembled into its current
% form by the action of the shells.
% 
between the Upper-Scorpius (USco) \ion{H}{I} shell and the Upper Centaurus-Lupus (UCL) wind bubble located in the 
Scorpius-Centaurus OB-association \citep[Sco-Cen]{sco-cen-blaauw-1964,sco-cen-hipparcos-dezeeuw-1999,usco-full-population-preibisch-2002,sco-cen-sfhb-preibisch-mamajek-2008}. 
This is the closest site of recent massive star formation to us, at a distance of 110-140 pc only.
% and it consists of three sub-groups with different ages and well known stellar populations down to $2\,M_\odot$
% \citep{sco-cen-stars-hipparcos-debruijne-1999}. The oldest one is the the Upper Centaurus-Lupus (UCL) sub-group with an age of $\sim 17$~Myr
% harboring 66 B-stars. With an age of $\sim 15$~Myr the Lower Centaurus Crux (LCC) sub-group is somewhat younger and contains 42 B-stars.
% The youngest sub-group is Upper-Scorpius (USco) with an age of $\sim 5$~Myr and consisting of 49 B-stars.
% 
% The ionizing radiation, the winds, and the
% supernova explosions of the numerous massive stars in Sco-Cen have strongly shaped the
% surrounding interstellar medium.
% 
These shells presumably were created by the feedback of the numerous massive stars in Sco-Cen.
In-between those two expanding structures there is the \lupus{} cloud which represents
% is found between the USco and the UCL sub-groups at a location where
% the expanding USco shell appears to interact with a bubble currently powered by the winds of the remaining
% B-stars of UCL. 
% With its close distance of $\sim150$\,pc Lupus~{\rm I} represents 
a good candidate where we can study how such
a collision process forms and influences new dense clouds in the ISM.\\
The distance of the Lupus cloud complex is %is discussed in detail in . The best estimate is 
about 140 pc, as discussed in detail by \citet{lupus-clouds-sfhb}. 
Additional details about the region and the \lupus{} cloud can be found in \cite{lupus-paper-gaczkowski-2015} (Paper~I hereafter).\\ %,
%the first paper in this series.
Molecular line observations of \lupus{} with different CO isotopes and other high-density tracers 
have been performed by several groups over the past two decades. Most
of them have spatial resolutions of several arcminutes, i.e. $\sim0.1$\,pc at the distance of \lupus{} of $\sim140$\,pc.
\cite{lupus-clouds-13co(1-0)-Tachihara-1996} observed the entire \lupus{}
cloud in the \cdo(1--0). %line with the 4\,m telescope at Nagoya University.
%Their observations had a half power beam width (HPBW) of $2.7\arcmin$ and they chose a grid
%spacing of $8\arcmin$. 
They estimated the cloud mass to be $\sim1200\modot$
and discovered the molecular outflow of IRAS source 15398--3359.
\cite{lupus-clouds-c18o-Hara-1999} have studied 12 cores
in \lupus{} in the \cao(1--0) line. The spatial resolution of the NANTEN telescope is $2.7\arcmin$, while their grid 
spacing was $2\arcmin$. They derived column densities, temperatures,
and sizes for the cores, and estimated their total mass to be $152\modot$. 
 This represents 46\% of their total \cao(1--0) emitting gas mass.
\cite{lupus-clouds-13co-cores-vilas-boas-2000} found 
15(14) condensations in $\rm C^{18}O$(1--0)($\rm^{13}CO$(1--0)) %with
%the 15\,m Swedish-ESO Submillimeter
%telescope (SEST). Their angular resolution was $48\arcsec$.
% They conclude recent star formation activity in the cloud.
%They 
highlighting the recent star formation activity in \lupus.
A large-scale $^{12}\mathrm{CO}$(1--0) survey of \lupus{}
has been performed by \cite{lupus-clouds-12co-Tachihara-2001}.  
%They mapped the cloud with a grid-spacing of $8\arcmin$ and a HPBW of $2.7\arcmin$ with NANTEN.
%They reported 
They find a velocity gradient along the long axis of \lupus{} from $4\kms$ in the south
to $6\kms$ in the north.
\cite{lupus-clouds-13co(2-1)-Tothill-2009} analyzed \lupus{} in both \hbox{CO(4--3)} and \cdot{}. 
%They used the Antarctic~Submillimeter
%Telescope and Remote Observatory (AST/RO) that allowed an angular resolution of $1.7\arcmin$ (CO)
%and $3.3\arcmin$ (\cdo), respectively. 
They found several possible signs of interaction between
\lupus{} and the USco \ion{H}{I} shell.
\cite{lupus-clouds-spec-cores-Benedettini-2012} found eight dense cores in Lupus~{\rm I} using high-density
molecular tracers at 3 and 12\,mm. % with the Mopra telescope.
They reported the presence of velocity gradients in the cloud and several
velocity components along the line-of-sight.
Finally, \cite{lupus-filaments-herschel-benedettini-2015} observed
the southern part of the main filament in CS(2--1) and they found several
sub-filaments in velocity space that correspond to sub-filaments
they have identified in their \herschel{} column density maps.

In our dust analysis presented in Paper~I we found a double peaked PDF which suggests 
%The results of our dust analysis presented in Paper~I 
a scenario of external compression of \lupus{} 
from USco and/or UCL.\\
%To further investigate this aspect, we proposed follow-up molecular line observations to study the kinematics of the gas in \lupus.
% Therefore, we proposed follow-up molecular line observations to further investigate this aspect because
% molecular lines are well suited to obtain kinematic information about the gas in \lupus{}.
%Systematic differences in velocities within the cloud or between one side of the cloud and the other
%could reveal signs of interaction between the cloud and the surroundings, 
%i.e. the HI shell of USco and the wind-blown bubble of UCL.
%If there were a sign of compression on one side of the cloud or the other it should be imprinted
%in the CO gas. One would expect to see differences in the line profiles (e.g. linewidth)
%between the edges of the cloud and the central parts.
In this work, we follow up on this suggestion, and present \cdot{} and \caot{}
observations of three distinct regions within the \lupus{} cloud with a high angular and spectral resolution of $\approx30\arcsec$
and $\approx0.1\kms$, respectively.
% 
% These regions will be called cut~A, cut~B, and cut~C. They are marked in the right panel of
% Figure~\ref{img:laboca-apex1-obs-setup}
% on top of the LABOCA continuum map. Cuts A and C cross the main filament perpendicularly, cut~B is more
% parallel to it but goes through the prominent class~0 protostar IRAS~15398--3359 
% in the center of the filament.
We present the analysis of the observational data and the comparison to the available \ion{H}{I} data.
From those we create a geometrical model for the interaction zone of shell and bubble. We show that the collision scenario is a plausible explanation of the observational data and the creation of \lupus ~with other factors likely having played a role as well. 
%
%From those we create a geometrical model for the interaction zone of shell and bubble and show
%that the collision scenario is a plausible explanation of the observational data and the creation of \lupus.
% The analysis of the CO and \ion{H}{I} data was done by myself. The \ion{H}{I} data was provided by our collaborator Daniel Kr\"oll but analyzed by myself. We am also 
%using the results from D.\,Kr\"oll's model fit 
%of the expanding USco shell to the \ion{H}{I} data.
% using the results from D.~Kr\"oll's fit of an expanding homogeneous spherical shell model 
% to the \ion{H}{I} data of the USco region.

%__________________________________________________________________

\section{Observations and data reduction}
\label{sec:observations}

\subsection{APEX CO observations}
\label{sec:data-reduction-co}

\begin{figure}[htb]
\centering
\includegraphics[height=0.41\textheight, keepaspectratio]{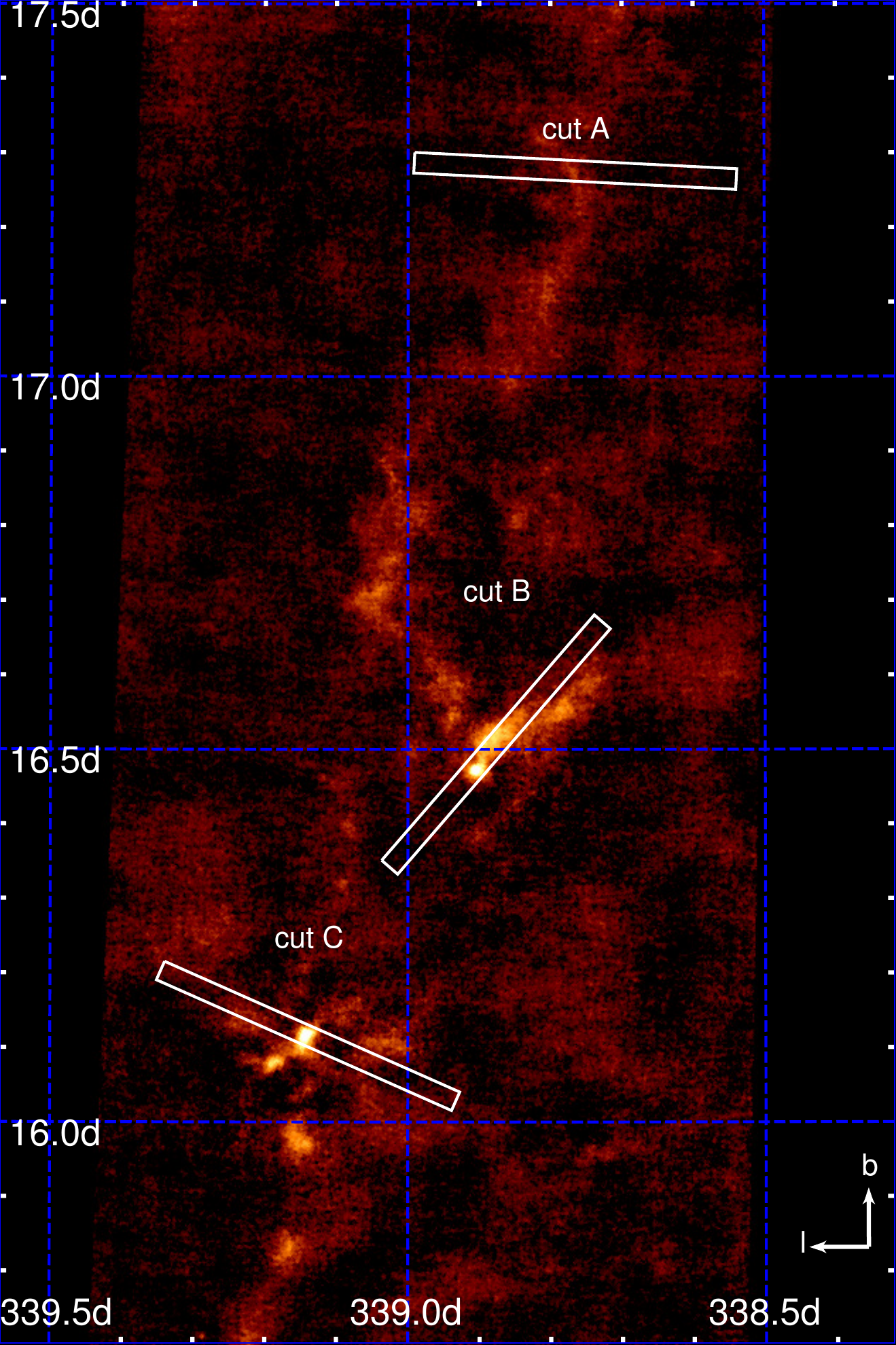}
 \caption{Setup of the APEX-1 \cdot{} and \caot{} observations of \lupus. Along each of the three cuts A, B, and C
  an on-the-fly map was obtained simultaneously in both lines. The size of each map is marked with a white box.
  The background image is the LABOCA continuum map (see Paper~I). The Galactic coordinate system is
 indicated.}
 \label{img:laboca-apex1-obs-setup}
\end{figure}

 The CO line observations of \lupus{} were performed with the APEX telescope \citep{apex} using
the APEX-1 receiver of the Swedish Heterodyne Facility Instrument \citep[SHeFI;][]{apex-shefi}.
The observations were  carried out on 18th and 20th August 2014 (PI: B.~Gaczkowski; Program ID: \hbox{093.F-9311(A)}).
\lupus{} was observed simultaneously in the \cdot{} (rest frequency $\nu^{13}_0=220.398677\,\mathrm{GHz}$) and \caot{} ($\nu^{18}_0=219.560357\,\mathrm{GHz}$)
line.  %at three different cuts across the cloud
 Three scans across the cloud were performed, as shown in Figure~\ref{img:laboca-apex1-obs-setup}. %in Fig.1. 
These were chosen based on our previous dust analysis 
to obtain representative samples of the conditions along the whole filament.
%The location of the observed regions is shown in Figure~\ref{img:laboca-apex1-obs-setup}.
The observing mode was {\it ``on-the-fly''} and the map parameters for each cut are listed in Table~\ref{tbl:apex-1-maps}.
The scanning direction was along the long side of a cut. The velocity resolution of 
the spectra is $\Delta v = 0.1\kms$ and the rms noise ranges between $\approx0.05$--0.1\,K for both lines.
The perceptible water vapor (PWV) during the observations was between 1.5 and 2\,mm. Each cut contains about 620 spectra.

The data reduction was done using the {\tt CLASS} package of the IRAM {\tt GILDAS} software\footnote{\url{http://www.iram.fr/IRAMFR/GILDAS}}.
The antenna temperature $T_\mathrm{A}$ was converted into a main-beam temperature $T_\mathrm{mb}=T_\mathrm{A}/\eta$, adopting
a main-beam efficiency of $\eta=0.75$ for APEX-1\footnote{\url{http://www.apex-telescope.org/telescope/efficiency}}.
All spectra were baseline subtracted. %{\bf Further, the \cao{} spectra were shifted into the same velocity range as the \cdo.}
% First, with a 0th order baseline and where 
% needed also with one of high polynomial
% order. 
Finally, for each line and cut, a map (or {\tt lmv} cube) was produced 
with the {\tt XY\_MAP} task which convolves the data with a Gaussian
of one third of the beam. The final angular resolution is $30.1\arcsec$ and the pixel size is $14.3\arcsec$.
% With the task {\tt GILDAS\_FITS} the {\tt lmv} cube can easily be converted into a {\tt FITS} file for 
% further analysis in
% {\tt IDL} or inspection with {\tt ds9}.

\begin{table*}[htb]
 \centering
 \caption[Details of the molecular line maps obtained for \lupus{} with APEX-1]{Details of the molecular line maps obtained for \lupus{} with APEX-1. The first column gives the name of the cut,
the second and third column give the x and y size of the whole map. 
% % 
Column four gives the length of the cut, columns five and six the
 position of the center of the map, and the last column gives the number of pixels that contain a spectrum.
% % 
These parameters are valid for both the \cdo{} and \cao{} maps of one cut. The angular resolution of the maps is $30.1\arcsec$ and the pixel size $14.3\arcsec$.
% % 
 The three cuts share a common OFF-position at $\rm RA=15$:44:23.605 and $\rm Dec=-$33:39:00.68.}
% % 
 \begin{tabular}{c c c c c c c}
 \hline\hline
 \noalign{\smallskip}
  Cut & Map size X  & Map size Y &  Length & Center RA$_{\rm J2000}$   & Center Dec$_{\rm J2000}$  & No. of spectra \\
%    $\rm [cm^2\,g^{-1}]$ & \multicolumn{4}{c}{$[\times10^{21}\,\mathrm{cm^{-2}}]$} & $[\rm M_\odot]$\\
 \noalign{\smallskip}
 \hline
 \noalign{\smallskip}
   A  &  $21\arcmin$ &  $18\arcmin$  &   $26\arcmin$  &  15:40:05.35  & $-$33:36:44.4 & 617 \\    
   B  &  $26\arcmin$ &  $6\arcmin$   &  $24\arcmin$   & 15:42:49.80  & $-$34:08:37.8 & 618 \\    
   C  &  $13\arcmin$ &  $24\arcmin$ &   $26\arcmin$  & 15:44:57.60  & $-$34:17:13.3 & 620 \\    
   \noalign{\smallskip}
 \hline
 \end{tabular}
\label{tbl:apex-1-maps}
\end{table*}

\subsection{GASS \ion{H}{I} data}
\label{sec:data-reduction-hi}

%The atomic hydrogen data are part of the second data release
%of the GASS H\,{\rm I} survey
 The atomic hydrogen data of the entire Sco-Cen region
used in our study  
are part of the second data release of the
GASS H I survey  \citep[][]{gass-hi-kalberla-2010}. 
%We retrieved them from the GASS data archive.\footnote{\url{https://www.astro.uni-bonn.de/hisurvey/gass/}}
%They have a spatial resolution of $14.4\arcmin$ which corresponds to a length scale 
%of 0.6\,pc at the distance of \lupus. The velocity resolution is $1\kms$.
%The details on this data-set are presented and analysed in a forthcoming publication. %Krause et al. 2016, submitted).}
%We retrieved them from the
%GASS data archive. 
This has a spatial resolution of 16.2$'$ which
corresponds to a length scale of 0.7 pc at the distance of Lupus
I. The velocity resolution is 1 km s$^{-1}$ and the rms noise
is 55 mK.\\
In order to characterise the HI shell around Upper Scorpius (the Upper Scorpius loop) and relate it to the Lupus I cloud, which forms part of the filaments 
of the shell rim, we fitted an expanding homogeneous spherical shell
to the HI data. In addition we used Na I absorption data  in the foreground of stars
with know distances from Hipparcos measurements, using the same shell model as for the HI data. This follows an analysis approach presented elsewhere \citep{Poppeletal2010, Welshetal2010}. % (P\"oppel+ 2010,
%Astronomy and Astrophysics, Volume 512, id.A83, 26 pp., Welsh et al,
%Astronomy and Astrophysics, Volume 510, id.A54, 2010). 
The parameters of our expanding-shell model are:
\begin{itemize}
\item distance to the center of the shell $D_0$
\item galactic longitude $l_0$ of the shell center 
\item galactic latitude $b_0$ of the shell center 
\item  radius of the inner edge of the shell $r_{in}$ 
\item thickness of the shell $\Delta$r
\item density of hydrogen atoms $n_H$ in the shell assumed to be spatially constant
\item expansion velocity of the shell $V_{\rm exp}$
\item bulk velocity of interstellar medium in the Sco-Cen region $V_0$.
\end{itemize}
The distance to the shell centre $D_0$ was determined from a
multi-parameter fit to the absorption data, yielding \\
$D_0 = 138\pm26$ pc.  
Here,  the uncertainties have been propagated from the uncertainties in the
distance measurements of the stars.\\
For the fit to the HI data, we transformed this model from 3D Cartesian
space $(x,y,z)$  into $(l,b,v)$ space, added the HI Milky
Way model of \citet{Kalberlaetal2007} as a large-scale background, 
%Kalberla et al. (Astronomy and Astrophysics, Volume 469,
%Issue 2, July II 2007, pp.511-527),
and optimised model parameters fitting to the HI data by a $\chi^2$ minimization. %Automated fit
%methods proved to be
%inefficient for the given data structure and model. We therefore
%varied the parameters
%individually to determine the global minimum in $\chi^2$. In this way we were able to determine a 
The set of best fitting parameters are:\\
$l_0 =347\pm 0.5 \deg, b_0 = 25\pm 0.5 \deg, r_{\rm in} = (12.3 \pm 0.7) \deg, \\
\Delta\,r= 2.7 \pm 0.3 \deg, n_{\rm H} = 6 \pm 2 {\rm cm^{-3}}, {\rm V_{exp}} = (7\pm 1) {\rm km/s}, \\
{\rm V_0} = 8 \pm 1 {\rm km/s}$.\\
Uncertainties are quoted as 1 $\sigma$. % errors of the linear $\chi^2$ distributions.\\
The total mass of hydrogen in the shell is 11\,000 $\pm$ 6000 M$_\odot$. This is similar to the value determined by \citet{Poppeletal2010} from lower quality data.\\
The mass determination has additional significant systematic uncertainties, as the shell is
not detected on the far side. On the near side, a spherical shape is plausibly seen, but the galactic foreground adds uncertainties. %complicates the analysis.

\section{Results}
\label{sec:methods}

%\subsection{Spectra}
%\label{sec:lupus-co-data-spectra}
%Throughout the following parts of this article we will use the simplified notation \cdo{} for \cdot{} and \cao{} for \caot{}, respectively.
The individual spectra are complex, with a variety of different line profiles.
%There are spectra with one, two or even three components. Many have asymmetric
%and non-Gaussian profiles. %In the following subsections we present the description of the velocity channel
%maps and the average resulting spectra through each cut. 
%\subsubsection{Velocity channel maps}
%\label{sec:velo-maps}
%Velocity channel maps for each cut and molecular line are shown in Figures~\ref{img:cutA-velochan} to \ref{img:cutC-velochan}.
 The velocity of the emitting gas ranges from $\vlsr=3.2$ to $6.8\kms$. % and the channel resolution is $\approx0.1\kms$.
The maps have been convolved with a Gaussian of two pixels or approximately one beam. 
All observed spectra can be found in Appendix~\ref{sec:appendix-lte} in Figures~\ref{img:cutA_C18O_Gauss}-\ref{img:cutC_13CO_Gauss}.  
%in the online version of this article. \\
%\paragraph{Cut A}
The main \cdo{} emission in {\bf cut~A} is at velocities of $\vlsr=5.5$--$6.2\kms$ with the highest
intensities between 5.9 and 6.1\kms. The emission is concentrated to the center of the cut and
extends to about $10\arcmin$
around the center in the channels with highest intensities. %At $\vlsr=6.4\kms$ the spatial
%extent of the emission is decreased to $\sim2\arcmin$ around the center.
%Between $\vlsr=3.4$--$2.8\kms$ a narrow emission patch appears north-east of the center.
%Its peak intensity is about one third ($\approx 2 $\,K) of the maximum intensity of the entire
%cut (at 6\kms).
%This suggests the presence of another gas component in the cloud
%with a different velocity.
%The \cao{} emission appears only in the channels \hbox{$\vlsr=5.1$--$6.3\kms$} with a peak intensity of
%$\approx2$\,K at 5.9\kms. 
Detectable \cao{} emission region is limited to about $2\arcmin$
around the very center of the cut.\\
%\paragraph{Cut B}
The highest intensities of the three cuts are found in {\bf cut~B}. %Most of the emission is at
%$\vlsr=4.7$--$5.8\kms$ with peak intensities of up to $\approx7.5$\,K between 4.9 and 5.1\kms.
%Between $\vlsr=4.2$--$6\kms$ a second emission region at the south-eastern edge
%is apparent and is spatially separated from the center. This second region
%has its emission peak in the 4.8\kms{} channel.
The bulk of the \cao{} emission lies %between $\vlsr=4.9$--$5.4\kms$ and hence in 
at the same velocity range
as the \cdo{} emission. It is concentrated on the center of the cut and no emission is found at either edge.\\
%\paragraph{Cut C}
Within {\bf cut~C} several spatially distinct
emission regions can be seen in \cdo{} and \cao.
%The \cdo{} peak intensities are lower than those of cut~B.
%the emission is distributed more evenly throughout the cut.
%Between $\vlsr=3.3$--$6.1\kms$ \cdo{} emission can
%be seen in at least one spot of the cut.
%The highest intensity of $\approx6$\,K is seen at a velocity of $\vlsr=5.1\kms$
%at a location north-east of the center. A second intensity peak of $\approx5$\,K at 4.2\kms{}
%can be found in the center of the cut.
In contrast to the other two cuts, the \cdo{} and \cao{} emission peaks do not overlap,
but are located at opposite ends of the cut.
\cao{} emission is found at three positions within the cut.
%For $\vlsr=4.2$--$5.0\kms$ there is emission in the center.
%For $\vlsr=4.8$--$5.3\kms$ emission appears south-west of the center where
%it also has its peak with an intensity of about 4\,K.
%At the position where \cdo{} has its highest emission,
%\cao{} emission is in the velocity range of $\vlsr=5.1$--$5.5\kms$ with a peak intensity
%of about 3\,K at 5.3\kms.
%The different velocity components in Section~\ref{sec:pv-diagrams}.
%
%\subsubsection{Average Spectra through the cuts}

\begin{figure}[htb]
 \centering
 \includegraphics[width=0.52\textwidth, keepaspectratio]{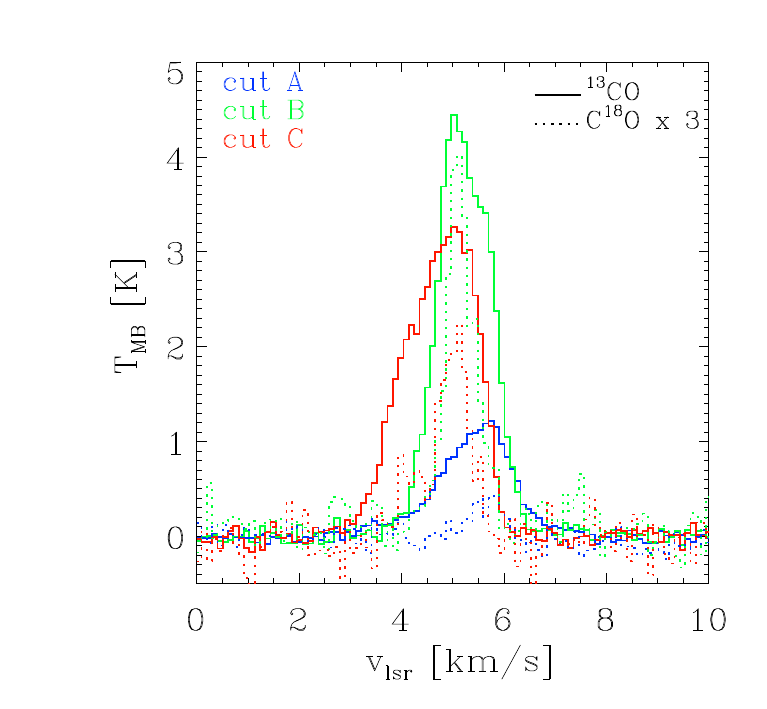}
 \caption[Average \cdo{} and \cao{} spectra of the three cuts]{Average \cdo{} and \cao{} spectra of the three cuts. 
The integrated intensities (for $\vlsr=0$--$10\kms$) of the \cdo{} lines are 
1.96, 5.69, and 5.55\kkms{} for cut~A, B, and~C, respectively. Those of the \cao{} lines are 1.22, and 0.50\kkms{} for 
cut~B and cut~C, respectively.}
\label{img:average-spectra}
\end{figure}

 The average \cdo{} and \cao{} spectrum has been then computed for each
cut and they are shown in Figure~\ref{img:average-spectra}.
The solid lines represent the average \cdo{} spectrum and the dotted lines the average
 \cao{} spectrum multiplied by three. 
Cut~A shows the lowest intensities of all three cuts in both lines. The averaged \cdo{} line is very 
broad and asymmetric towards velocities lower
than the velocity of the peak, i.e. $\vlsr=5.75\kms$. The peak intensity of the \cdo{} line
is 1.22\,K.
The \cdo{} line of cut~B is the most narrow and has
the highest peak intensity of all three cuts with a value of 4.44\,K at $\vlsr=5.02\kms$.
%Another component might be present in this cut
%since a second peak
%appears around 5.5\kms. 
For cut~C the \cdo{} line also peaks at $\vlsr=5.02\kms$ with a peak
intensity of 3.26\,K. 
%Also here the line might consist of two components. A second peak is seen around 4.25\kms.
The averaged \cao{} lines of cut~B and cut~C both peak at $\vlsr=5.13\kms$ and 
have a peak intensity of 1.34 and 0.74\,K, respectively.
%Both lines seem to support the idea of another component 
% in both cuts as there might be a second peak in each line
%roughly at the same position as in the \cdo{} spectra. But if there is a second component it is much weaker than the main one.
%I will further investigate this in Sections~\ref{sec:pv-diagrams} and~\ref{sec:velo-components}.
% The averaged spectra suggest the presence of a possible large-scale 
% velocity gradient running along the filament's
% major axis from north-west (cut~A) to south-east (cut~C) of almost 1\kms because
% the northern part of \lupus{} has on average higher \vlsr{} velocities than the central and
% southern part. However, the behavior of this gradient throughout the cloud cannot 
% be determined from these three cuts alone.
% and peak positions may shift across one cut.

\begin{figure*}[!htb]
    \centering
%     \subfigure{
\includegraphics[width=9cm]{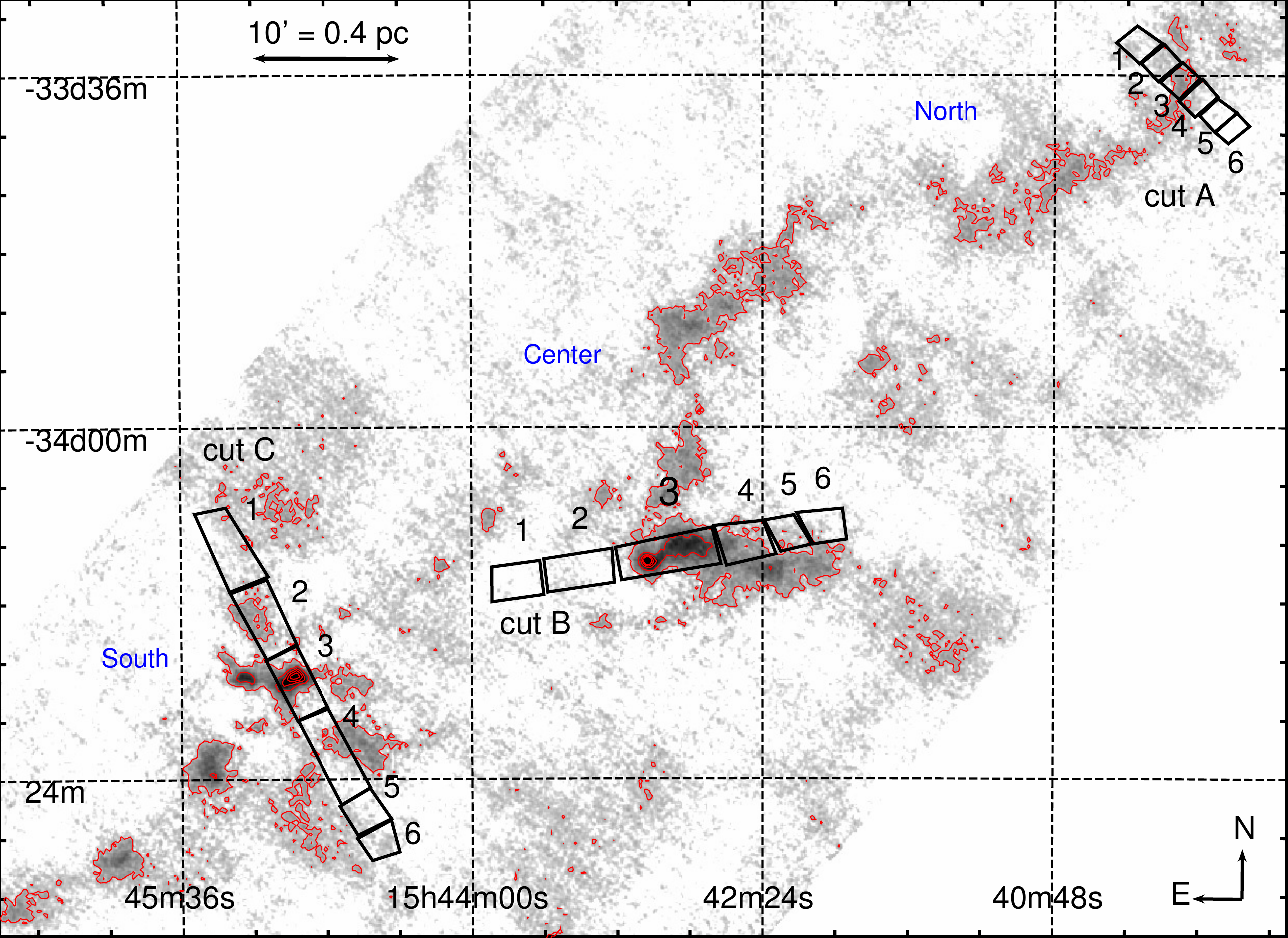}
\includegraphics[height=7cm]{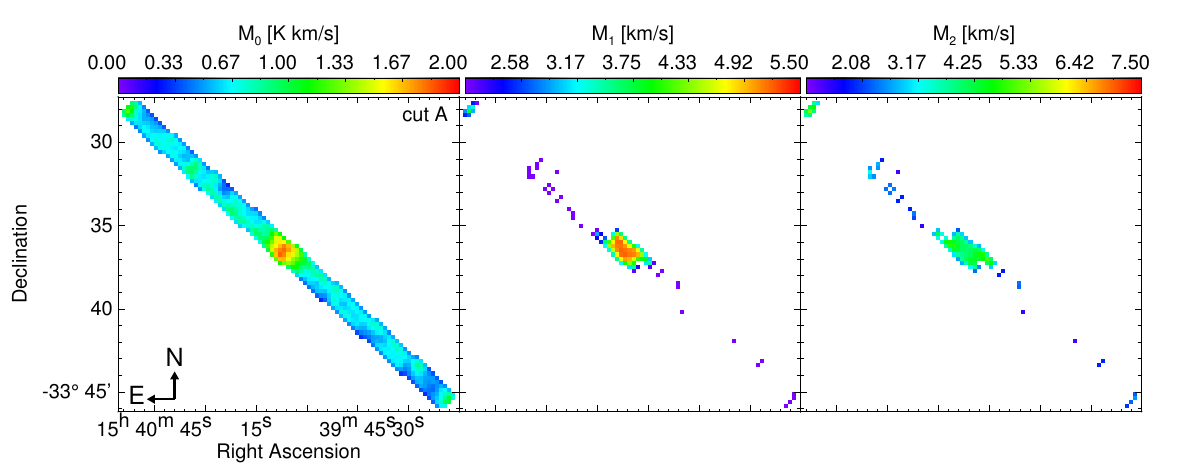}
%\includegraphics[height=7cm]{Lupus_I_cutA_C18O_moments_boxes.pdf}
% }
% 
%     \subfigure{
\includegraphics[height=7cm]{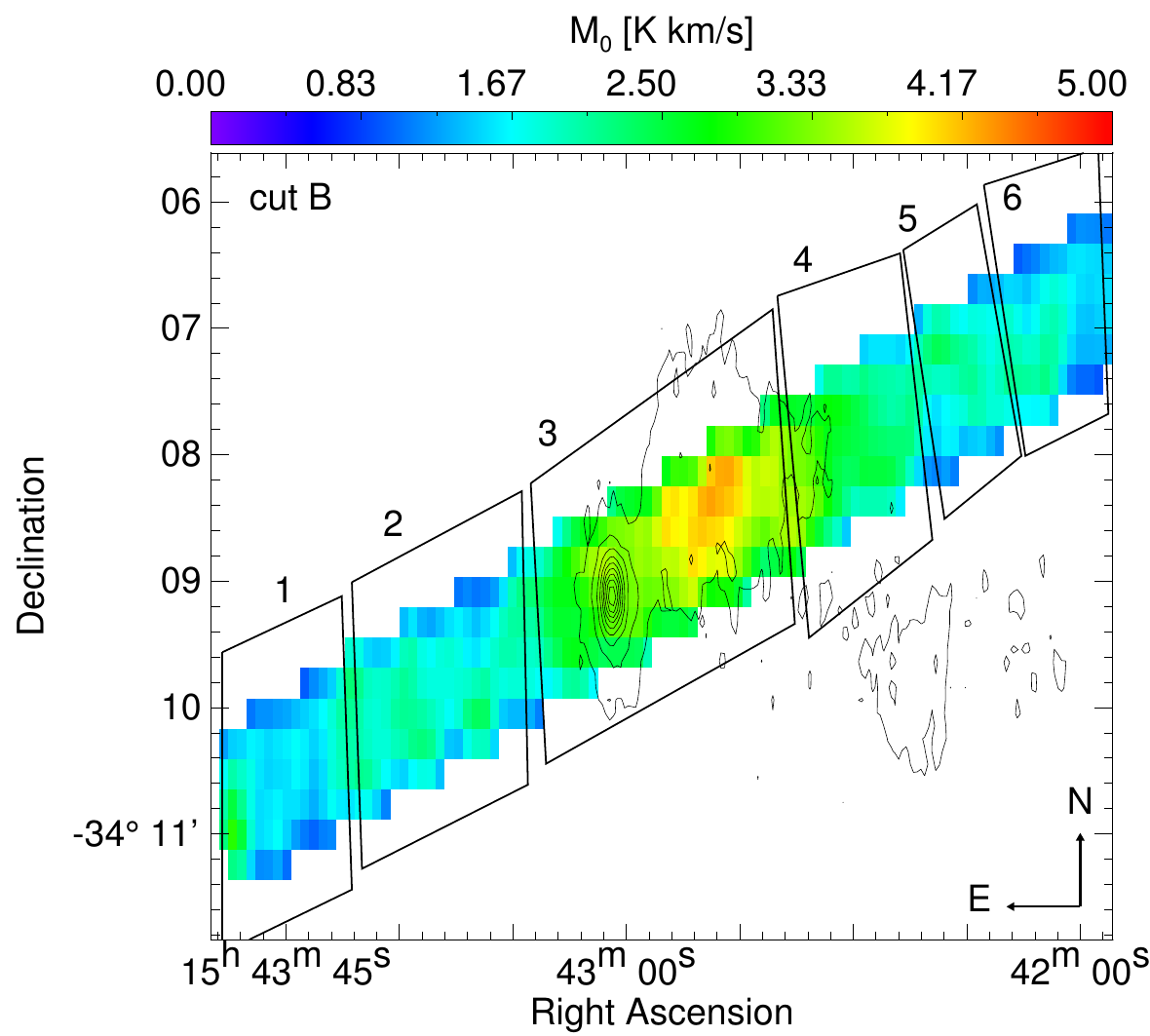}
%\includegraphics[height=7cm]{Lupus_I_cutB_C18O_moments_boxes.pdf}
% }
%     \subfigure{
\includegraphics[height=7cm]{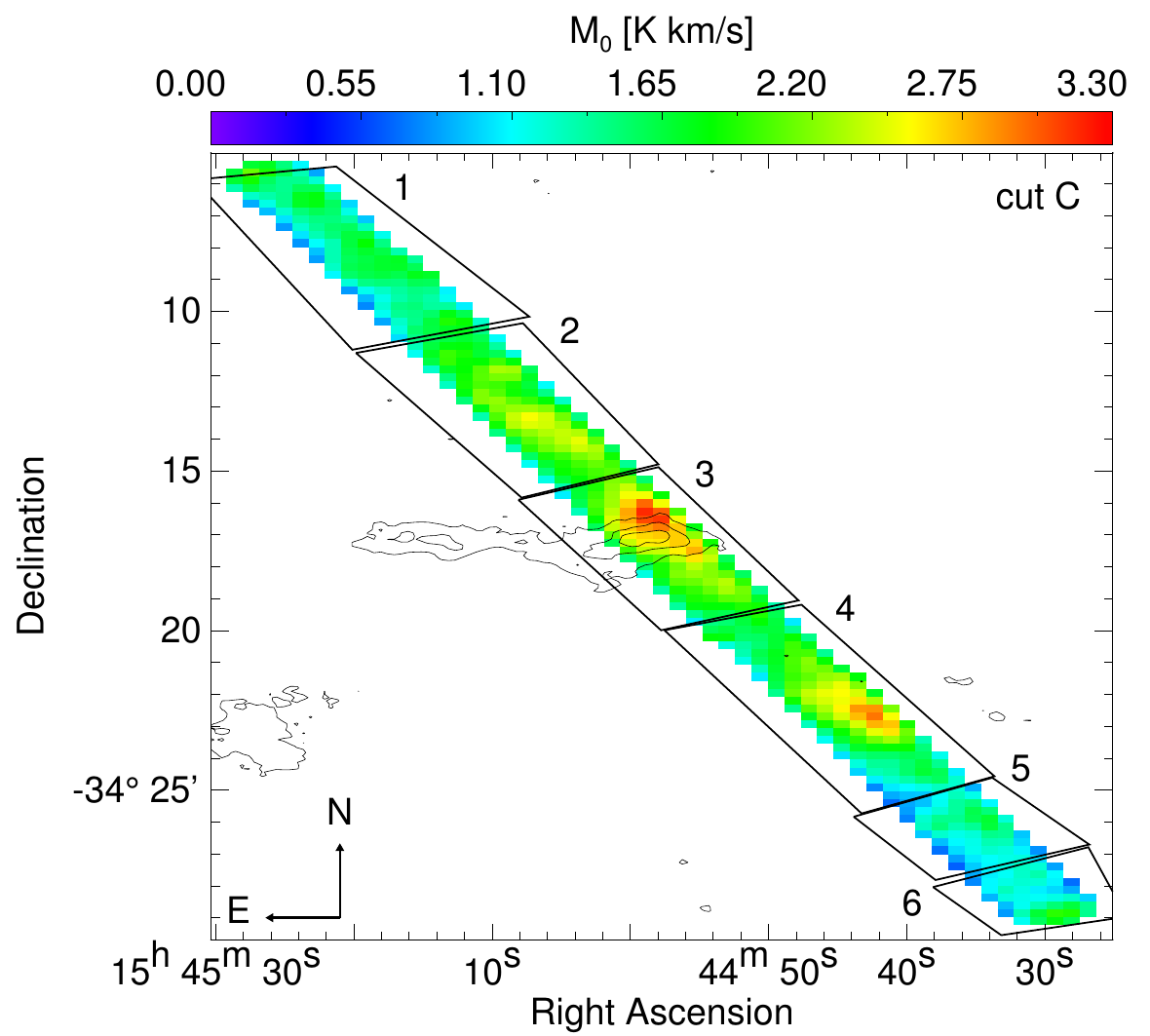}
%\includegraphics[height=7cm]{Lupus_I_cutC_C18O_moments_boxes.pdf}
% }
%     \subfigure{
% }
\caption[Division of each cut into boxes]{{\it Top Left:} Dust
column density map from LABOCA with red contour levels of 0.3, 1, 1.7, 2.3, and \nhz{3}{22}. Integrated \caot{} intensity maps of cut~A, B, and C ({\it Top Right } and {\it Bottom} figures, respectively). The six boxes into which each
cut was divided are shown, as well as the contours of the LABOCA $870\,\mu$m continuum emission.
Levels are from 10\%--100\% of the peak intensity of 1.37\,Jy/beam.}
\label{img:cuts-boxes}
\end{figure*}

\section{Analysis}
\label{sec:analysis}
%The analysis proceeds with the following calculations of
We first calculate the optical depth, excitation temperature, and column density under the assumption of local thermodynamical equilibrium (LTE). A multi-gaussian fit  determines the different velocity components of the observed spectra. We finally search for a non-thermal contribution to the spectra.
\subsection{LTE density analysis}
\label{sec:lupus-lte-analysis}
 Spectra with a $\rm S/N > 4$ were selected to compute optical depth, excitation temperature, and column density. 
This assures that both lines are clearly detected.
%  and reduces possible confusion.
%Constants of the \cdo{} and \cao{} line that are needed for these calculations are listed in Table~\ref{tbl:co-lines}. 
For our analysis we assume LTE conditions: The excitation temperature \tex{} is equal to the kinetic temperature of the gas and the same
for all transitions and both CO isotopologues. This assumption should hold when the temperature distribution along the line-of-sight is uniform. \\
% Typically, this is not the case and will thus lead to errors that have to be taken into account. 
The equations and constants used for the calculations of the excitation temperature and the column density 
in this section can be found in Appendix~\ref{sec:appendix-lte}. \\
%\subsubsection{Optical depth}
The $\mathrm{H_2}$ column density map calculated from dust emission (see Paper~I) showed
that almost all of \lupus{} lies below \nhz{1}{23}. 
% % 
Therefore, \cao{} is optically thin \citep{co-observations-w43-carlhoff-2013}. 
% % 
The optical depth of the \cdo{} gas is computed %using constant abundance ratio of \cdo{} to \cao{} %one can estimate 
\citep[see e.g.][]{co-observations-column-densities-high-av-myers-1983, co-observations-column-densities-ysos-taurus-ladd-1998} using  the relation: %for the optical depths %of both lines are connected via
\begin{equation}
\tau_{13} =\Gamma\times\tau_{18}
\end{equation}
with $\tau_{13}=\tau(\cdo)$, $\tau_{18}=\tau(\cao)$ and 
$\Gamma= [\cdo]/[\cao]$ is the relative abundance ratio of the given isotopologues. Here we adopt a constant abundance ratio of $\Gamma=7.3$
 \citep[][local ISM]{abundances-isotopologues-ism-wilson-rood-1994}. Then the optical depth $\tau_{13}$ can be approximated by the intensity ratio of the lines
\begin{equation}
\frac{T^{13}_\mathrm{mb}(v)}{T^{18}_\mathrm{mb}(v)} = \frac{1-\exp({-\tau_{13}})}{1-\exp({-\tau_{13}/\Gamma})}
\label{eq:tau13_18}
\end{equation}
with $T^{13}_\mathrm{mb}(v)$ and $T^{18}_\mathrm{mb}(v)$ the main-beam temperature of the corresponding line at velocity
channel $v$. In principle this relation is valid for every velocity channel. Here, we chose $T^{18}_\mathrm{mb}(v)=T^{18}_\mathrm{peak}$
the peak brightness temperature of the \cao{} line. Then $T^{13}_\mathrm{mb}(v)=T^{13}_\mathrm{peak}$ is the corresponding brightness temperature of
\cdo{} at the velocity channel of the \cao{} peak intensity. 
% we will use the notation $T^{13}_\mathrm{peak}$ for this value.
% The resulting equation for the optical depth of \taud{} is transcendent and can only
% be solved numerically. It is illustrated in the top panel of Figure~\ref{img:tau}. This method breaks down when the line ratio approaches
% unity which we also find at a few pixels of cut~C. There the \cao{} line is very strong compared to the rest of the cut and the resulting \taud{} 
% is even ten times higher than in the rest of the cut. 
The resulting optical depth maps for \cao{} are shown in the left panels of Figure~\ref{img:lte-results}. The optical depth $\tau_{\rm 18}$ is in the range 0.3 to 1 for most of the cloud, with isolated peaks up to about 1.5. 
\\
The {\bf excitation temperatures} are computed %from the optical depth $\tau$ 
using the optically thin \cao{} line (equations \ref{eq:tex} and \ref{eq:j(t)} in the appendix).  
%, one can calculate the corresponding 
%via Equation~\ref{eq:tex}.
%\begin{equation}
%T_\mathrm{ex}(\tau)=\frac{T^{18}_\nu}{\ln\left\{1+\left[  \frac{T^{18}_\mathrm{peak}}{T^{18}_\nu}\left(1-\exp(-\taua)\right)^{-1}+J(T_\mathrm{bg})/T^{18}_\nu \right]^{-1}\right\}}
%\label{eq:tex}
%\end{equation}
% 
%where
%\begin{equation}
%J(T)\equiv\frac{c^2}{2k_B\nu^2}\,B_\nu(T)=\rm T_\nu\,[\exp(\rm T_\nu/ T)-1]^{-1}
%\label{eq:j(t)}
%\end{equation}
%$\rm T_\nu=\frac{h\nu_0}{k_B}$, and $T_\mathrm{bg}=2.73\,\mathrm{K}$ the temperature of the cosmic
%microwave background (CMB). 
%After determining the optical depth, one can calculate the corresponding excitation temperature
%via Equation~\ref{eq:tex}.
% by reorganizing Equation~\ref{eq:Ionofftemp} to
% 
%We used the optically thin \cao{} line.  %for the calculation of \tex{} because it is more reliable to use the optically thin line.
% Here we only include the CMB, neglecting contributions from infrared dust emission.
% The behavior of \tex{} in dependence of $\tau$ is shown
% in Figure~\ref{img:tau}. \tex{} becomes independent of \taua{} for optical depths of more than $\approx4$--5.
% This is expected as one probes just the outermost layer of a cloud once it is dense enough
% and becomes optically thick. 
The resulting \tex{} maps are displayed in the middle panels of Figure~\ref{img:lte-results}. 
The mean excitation temperatures for cuts A, B, and C are 16.8, 14.7, and 14.7\,K. % for 
%cuts A,B,C correspond to mean excitation temperatures 
%cut~A, cut~B, and cut~C, respectively.
%These values are 
%5.5, 3.6, and 
%6.2\,K lower than these. 
 %higher than the corresponding mean \tex. 
 %{\bf The dust temperature of cut~A is
%temperatures, since cuts~B and~C have similar dust temperatures but very different
%excitation temperatures.}\\
Our excitation temperatures calculated in our analysis are comparable to previous findings from different
CO and molecular observations of \lupus. 
With Equation~\ref{eq:nco}  we compute the {\bf \cao{}
 column density},  %using Equation~\ref{eq:nco}.
% 
% 
%Line constants are listed in Table~\ref{tbl:co-lines}.
% 
and find $\mathrm{H_2}$ column density %{\bf is then} computed %density simply follows,
assuming a constant abundance ratio of $[\cao]/[\mathrm{H_2}]=1.7\times10^{-7}$
\citep{molecular-lines-triggered-sf-haworth-2013} for the three column density maps shown in 
the right panels of Figure~\ref{img:lte-results}.\\
% \paragraph{Cut A}
% 
% The \hzw{} column density in cut~A is the lowest of the three cuts. The mean and
% median are 3.4 and \nhz{3.5}{21}, respectively. The maximum column density
% is \nhz{5.5}{21}. The median is about 2.5 times
% smaller than that of cut~B and that of cut~C.
% 
% \paragraph{Cut B}
% 
% The values of the column density within cut B range between \nhz{2.2}{20} and 
% \nhz{2.2}{22} with a median of \nhz{9.5}{21}. These high values might
% be due to the protostellar core present in the cut.
% 
% 
% \paragraph{Cut C}
% 
% This cut shows column densities similar to cut B. The minimum,  
% maximum, and mean of \nhz{6.2}{20}, \nhz{4.3}{22}, and \nhz{1.0}{22}, respectively, 
% are higher than in cut B, but the median of \nhz{8.8}{21} is again lower by $\approx6\%$. % \\ \\
% 
% \paragraph{}
 The column density distribution (called {\it N$_{H[CO]}$}) has been compared to the total column density obtained from the dust ({\it N$_{H[FIR\,dust\,em.]}$} from Paper~I). % computing the ratio of those values in the cuts. 
\\
We fit a gaussian to the distribution of the ratio between the  {\it N$_{H[CO]}$} and %the dust 
{\it N$_{H[FIR\,dust\,em.]}$} %column density 
first considering all the cuts together, then among each of the 3 cuts. 
The distribution peaks at 0.8 with a standard deviation of 0.3. In general, the column density from CO gas emission is found lower than the
column density from dust emission. %the gas column density is lower then the dust ones.  
Along cut~A the ratio between gas and the dust column density peaks at 1 with a  standard deviation ($\sigma$) of 0.7, $\sigma$=0.3 for cut B, while for cut C the 
peak of the ratio is not clearly identified as the other 2 cuts. 

\begin{figure}[!htb]
\centering
%     \subfigure{
    \includegraphics[width=9cm]{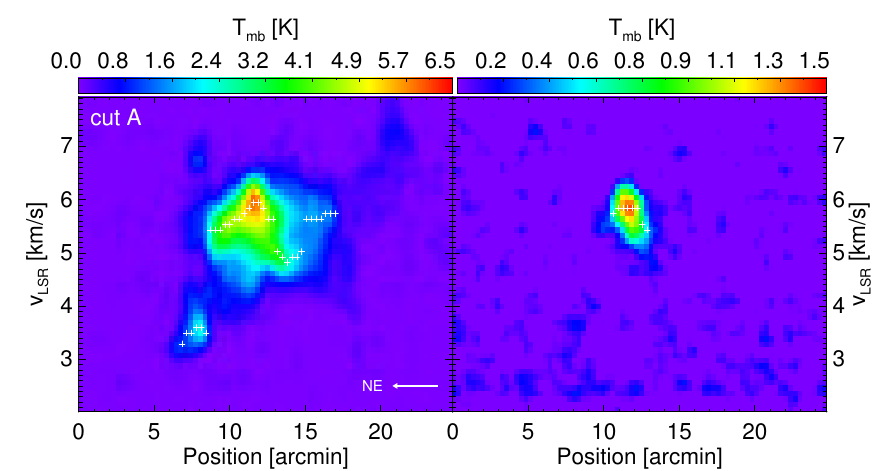}
%    \includegraphics[width=9cm]{Lupus_I_cutA_PV}
%     }
%     \subfigure{
    \includegraphics[width=9cm]{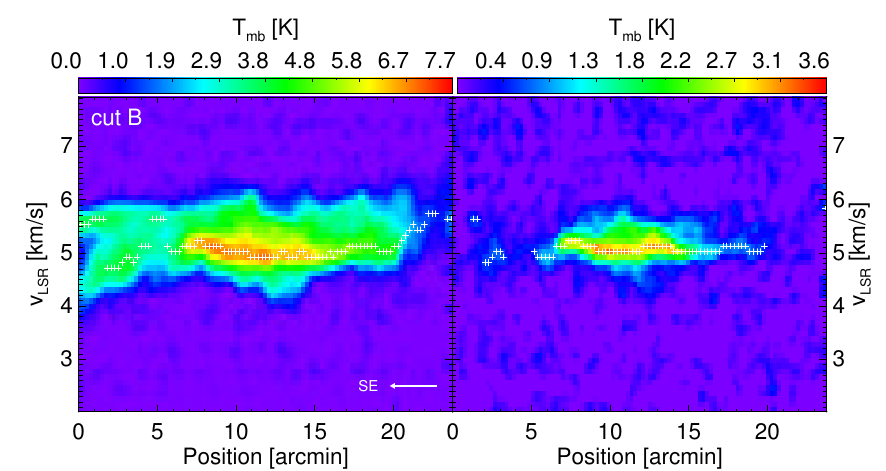}
%   \includegraphics[width=9cm]{Lupus_I_cutB_PV}
%     }
%     \subfigure{
    \includegraphics[width=9cm]{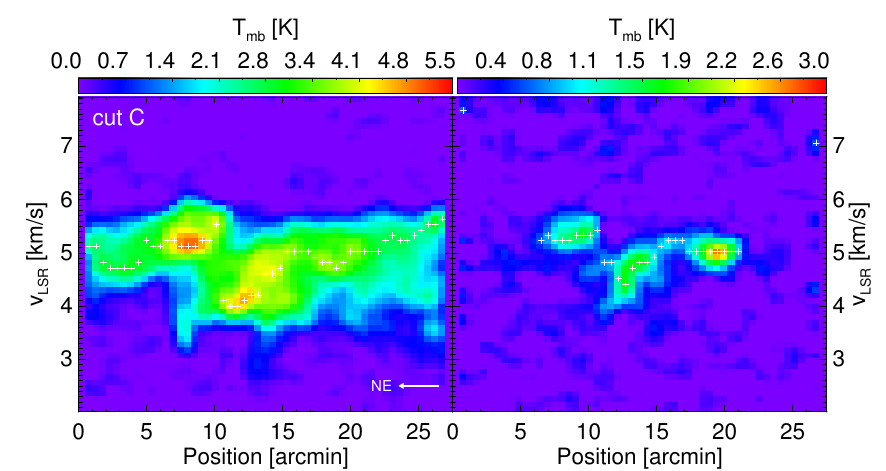}
%    \includegraphics[width=9cm]{Lupus_I_cutC_PV}
%     }
\caption[PV diagrams of the three cuts]{Position-velocity diagrams of cut~A, B, and C 
(from top to bottom) in \cdot{} (left panel) and \caot{} (right panel)
along a line through the middle of the cut. The x-axis gives the position along this line.
The white arrow in the bottom right corner of the left panels indicates the orientation of the cut in the
Celestial coordinate frame.
The white crosses mark the velocity
of the peak intensity at each x-position. 
The cutoff is set to 1\,K in \cdot{} for all three cuts and 0.5\,K in \caot{} for cuts~B and C.
The \caot{} cutoff for cut~A is 0.3\,K.
All maps have been convolved with a Gaussian of two pixels or approximately one beam.}
\label{img:cuts-pv}
\end{figure}
\subsection{Determination of different velocity components and non-thermal motions}
\label{sec:app-velocities}
\begin{figure*}[!htb]
 \centering
 \includegraphics[width=13cm]{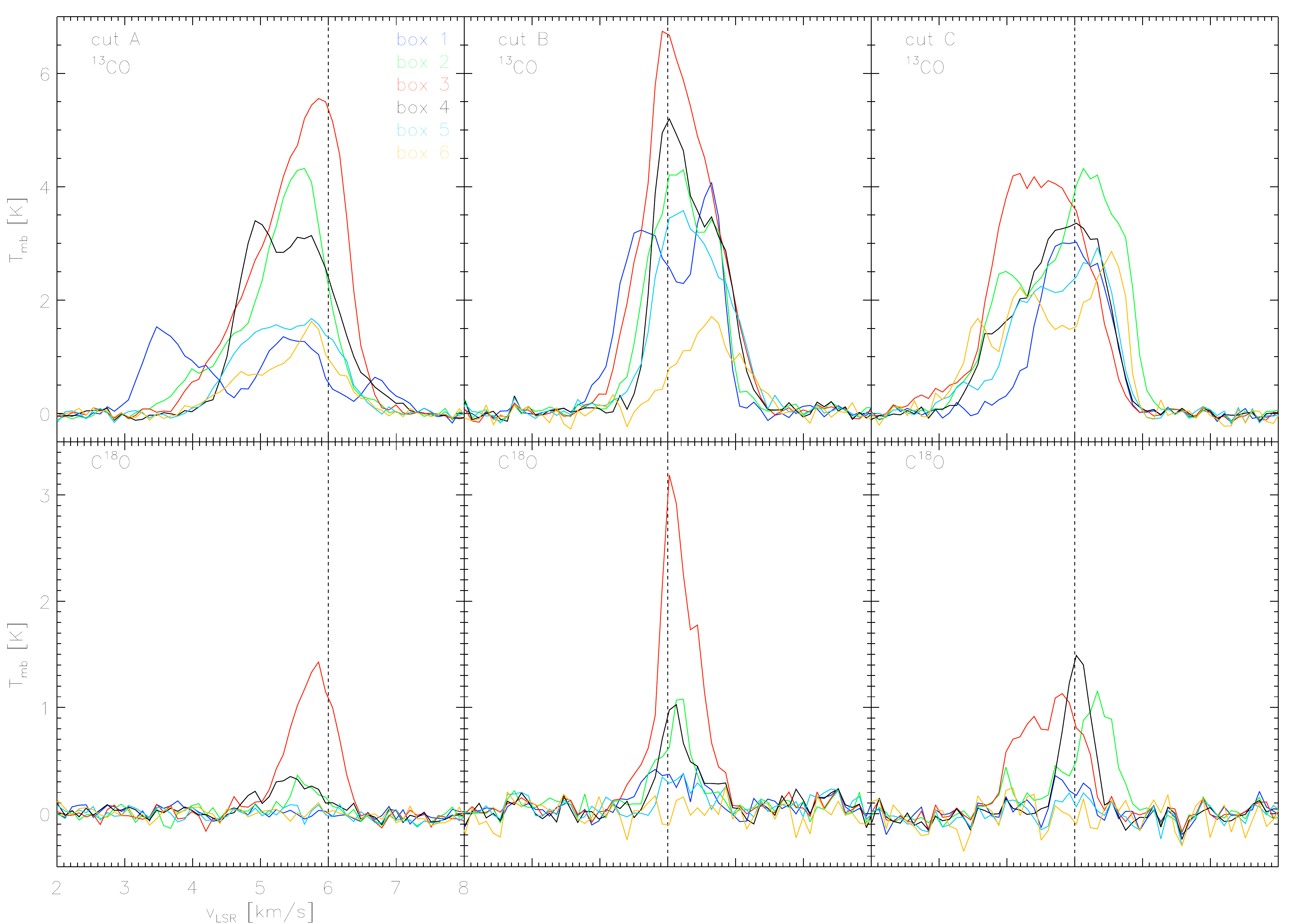}
 \caption[Averaged spectra within the six boxes into which each cut was divided]
 {Averaged spectra within the six boxes into which each cut was divided. The top row shows the average \cdot{}
spectra; the bottom those of \caot. Each column represents one of the three cuts. 
The different colors denote the six boxes from 1 to 6.
The vertical dashed line marks for orientation $\vlsr=6\kms$ for cut~A and $\vlsr=5\kms$ for cuts~B and~C.}
\label{img:average-spectra-boxes}
\end{figure*}

In this section we present a detailed analysis of the different
CO velocity components. \\% present in the three cuts.\\
Each cut was divided into six boxes as shown in Figure~\ref{img:cuts-boxes}.  The derived position-velocity (PV) diagrams %revealed that there are several velocity components present in the three cuts through Lupus I 
are shown in Figure~\ref{img:cuts-pv}. \\
Figure~\ref{img:average-spectra-boxes} shows the average \cdo{} (top row) and \cao{} spectrum (bottom row). \\
Using a multi-component (up to three) Gaussian fit of the spectra (Figures~\ref{img:cutA_C18O_Gauss}~to~\ref{img:cutC_13CO_Gauss}), we computed the peak position and FWHM for each component fo the  \cdo{} (top row) and \cao{} %present in the averaged spectrum of one box its 
% was determined %in two ways. First empirically from the spectra (see resuls in Table~\ref{tbl:peaks_fwhm}) 
%and then
%All the details about the velocity peak and broadening of \cao{} and \cdo{} are summarized in 
 (see Table\,\ref{tbl:peaks_fwhm_gauss} for more details). 
%All the results are in Table~\ref{tbl:peaks_fwhm_gauss}. 
%These peak positions and FWHM are shown in Figure~\ref{img:peaks_fwhm}.
% what made it easier to find and group similar velocity components. 
%In the next step,
%a mean peak position and a mean FWHM was determined for each group of velocity components. 
%The Gaussian fits to the spectra are presented in .\\
\begin{table*}
 \centering
% \begin{footnotesize}
 \caption[Peak positions and FWHM of the different components in each averaged
  spectrum of a box within a cut from the Gaussian fit]
  {Values of peak positions ($v^\mathrm{G}_{\mathrm{p}_i}$) and FWHM ($\Delta v^\mathrm{G}_i$) 
 of the different components in each averaged spectrum of one box within a cut determined from 
 multi-component Gaussian fitting
 (see Figures~\ref{img:cutA_C18O_Gauss}~to~\ref{img:cutC_13CO_Gauss}).
 	   % Velocity peaks are grouped in the same way as in the empirical analysis (see Table~\ref{tbl:peaks_fwhm})
 	     %and thus not always ordered by increasing velocity.
	     } 
\begin{tabular}{c  c c  c c    |c c  c c  c c}
%  \begin{tabular}{c  c c  c c    |c c  c c  c c}
 \hline\hline
 \noalign{\smallskip}
  Box & \multicolumn{4}{c|}{\cao}  & \multicolumn{6}{c}{\cdo} \\
      & \multicolumn{4}{c|}{[\kms]}  & \multicolumn{6}{c}{[\kms]} \\
  \# &  $v^\mathrm{G}_{\mathrm{p}_1}$ &  $\Delta v^\mathrm{G}_1$ &  $v^\mathrm{G}_{\mathrm{p}_2}$ &  $\Delta v^\mathrm{G}_2$ &  $v^\mathrm{G}_{\mathrm{p}_1}$ &  $\Delta v^\mathrm{G}_1$ & $v^\mathrm{G}_{\mathrm{p}_2}$ & $\Delta v^\mathrm{G}_2$ & $v^\mathrm{G}_{\mathrm{p}_3}$ & $\Delta v^\mathrm{G}_3$ \\
%  \noalign{\smallskip}
 \hline
 \noalign{\smallskip}
  \multicolumn{10}{c}{cut~A} & \\
 \noalign{\smallskip}
 \hline
  1 &  --   &  --   &  -- &  --      &   5.45 & 1.03 & 3.68  &  0.97 & 6.76 & 0.53 \\
  2 &  5.65 &  0.55 &  -- &  -- &   5.60 & 0.83 & 4.96  & 1.73 &-- & -- \\
  3 &  5.77 &  0.76 &  -- &  -- &   5.95 & 0.76 & 5.27  &  1.44 & --   & -- \\
  4 &  5.45 &  0.94 &  -- &  -- &  5.63 & 1.22 & 4.86  &  0.56 &  --   & -- \\
  5 &  --   &  --   &  -- &  --      &  5.86 & 0.88 & 4.96  &  1.00 &  --   & -- \\
  6 &  --   &  --   &  -- &  --      &  5.76 & 0.87 & 4.75  &  0.87 &  --   & -- \\
 \hline
  {\O}  & 5.62   &  0.75   &  -- &  -- &  5.71 &  0.93 & 4.75   & 1.10   &  6.76   & 0.53 \\
\hline
 \noalign{\smallskip}
  \multicolumn{10}{c}{cut~B} & \\
 \noalign{\smallskip}
 \hline
%  \noalign{\smallskip}
  1 &  5.03   & 1.08   &  -- &  -- &  4.71 &  0.97 & 5.62 & 0.48 & --   & -- \\
  2 &  5.14   & 0.63   &  -- &  -- &  5.13 &  1.00 & 5.75 & 0.30 & --   & -- \\
  3 &  5.14   &  0.66  &  -- &  -- &  5.15 &  1.19 & --   & --   & --   & -- \\
  4 &  5.13   &  0.63  &  -- &  -- &  5.00 &  0.53 & 5.64 & 0.74 & --   & -- \\
  5 &  5.25   &  0.81  &  -- &  -- &  5.34 &  1.21 & --   & --   & --   & -- \\
  6 &  --        &  --      &  -- &  -- &   -- & -- & 5.62 &  1.10  &--   & -- \\
\hline
 {\O} &  5.14 &  0.76 &  -- &  -- &  5.07 &  0.98 & 5.66 & 0.66 & --   & -- \\
\hline
 \noalign{\smallskip}
  \multicolumn{10}{c}{cut~C} & \\
 \noalign{\smallskip}
 \hline
%  \noalign{\smallskip}

  1 &   4.93 & 0.55 &  --   &  --  &    4.97 &  1.09 & --     & --   & --   & --   \\
  2 &  5.35 &  0.57 & 4.51 &  1.31 &  5.25 & 1.12 & 4.02 &  0.90 &  --   & --  \\
  3 &   4.86 &  0.65 & 4.22 &  0.55 &    --    & -- & 4.50 &  1.51  & --   & --   \\
  4 &   5.04 &  0.45      & --   &  --  &  5.15  & 0.88 & 4.34 &  1.31 &  -- & --   \\
  5 &  --     &  --   &  --   &  --    &  5.35 & 0.63  & 4.50 & 1.12 & 3.28 &  0.44 \\
  6 &  --     &  --   &  --   &  --    &  5.51 & 0.60  & 4.35 & 1.41 & 3.51 &  0.29  \\
\hline
  {\O} &  5.05   &  0.56   & 4.37   &  0.93  &   5.25 &  0.86 & 4.34 & 1.25 & 3.40 & 0.37 \\
%  \noalign{\smallskip}
 \hline
%  \end{tabular}
\label{tbl:peaks_fwhm_gauss}
\end{tabular}
% \end{footnotesize}
\end{table*}
 %\paragraph{Cut A}
%In cut~A the \cao{} line is detected in the central boxes (2--4). 
%In cut~A there is one group
%of \cdo{} peak velocity components with a mean value of 5.7\kms. 
%In box~1 there are also two other \cdo{} components. In those boxes, where both a \cao{} and a \cdo{} line are present, the peak positions
%agree well with each other whereas the \cdo{} lines are roughly twice as wide as the \cao{} lines.
%The \cdo{} peak velocities at the two edges of the cloud (box~1 and~6) are significantly
%different, while the FWHMs are similar. 
%Both edges do not show significant amounts of \cao{} emission.
%For boxes 4, 5 and 6 the double-peak is likely due to the fact that the line is becoming optically thick due to 
%self-absorption. 
% In cut~B there is one component of \cao{} peak velocities. The mean FWHM of the 5 cuts is of the same order of magnitude
%of cut~A.
%In \cdo{} 3 boxes show two peak velocities. In the first two boxes, in particular, 
%the highest velocity component has a FWHM 2 times narrower than the low velocity 
%component. 
%In those boxes where both a \cao{} and a \cdo{} line are present the peak positions
%agree well with each other. \\
%Within cut~C are present two peak velocity components of \cao{}.
%In these boxes where both lines are present, the peak velocities
%of \cao{} and \cdo{} agree with each other. In these cases the FWHM of the \cao{} line is lower than that of \cdo. 
 %\cdo{} covers also boxes 5 and 6 and three velocity components are present. \\
 We conclude from this 
 %This analysis leads to the main conclusion 
 that a velocity decrease from cut A to cut C is seen. 
 %gradient is present in the cloud, decreasing from cut A to cut C.
No clear trend is seen in linewidths, suggesting a homogeneous level of turbulence. \\
%\begin{figure*}[!htb]
%\centering
% \includegraphics[width=0.75\textwidth, keepaspectratio]{Peaks_FWHM_all_cuts.pdf} 
%\caption[Peak velocities and FWHM for each component in a box of one cut]{Peak velocities and FWHM for each component in an averaged spectrum of one box (1 to 6) within one cut.
%	 Blue crosses and cyan diamonds mark the empirically determined values of peak position and FWHM directly from the spectra
%	 for $\rm ^{13}CO$ and $\rm C^{18}O$, respectively. Their values can be found in Table~\ref{tbl:peaks_fwhm}. 
%	 The red triangles and the orange boxes mark those values as determined from
%	 multi-component Gaussian fits to each spectrum 
%	 (see Figures~\ref{img:cutA_C18O_Gauss} to \ref{img:cutC_13CO_Gauss} and Table~\ref{tbl:peaks_fwhm_gauss}).
%	  The dotted, the dashed and the
%	 dashed-dotted lines mark the average of the empirically determined component 1, 2, and 3, %respectively. 
%	 Error bars are $\pm 0.2\kms$.}
%\label{img:peaks_fwhm}
%\end{figure*}
%% \newpage
%% \clearpage
%\subsection{Non-thermal motions}
The {\bf non-thermal velocity dispersion} of the gas has been then derived from the measured
linewidth, subtracting the thermal component. %, t can be derived, and hence the non-thermal motion of the gas. 
%, which in general is associated to
%turbulence, 
%one can calculate the from . 
%For this, we 
Following the standard assumption that the thermal and non-thermal
components are independent \citep[see e.g.][]{subsonic-turbulence-cores-myers-1983}, 
the non-thermal velocity dispersion $\sigma_\mathrm{NT}$ is
\begin{equation}
\sigma_\mathrm{NT}=\sqrt{ \frac{\Delta v^2}{8\ln2} - \frac{\mathrm{k_B}\,T_\mathrm{kin}}{m} }
\end{equation}
with $\Delta v$ the measured linewidth, $\mathrm{k_B}$ the Boltzmann constant, $T_\mathrm{kin}$ the
kinetic gas temperature (here we adopt 10\,K), and $m$ the mass of the observed molecule 
($m_\mathrm{^{13}CO}=29\,\mathrm{u}$, $m_\mathrm{C^{18}O}=30\,\mathrm{u}$).
Dividing $\sigma_\mathrm{NT}$ by the isothermal sound speed $c_\mathrm{s}=0.19\kms$ for a 10\,K ISM gas,
 gives a direct measure of non-thermality of the component. For the measured linewidths of \cdo{} 
 in the three cuts, 
%the ratio $\sigma_\mathrm{NT}/c_\mathrm{s}$
this indicator is between 2.0 and 4.5, suggesting that the gas would be supersonic.
 However, correcting the \cdo{} linewidth %is known to have a significant contribution from 
for opacity broadening due to its higher 
optical depth $\taud>1$--2 %, while  and should therefore be corrected 
\citep{co-opacity-broadening-philips-1979}, while 
 \cao{} $\taua<1$, % in most cases.%
%For this molecule the measured
%linewidths are between 0.4 and 0.8\kms. This 
this yields a lower indicator value of $\sigma_\mathrm{NT}/c_\mathrm{s}$ 
between 0.9 and 1.8.
Hence, most of the \cao{} gas within the three cuts is in the trans-sonic regime 
($1 < \sigma_\mathrm{NT}/c_\mathrm{s} < 2$). 
%  what means that
% these regions of \lupus{} are not quiescent, but have a significant level of turbulence. 
%In cuts~B and~C this is probably because of the pre- and protostellar cores. 
%But also the gas in cut~A shows these high linewidths although no star-formation is seen there. 
Typically, the gas of a quiescent cloud resides in the sub-sonic regime as found e.g. by 
\cite{dense-cores-fragmentation-L1517-hacar-2011} for L1517. 
In \lupus{} this signature of turbulence might have been imprinted on the gas by the large-scale interaction 
with the \ion{H}{I} shell of USco and the wind-blown bubble of UCL. 
\section{Comparison between CO and HI}
\begin{figure*}
% [!htb]
\centering
       \includegraphics[width=12cm]{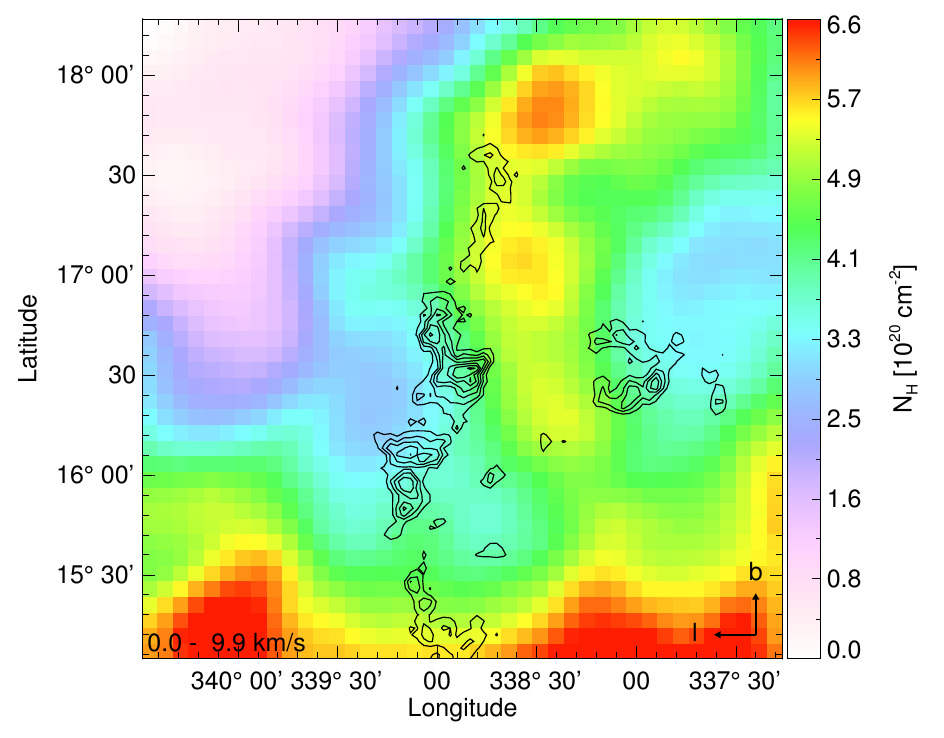}
\caption{\ion{H}{I} column density map calculated from the 
integrated emission for $\vlsr=0$--$10\kms$. Black contours show the integrated \cao(1--0) emission
observed by \cite{lupus-clouds-c18o-Hara-1999}. 
% , i.e. 
% within the same velocity range as for our CO data.
}
\label{img:hi-lupus}
\end{figure*}

In this section we present a comparison between CO and \ion{H}{I} data and a geometrical model to check the scenario that \lupus{} is located in the interaction zone of the USco \ion{H}{I} shell
and the UCL wind bubble. \\
Figure~\ref{img:hi-lupus} shows %an \ion{H}{I} velocity channel map 
for $\vlsr=0$--$8\kms$ an \ion{H}{I} column density map of \lupus, together with 
%are shown in Figure~\ref{img:hi-lupus}. 
%Black contours represent the integrated 
\cao(1--0) emission
\citep{lupus-clouds-c18o-Hara-1999} 
% The black contours show
% the \herschel{} \twof{} emission to mark the outlines of \lupus.
% 
%The \ion{H}{I} emission lies in 
for the same velocity range as the CO emission. %, i.e. mainly
%between \hbox{$\vlsr=3$--$7\kms$}. The peak emission is at $\vlsr=5\kms$ which agrees
%with the velocity of the CO emission peaks in cuts~B and~C.  
%But the column density map reveals that the 
Although the peak velocities of HI and CO agree, the  \ion{H}{I} and CO emission do not 
spatially agree with each other: 
the highest \ion{H}{I} column density is located at the north-western end of \lupus, %Two %other
%slightly 
%lower peaks are found further south-west {\bf BE QUANTITATIVE!}.
while in the central and the southern part of \lupus{} there is less \ion{H}{I} emission, 
%than in the north, 
roughly by a factor of 1.5. Between $b=15\degr30\arcmin$--$17\degr$ and $l=338\degr30\arcmin$--
$339\degr30\arcmin$ the deficit in \ion{H}{I} emission around the center-south of \lupus{}
has the form of a little hole.
The \ion{H}{I} emission, in general, is enhanced
at the Galactic west side of \lupus{} and has a deficit on its east side, with an average 
%\ion{H}{I} column density , with an for \lupus{} is 
$\langle N_\mathrm{H}\rangle=\nhz{5.5}{20}$. 
%This means that t
 The atomic to molecular ratio is %less than unity for both parts
%of the cloud 
% assuming the averaged 
%using the $\mathrm{H_2}$ column densities
%derived from our CO analysis of
%$\nhz{3.4}{21}$ and $\sim10^{22}\,\mathrm{cm^{-2}}$
%% dust emission of $\nhz{1.10}{21}$ and $\nhz{1.72}{21}$ for
%for the north and the center-south, respectively.
% Using the mean \hzw{} column density from 
% would decrease the ratio even further.
%However, 
poorly-defined, given the large uncertainties in the these values. %, the ratio
%could also well be around or above unity.
\cite{lupus-filaments-herschel-benedettini-2015}, for example, reported a ratio of
 $\langle N_\mathrm{H}\rangle/\langle N_\mathrm{H_2}\rangle\approx1.6$
calculated from their \herschel{} column density maps and the \ion{H}{I} data of \cite{sco-cen-HI-degeus-1992}.\\
The particular distribution of CO and \ion{H}{I} shows that \lupus{} marks a transition region between the molecular and atomic
phase of \ion{H}{I}. 
% It is likely that \lupus{} is embedded into the wall
% of the USco \ion{H}{I} shell. The configuration of \ion{H}{I} and CO does support this view.
% First the shell would have to compress the atomic hydrogen so it can become
% molecular. The front of the expanding shell compresses the atomic hydrogen
% so it can form molecular hydrogen. Then while the front is further expanding,
% the molecular material is left behind as can be seen here.
At those places in the cloud where the atomic hydrogen emission is decreased, 
the molecular emission is enhanced and vice versa (see Figure~\ref{img:hi-lupus}).
Our analysis shows that the \cao{} emission is %almost
%negligible in cut~A (north) and 
significantly higher %in cuts~B and~C,
%i.e. 
in the center and the south of \lupus.
% This means that part of the cloud where substantial amounts of atomic hydrogen have already become molecular,
% the \ion{H}{I} emission is decreased and the molecular CO emission is enhanced.
% Where the \ion{H}{I} emission is highest there is also the highest \ion{H}{I}
% column density. Where the density is high enough the atomic hydrogen can convert
% into molecular hydrogen.
%This can explain why the northern part of the cloud is different from the center
%and the south. The 
 We conclude that the north has not built enough dense molecular material yet 
to start the star formation process. %Therefore, the northern part of \lupus{}
%shows no current star formation activity.
The lower dust column densities seen there 
coincide with
%might be another reason for 
a higher fraction
of atomic material. 
%As $\mathrm{H_2}$ forms predominantly on the surface
%of dust grains the formation efficiency is lower where 
%% less dust is available within the cloud.
%the dust density is lower within the cloud.
%In the central and southern part of \lupus{} the molecular to atomic portion of the material is significantly higher.
%The 
Correspondingly, \cao{} emission is strongest in the central and southern parts of Lupus I where the \ion{H}{I} emission has a deficit and 
active star formation  and high CO column densities 
($N(\mathrm{H_2})\approx10^{22}\,\mathrm{cm^{-2}}$) can be seen.\\
\cite{mc-formation-synthetic-HI-CO-heiner-2015} have created synthetic
\ion{H}{I} and CO observations from a numerical simulation of decaying turbulence
in the thermally bistable neutral medium. They found a power-law tail in the column
density PDF when molecular clouds have formed. This tail, however,
only appears in the PDFs of the molecular material and not the \ion{H}{I}. Hence, they conclude
that the molecule formation is directly correlated with gravitational
infall. As the dust PDF of the northern part of \lupus{} (see Paper~I) does not show
a power-law tail, the gravitational infall might not yet be strong enough
to convert as much atomic material into molecular %one as is apparent 
in the center-south. 
%Another explanation might be that the molecular
%material that has already formed in the north is still CO dark because there is not yet
%enough column density to provide sufficient shielding for CO from destructive UV
%radiation. 
%The average $\mathrm{H_2}$ column density in the north
%as calculated from the \herschel{} column density map is \nhz{1.1}{21}
%and thus lower than the required value for CO shielding ($\sim10^{22}\cmmz$).
%Our CO data yielded an $\mathrm{H_2}$ column density 
%of \nhz{3.5}{21} for cut~A. Somewhat higher than that from
%the dust analysis, but still lower than the required $10^{22}\,\mathrm{cm^{-2}}$.
% 
% 
% We conclude, that this particular configuration of the atomic and molecular material
% in and around \lupus{} and the co-moving CO and \ion{H}{I} gas velocities
% suggest that the formation of the cloud
% is linked to the large-scale feedback of the USco and the UCL subgroups
% of Sco-Cen. 
% Hence, the next chapter will shed light on \lupus{} in the context
% of its surrounding and the history of the star formation complex it is part of.
%\cite{w43-atomic-molecular-transition-motte-2014}
%\begin{figure*}
% [!htb]
%\centering
  %      \includegraphics[width=0.85\textwidth]{Lupus_I_HI_channel_map.pdf}
%\caption{ \ion{H}{I} channel maps for a velocity range of $\vlsr=0.8$--$7.4\kms$. 
%Black contours show the integrated \cao(1--0) emission
%observed by \cite{lupus-clouds-c18o-Hara-1999}. 
%They start at 0.3\kkms with a step size of 0.3\kkms. 
%}
%\label{img:hi-lupus}
%\end{figure*}
%\section{Results}
%\label{sec:results}
\subsection{Geometrical model for the interaction zone}
\label{sec:usco-ucl-model}
\begin{figure*}[!htb]
    \centering
    \includegraphics[width=\textwidth]{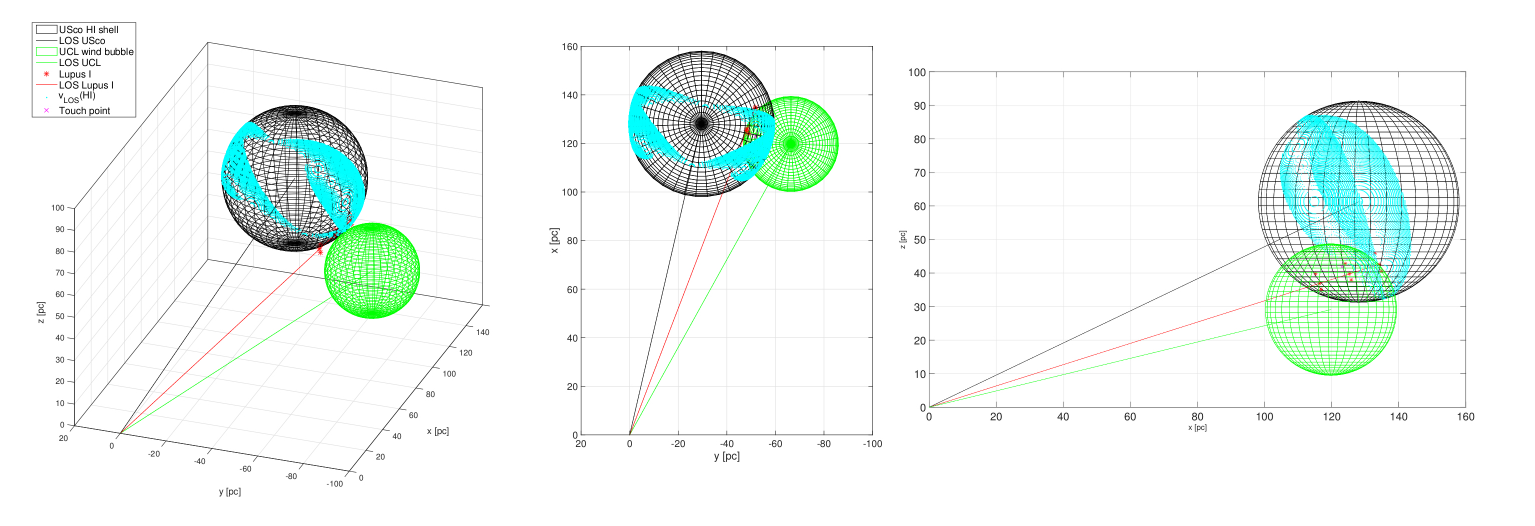}

\caption[Perspective view of the geometrical model of interaction zone USco - UCL]
{Geometrical model of the possible interaction zone between the USco \ion{H}{I} shell (black sphere) and the UCL wind bubble (green sphere).
The observer is located in the origin of the coordinate system. The position of \lupus{} is marked by three red asterisks that represent the top, middle, 
and bottom of the main filament, respectively. The cloud is plotted at three different distances of 150, 140, and 130\,pc
along the red line which is the line-of-sight from us to \lupus.
The black and the green solid lines are the line-of-sight from the observer to the corresponding sphere.
The cyan dots mark those positions on the USco shell that have line-of-sight velocities between 3 and $6\kms$,
i.e. similar to \lupus{} (the white gaps are an effect of the grid model used in the calculations).
They were calculated from the model fit to the \ion{H}{I} data of USco. The magenta cross marks the touch point
between the USco and the UCL sphere.  The projections are show in the different panels.}
\label{img:usco-ucl-model}
\end{figure*}

% \nomenclature{Hipparcos}{Hipparcos Space Astrometry Mission}
% \nomenclature{GMC}{Giant Molecular Cloud}
 %In this section we present a 
 We construct a geometrical model with the %to check the scenario that \lupus{} is located in the interaction zone of the USco \ion{H}{I} shell
%and the UCL wind bubble. %, we have created a simple geometrical model.
%The 
observer located in the origin of the coordinate system which is a 3D Cartesian
 Galactic coordinate system with
\begin{equation}
\begin{pmatrix} x \\ y \\ z \end{pmatrix} = \begin{pmatrix} r\,\sin\theta\,\cos\phi \\  r\,\sin\theta\,\sin\phi \\ r\,\cos\theta \end{pmatrix}
% 
% [ 
% \left( 
% \begin{array}{ccc}
% a & b & c \\ [3ex]
% d & e & f \\
% g & h & i 
% \end{array}
%
% \renewcommand*{\arraystretch}{2} 
% \vec{\mathcal{U}} := \gamma^{1/2}\!\left[
% \begin{array}{c}
% D \\
% 
% S_j
% 
% E
% \end{array}\right]
% ,~~~
% \vec{\mathcal{F}}^i := \gamma^{1/2}\!\left[
% \begin{array}{c}
% \! (\alpha v^i - \beta^i) D \! \\
% \!\alpha S^i_j-\beta^i S_j\! \\
% \!\alpha S^i-\beta^i E\!
% \end{array}\right],
% 
% \right)]
% 
\end{equation}
where $\theta=90\degr-b$, $\phi=l$, $r$ is the distance, $b$ and $l$ the galactic latitude and
longitude, respectively. The x-axis points %to is defined by the direction
to the Galactic center.
Both, \ion{H}{I} shell and wind bubble are represented by a sphere. We place the center of the USco sphere
at $(l,b)=(347\degr,+25\degr)$ and a distance of 145\,pc.
A radius of $\sim30$\,pc is the inner shell radius
from our \ion{H}{I} model fit  (details in Kr{\"o}ll 2018) %Krause et al. 2016, submitted)} 
%of Daniel Kr\"oll, 
as observations suggest that \lupus{} is located closer to the inner edge of the shell.\\
% we assume a radius of $\sim30$\,pc which is an average
% between the inner and outer shell radius from the \ion{H}{I} model fit of Daniel Kr\"oll.
We estimate the average projected position of the B-stars southwest of \lupus{}
that might be responsible for the wind bubble (see Paper~I) to be $(l,b)=(331\degr,+12\degr)$.
This also agrees with the center of a circle in the ROSAT image enclosing the X-ray emission
in this region (c.f. Fig.~7 in Paper~I).
The average distance to the stars of the UCL subgroup is about 5\,pc smaller
than that to USco \citep{sco-cen-hipparcos-dezeeuw-1999} and thus we place the center
of our UCL sphere model at 140\,pc.\\
Our UCL wind bubble is thus on the near side %placed below and slightly in front 
of the USco \ion{H}{I} shell. % with respect to us.
%We associate a sphere of influence of the massive stars in UCL in an ad hoc manner:
%by increasing the radius of the UCL sphere 
%its expansion toward USco and \lupus{} is simulated.
Figure~\ref{img:usco-ucl-model} shows the %configuration
of the model in three perspectives, at %a perspective view %. Figures~\ref{img:usco-ucl-model-projections}-\ref{img:usco-ucl-model-projections1} show 
%the model 
%{\bf and in a side and 
%a top view}. The black sphere in the figures
%represents the USco shell, while the green sphere represents 
the stage at which the UCL bubble touches 
the USco sphere. % (touch point is marked by a magenta cross). 
% The yellow sphere represents an expansion up to a radius equal to the distance
% The position of \lupus{} is marked by three red asterisks that represent the top, middle, 
% and bottom of the main filament, respectively. We have plotted the cloud at three different distances of 150, 140, and 130\,pc
% along the red line which is the line-of-sight between observer and \lupus.
% The black and the green solid lines are the line-of-sight between the observer and the corresponding sphere.
% % hereby we assume \lupus{} to be at $150\pm20$\,pc.
% % The touch point between the two spheres is at a distance of $\sim140\pm6$\,pc to us and 
% % $\sim13\pm6$\,pc from \lupus{}. The UCL sphere radius at that stage is $\sim15\pm3$\,pc.
% % The final expansion radius of UCL would be 25\,pc corresponding to the distance between
% % \lupus{} and the center of UCL.
% The model shows that the expansion of the UCL sphere goes in the 
% direction of \lupus{} which would eventually be crossed by the sphere.
% 
% In a next step the geometrical model is related to the observed velocities
% of atomic \ion{H}{I} (shell) and molecular CO (\lupus). The model used to fit 
%of D.\,Kr\"oll 
%  the \ion{H}{I} data of the USco shell  
% yields the line-of-sight velocity at each position on the surface of the shell 
% as would be observed from earth. 
%(Krause et al. 2016, submitted).
In Figure~\ref{img:usco-ucl-model} all points on the surface of the USco sphere that have a model velocity 
between \hbox{3--$6\kms$} are marked in cyan. %dots 
%(the white gaps are an effect of the grid model used in the calculations).
This represents the CO velocity range we have observed in \lupus.
The %green UCL sphere intersects these points on the black USco
%sphere. This means that the 
UCL wind bubble 
(i.e. the sphere of influence of the massive stars in UCL) %as defined above)
interacts with the USco \ion{H}{I}
shell at positions that have a similar gas velocity to \lupus.\\
The \lupus{} cloud and the USco \ion{H}{I} shell have similar observed velocities and thus 
it seems reasonable that \lupus{} is expanding with the USco shell.\\
In our purely geometrical model, 
which does not accurately reflect the physics 
and the resulting complex morphologies present in the Sco-Cen region,
the UCL wind bubble and the USco \ion{H}{I} shell nevertheless intersect at %the right time and the right place with 
velocities consistent with our observations of \lupus.
Therefore, the observational data are in agreement with the idea that \lupus{} is in an interaction 
zone between the USco \ion{H}{I} shell and the UCL wind bubble.
Moreover, this model comparison indicates that \lupus{} is expanding with the USco shell and that the cloud
 is located at the inner edge of the \ion{H}{I} shell.

%\begin{figure}[!htb]
%    \centering
%%     \subfigure{\includegraphics[width=0.8\textwidth]{USco_UCL_model_plot_legend.pdf}}

% \subfigure{
%\includegraphics[width=9cm]{USco_UCL_model_plot_xz.pdf}
% }
%\caption[Projections of the geometrical model of interaction zone USco - UCL]{Same as Figure~\ref{img:usco-ucl-model} in a side projection.}
% \vspace{14cm}
%\label{img:usco-ucl-model-projections}
%\end{figure}

%\begin{figure}[!htb]
%    \centering
%% \subfigure{
%\includegraphics[width=9cm]{USco_UCL_model_plot_xy.pdf}
% }
%\caption[Projections of the geometrical model of interaction zone USco - UCL]{Same as Figure~\ref{img:usco-ucl-model} in a projection from above.}
%\label{img:usco-ucl-model-projections1}
%\end{figure}

\begin{figure*}[htb!]
%  \subfigure{
  \centering
\includegraphics[width=9.1cm]{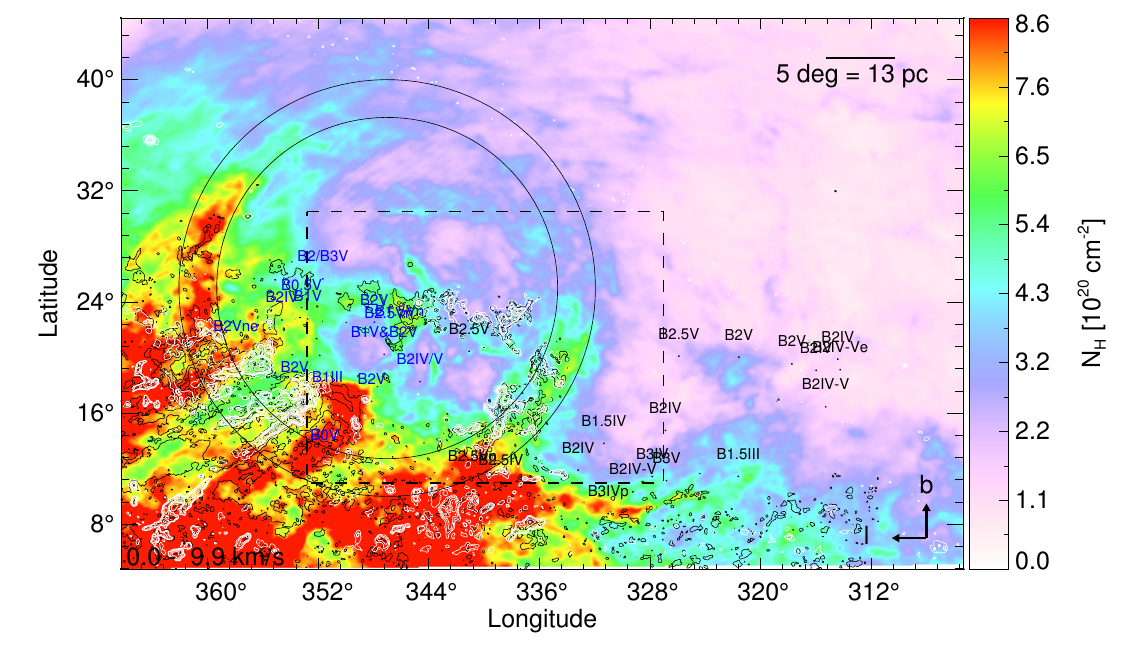}
%\includegraphics[width=9.1cm]{USco_Lup_UCL_HI_channel_map_integr.pdf}
% \caption{}
%     \label{img:herschel-temp-map}
% }
% \vspace{10cm}
%  \subfigure{
%   \centering
\includegraphics[width=9.1cm]{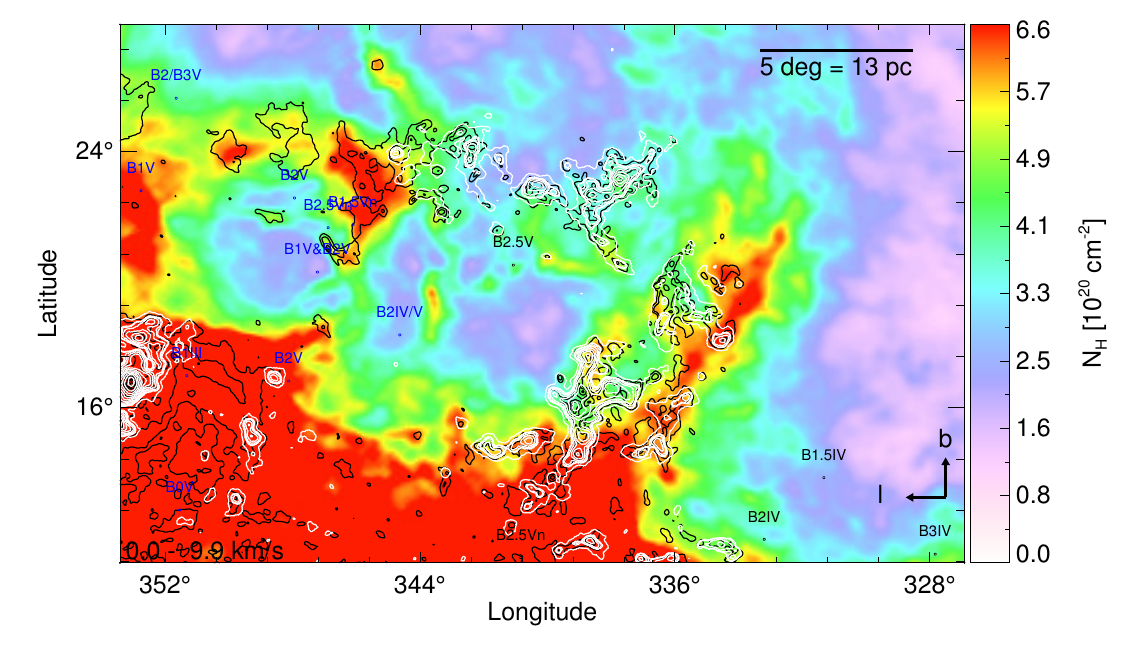}
%\includegraphics[width=9.1cm]{USco_Lup_UCL_HI_channel_map_integr_zoom.pdf}
%% \caption{}
%   \label{img:herschel-temp-histo}
% }
\caption{{\it Left:} \ion{H}{I} column density map of USco and the surroundings of \lupus{}
calculated from the integrated emission for $\vlsr=0$--$10\kms$.
The two black circles indicate the inner and outer radius of the USco shell
from the model fit.
The black and white contours represent
the dust emission at $850\,\mu$m and the \hbox{CO(1--0)} emission from \planck, respectively.
The blue and black labeled dots represent early B-type stars in USco and UCL, respectively.
{\it Right:} Zoom into the
region indicated as a dashed black box in the left panel.}
\label{img:usco-hi-mom0}
\end{figure*}

\section{The surroundings of Lupus~{\rm I} and the interaction with USco and UCL}
\label{sec:surrounding}

Figure~\ref{img:usco-hi-mom0} shows the USco \ion{H}{I} shell as observed today.
%We have created an 
A \ion{H}{I} column density map of USco and the surroundings of \lupus{} 
is shown in the top panel, 
%The black and white contours represent
together with the dust emission at $850\,\mu$m and the CO(1--0) emission from \planck, and %, respectively.
%The blue and black labeled dots represent 
early B-type stars in USco and UCL. %, respectively.
Our model fit to the \ion{H}{I} data 
%done by Daniel Kr\"oll 
yields a shell expansion velocity of $\sim7\,{\rm km\,s^{-1}}$
and a thickness of $\sim6$\,pc. The current inner and outer radius is $\sim30$\,pc and $\sim36$\,pc, respectively 
(%both radii of the USco shell are 
indicated by the two black circles in Fig.~\ref{img:usco-hi-mom0}).
East and west of \lupus{} there are 
clear \ion{H}{I} voids and the atomic material is concentrated in the same area as the dust.
A further zoom on \lupus{} (right panel of Fig.~\ref{img:usco-hi-mom0})
%i.e. the ring-like dust ridge and the areas east and west of it,
reveals a complicated distribution of atomic, molecular, and dust material.
Whereas the CO emission is always found in regions
of dust emission, there are several areas within the USco shell ridge 
where there is \ion{H}{I} emission but neither dust nor CO are detected.
The layering of atomic and molecular material, as can be seen
in \lupus{} (molecular gas, traced by CO, is surrounded by \ion{H}{I}), is also found in some other
locations in the shell around \lupus. %Although, there are certain positions at which
%\ion{H}{I} emission peaks coincide with CO emission peaks.
In the part of the shell
where \lupus{} is located, the molecular gas is found preferably behind the 
atomic gas, i.e. closer to the inner edge of the shell.
However, 
%there is no clearly separated and homogeneous layering of these two phases.
%this picture is probably not as homogeneous as it might seem at first.
%there seems to be no simple 
%homogeneous regarding 
%with clearly separated phases (e.g. shell is atomic at the outer edge and molecular
%behind).
%It is rather likely that 
local turbulence and inhomogeneities apparently lead to %mix
%the material into 
complex configurations. % leading to
%different atomic to molecular fractions inside the shell.

% which seems to be part of the USco \ion{H}{I} shell. 

The comparison of our CO to \ion{H}{I} data showed that both
fall into the same velocity range. If thus \lupus{} is co-moving with the atomic gas in the shell its %what is clearly indicating the 
formation within the shell wall  seems plausible. 
\cite{mcs-in-supershells-dawson-feb-2011} found that expanding supershells are capable of sweeping up %accumulating 
big amounts of gas %material when they sweep through 
from the ISM. %Thus, they have a greatly i
Increased density within their walls
 leads to very short chemical timescales. They estimate a transition
timescale from atomic to molecular material of \hbox{$\sim10^6$--$10^7$\,yr} \citep[see also][]{mc-formation-shocked-layers-koyama&Inutsuka-2000,
mc-formation-behind-shock-waves-bergin-2004,mc-formation-rapid-heitsch&hartmann-2008,mc-formation-timescale-clark-2012}.
The estimated age of the USco shell is $\sim5$\,Myr and hence it should have had sufficient time to
accumulate locally enough atomic gas to convert it into molecular gas at the position of \lupus{}.
From the dust analysis described in Paper~I we found that the total cloud (gas + dust) mass 
of \lupus{} is only $\sim170\modot$. This is at least
two orders of magnitude lower than the estimated total mass of the USco shell 
($\sim10^4\,\modot$) as derived from the \ion{H}{I} model fit.  %of D.\,Kr\"oll.
If the atomic gas was homogeneously distributed within the shell, then 
at least $\sim10^3\modot$ would have been swept up at the current position of \lupus{}, %which is
enough to create a $\sim100\modot$ molecular cloud, and similar to the estimate of the total cloud mass by \citet{lupus-clouds-13co(1-0)-Tachihara-1996}. 
Furthermore, the analysis of the LABOCA continuum data has revealed many pre- and protostellar cores
(see Paper~I).
This suggests that the formation timescale of the cloud and the onset of star formation should be on the 
order of even less than 2\,Myr which is still in agreement with the atomic-to-molecular transition timescale.
%\cite{lupus-submm-polarimetry-matthews-2014} have found that t
The large-scale magnetic field is 
perpendicular to the Lupus~{\rm I} filament, i.e. pointing in the direction of the USco shell expansion.  
%\citep{lupus-submm-polarimetry-matthews-2014}. 
This might
have favored the accumulation of cold, dense atomic gas along the field lines and promoted fast molecule 
formation \citep{lupus-submm-polarimetry-matthews-2014, rapid-formation-mcs-hartmann-2001,mc-formation-mhd-vazquez-semadeni-2011}.

The slightly supersonic irregular motions we find are in excellent agreement with predictions for clumps formed
 by the thin shell instability \citep{instabilities-shocks-vishniac-1983}.
 3D simulations explicitly predict such
  velocities in clumps formed dynamically in expanding supershells 
  \citep[][Fig.~9]{superbubbles-vishniac-krause-2013}. 
 We thus have three pieces of evidence, the correspondence between bulk HI and molecular velocities, the spatial anti-correlation between molecular and atomic gas, and the magnitude of the irregular motions % ...
%two pieces of evidence, the correspondence between bulk HI and molecular velocities, and the magnitude of the irregular motions 
 which supports the picture that the \lupus{} molecular cloud has formed
  dynamically in the expanding USco supershell. 
%On the western side of \lupus{} there is the UCL wind bubble.
%The 
Also X-ray observations suggest that the UCL wind bubble might be colliding with the
USco H\,{\rm I} shell right at the position of Lupus~{\rm I,} squeezing it in between (see Paper~I).
This wind bubble, probably still additionally pressurized by past supernova activity, 
might have provided a counter-pressure to the expanding USco shell and thus favored this 
position for an additional compression of the shell material.
In this way a new molecular cloud could have been created there and it 
would explain why not more very young star forming
clouds (except $\rho$~Ophiucus that is believed to be pre-existing \citep[e.g.][]{rho-oph-sf-motte-1998})
are seen distributed within the wall of the USco shell.\\
%The dust emission maps have revealed an elongated shape of \lupus{} which is
%mainly parallel to the edge of the USco shell.
This view is also supported by the column density PDFs of the dust emission (see Paper~I), with the 
 double-peak profile over the whole extent of the
cloud. Together with the dust column density map this shows that \lupus{} consists of a dense layer 
in the center of the filament which is surrounded by much more diffuse material. Simulations have shown
that this is consistent with the idea of a two-sided compression through colliding flows 
\citep[see][]{pdf-double-peaks-colliding-flows-matsumoto-2015}. Then the molecular cloud is created in a dense and thin sheet 
which is surrounded by the diffuse material of the colliding flows.\\
%A significant level of turbulence is also present in the cloud, as indicated by transonic linewidths
%in the \cao{} data of all three parts of \lupus. Hence, this turbulence might have been injected by 
%the collision process.
% \newpage
This picture does, however, not come without difficulties: A formation of molecular clouds in  colliding supershells requires that the shells have the same pressure evolution \citep{collision-bubbles-simulation}. Any pressure difference %- likely because stars as well as environment can never be the same in two given groups of stars - 
can lead to cloud ablation and might destroy such clouds within about 10$^6$ yr \citep{superbubbles-vishniac-krause-2013}. 
Since Lupus~I is located between USco and UCL, its position also agrees with the much more extended Sco-CMa stream \citep{BouyAlves2015}. It is therefore also possible that pre-existing density enhancements played a role in the formation of Lupus~I.\\
\citet[][]{thompsonetal2012} detected an increase in star-forming sources at the rim of the shells of HII regions (or IR/Spitzer bubbles). %HI is too diffused to form stars, you require molecular gas to form stars. 
% Other observations suggested that star formation near the rim of HI shells is frequent \citep[e.g.][]{thompsonetal2012} and it would be implausible to assume that all these shells have a similar counter-pressure at the locations where the stars form.  
%Several scenarios try to explain star formation associated with expanding shells such as the ``Collect and Collapse'' 
Shells are commonly observed in the Milky Way and other galaxies, but most diffuse shells seen in HI or the infrared do not have obvious triggered star formation. Dense molecular shells and pillars around HII regions often do have such triggering \citep[e.g.][]{Elmegreen2011}. 
Several scenarios try to explain star formation associated with expanding shells such as the ``Collect and Collapse'' \citep[e.g.][]{ElmegreenLada1977, Whitworthetal1994},   %Elmegreen & Lada 1977, ApJ, 214, 725; Whitworth et al. 1994, MNRAS, 268, 291), 
or the ``Radiative Driven Implosion'' \citep[][]{Deharvengetal2010} or even the Enhancement of pre-existing Density substructures and subsequent Global Implosion \citep[e.g.][]{Palmeirimetal2017, Walchetal2015}. %They have, however not addressed the specific observables we have obtained for Lupus~I (i.e. the double peaked PDFs and irregular sonic velocities). Lupus I is likely a piece of the HI shell around USco that is undergoing a phase transition to molecular gas
%and newly formed stars. At least, this is consistent with our molecular and HI spatio-kinematic analysis.
The trans-sonic irregular motion in Lupus I can be explained by the Vishniac thin shell instability. 
This supports a picture where clumping and possibly molecular gas formation was going on for a while, but star 
formation only set in when an external compression, quite possibly due to a supernova, triggered it. 
This is also consistent with the double-peaked density PDF in Lupus I and the coeval onset of star formation
in the nearby $\rho$~Ophiucus cloud, which could have been triggered by the same event. 
If the star formation was also related to gradual transformation of gas collected in a shell, then 
it would be more difficult to understand the sudden and coeval onset of star formation in both,
Lupus I and $\rho$~Oph. 
Thus, the data support a more complex, multi-stage formation process for Lupus I than a simple collect
and collapse scenario.

%It seems likely that our findings point to a more complex scenario for Lupus I with several stages of formation.

\section{Summary and Conclusions}
\label{sec:summary}

%Resuming our dust analysis of the \lupus{} molecular cloud and its surroundings of Paper~I,
%we have performed follow-up \cdot{} and \caot{} line observations of \lupus{}
%with the APEX telescope at three distinct cuts through the cloud.
 In this paper we presented new \cdot{} and \caot{} line observations of \lupus{}
with the APEX telescope at three distinct cuts at different parts of the cloud.

A comparison of our CO data to \ion{H}{I} data yielded that the emission
of both tracers lies in the same velocity range.
Moreover, \cao{} emission peaks located in the center-south
of \lupus{} coincide with deficits in \ion{H}{I} emission whereas the position of the \ion{H}{I} emission peak in the north 
shows a deficit in \cao{} emission. This resembles two different stages in the transition between
atomic and molecular material. In the north, a smaller fraction of atomic hydrogen has been
converted into molecular hydrogen and subsequently into dense CO. Therefore, this part
of the cloud is quiescent. The center-south of the cloud has instead enough molecular material
to actively form stars. This could also be a consequence of the lower dust column density
in the northern part of \lupus, because a smaller amount of dust particles reduces the formation
efficiency of molecular hydrogen and provides less shielding from destructive UV radiation.

A comparison of the large-scale \ion{H}{I}, CO, and dust emission in the USco shell and in the vicinity of \lupus{}
revealed that the molecular gas is always found in regions of dust emission and both of those
components are preferentially found behind the outer edge of the atomic shell, i.e. further inwards. 
This is the expected configuration for molecular gas being produced inside the expanding \ion{H}{I} shell.
% However, this picture is not found in all parts of the shell. Thus, it is rather likely that 
% local turbulence and inhomogeneities mix
% the material into complex configurations where each part of the shell wall may have different
% atomic to molecular fractions.
% 
The CO velocities of \lupus{} are in the same range as the \ion{H}{I} velocities of the USco shell and thus
the cloud is co-moving with the shell what would agree with the formation from the shell material.
The timescale of the transition between
atomic and molecular gas in such a massive shell with $M\approx10^4\,\modot$ (estimated from
our model fit to the \ion{H}{I} data)
 was estimated to be $\sim10^6$--$10^7$\,yr. This is consistent
with the age of the USco shell ($\sim5$\,Myr) and the age of the young stellar objects in \lupus{} ($<2$\,Myr).
%Furthermore, about ten times more material than the mass of \lupus{} should have been swept up in the shell
 %at the position of \lupus{} which should have allowed the creation of a $\sim170\modot$ cloud.
% On the western side of \lupus{} there is the 
The UCL wind bubble on the western side of \lupus{}
is colliding with the USco H\,{\rm I} shell at the position of \lupus{} squeezing it in between.
Thus, it acted as a counter-pressure to the expanding USco shell and 
favored this position for an additional compression of the shell material.
In this way a new molecular cloud was created there. This additional collisional pressure might 
explain why not more very young star forming
clouds are seen distributed within the wall of the USco shell.
We showed by a geometrical model that
the observational data are in agreement with this idea. % that \lupus{} is indeed in an interaction 
%zone between the USco \ion{H}{I} shell and the UCL wind bubble.
%%that confirmed the plausibility that \lupus{} might 
%%be located in the interaction zone of the USco \ion{H}{I} shell
%%and the UCL wind bubble 
%In the model two spheres represent the shell and the bubble. 
%The USco sphere was placed at 145\,pc, the UCL sphere
%5\,pc in front of it (with respect to us) what corresponds to the average distance to each subgroup. 
%The calculated touch point of the two spheres is close to the
%possible position of \lupus{} and is located in a region on the USco sphere that
%has similar predicted line-of-sight \ion{H}{I} velocities to that of the CO gas in \lupus.
This model also indicated that \lupus{} can be expanding with the USco shell and that the cloud 
is located at the edge of the inner shell. \\% which is consistent with observations.\\
%\lupus{} is also located in a place where gas leftover from the formation epoch of UCL and USco would be expected. A pre-existing density enhancement could also have promoted cloud formation at this location. 
%The results we have presented in this work 
We suggest that \lupus{} 
was and is strongly affected by large-scale external compression originating %compressing forces 
from the expansion of the USco H\,{\rm I} shell and the UCL wind bubble, and that the %bubble.
 cloud was formed out of the atomic material swept up by
the USco shell and is now expanding with the shell. 
Pre-existing gas structure and collision with the UCL wind bubble 
were likely decisive factors that enhanced a local inhomogeneity and the density
in the USco shell at a position that favored the formation of \lupus.
%To further test the here presented formation scenario of \lupus{},
%we will compare our observational results to dedicated numerical simulations performed
%in our group. 

\begin{acknowledgements}
We would like to thank the referee for his/her constructive comments which helped to improve this paper. 
This work was supported by funding from Deutsche Forschungsgemeinschaft under DFG project numbers PR~569/10-1 and PR~569/10-2
in the context of the Priority Program 1573 "Physics of the Interstellar Medium".

Additional support came from funds from the Munich Cluster of Excellence "Origin and Structure of the Universe".

% We thank Jean-Philippe Bernard for computing the {\it Planck} offsets for the {\it Herschel} maps.

% B.G. thanks Veronica Roccatagliata and Jim Dale for fruitful and helpful discussions.

The \textit{Herschel} spacecraft was designed, built, tested, and launched under
a contract to ESA managed by the Herschel/Planck Project team by an industrial
consortium under the overall responsibility of the prime contractor Thales
Alenia Space (Cannes), and including Astrium (Friedrichshafen) responsible for
the payload module and for system testing at spacecraft level, Thales Alenia
Space (Turin) responsible for the service module, and Astrium (Toulouse) responsible
for the telescope, with in excess of a hundred subcontractors.

Based on observations obtained with {\it Planck} (\url{http://www.esa.int/Planck}), an ESA science mission with instruments and contributions 
directly funded by ESA Member States, NASA, and Canada.

HIPE is a joint development by the {\it Herschel} Science Ground Segment Consortium, consisting of ESA, the NASA {\it Herschel} Science Center, and the HIFI, PACS and SPIRE consortia.

This research has made use of the SIMBAD database,
operated at CDS, Strasbourg, France.

We acknowledge the use of NASA's {\it SkyView} facility (http://skyview.gsfc.nasa.gov) located at NASA Goddard Space Flight Center.

\end{acknowledgements}
\begin{appendix}
\section{Calculations of gas temperatures and column densities}
\label{sec:appendix-lte}
\subsection{Excitation temperature}
The excitation temperature can be calculated via
\begin{equation}
T_\mathrm{ex}(\tau)=\frac{T^{18}_\nu}{\ln\left\{1+\left[  \frac{T^{18}_\mathrm{peak}}{T^{18}_\nu}\left(1-\exp(-\taua)\right)^{-1}+J(T_\mathrm{bg})/T^{18}_\nu \right]^{-1}\right\}}
\label{eq:tex}
\end{equation}
where
\begin{equation}
J(T)\equiv\frac{c^2}{2k_B\nu^2}\,B_\nu(T)=\rm T_\nu\,[\exp(\rm T_\nu/ T)-1]^{-1}
\label{eq:j(t)}
\end{equation}
$\rm T_\nu=\frac{h\nu_0}{k_B}$, and $T_\mathrm{bg}=2.73\,\mathrm{K}$ the temperature of the cosmic
microwave background (CMB). Eq.\,\ref{eq:tex}, \ref{eq:j(t)} are derived from the equations given in \citet{bookwilson}.
\begin{table}[htb]
 \centering
 \caption[Constants of the \cdot{} and \caot{} molecular line transitions]{Constants of the \cdot{} and \caot{} lines that are needed for the calculations in this chapter.
$\nu_0$ is the line's rest frequency, $T_\nu=h\,\nu_0 / k_B$, $\mu$ is 
the molecule's dipole moment, and $B_\mathrm{rot}$ its rotational constant.}
 \begin{tabular}{c c c c c}
 \hline\hline
 \noalign{\smallskip}
  Line &  $\nu_0$   &  $T_\nu$ & $\mu$ & $B_\mathrm{rot}$ \\
         &  [GHz]      &  [K]        & [D]      & $[10^{10}\,\mathrm{Hz}]$ \\
 \noalign{\smallskip}
 \hline
 \noalign{\smallskip}
  \cdot &  220.3986765   &  10.577469 & 0.122 & 5.509967 \\
  \caot &  219.5603568   &  10.537236 & 0.110 & 5.489009 \\

  \noalign{\smallskip}
 \hline
 \end{tabular}
\label{tbl:co-lines}
\end{table}
\subsection{$H_2$ column density}
For the calculation of the column density we used
\begin{equation}
N(\cao)=\frac{\taua}{1-\exp(-\taua)}f(T_\mathrm{ex})\int{T^{18}_\mathrm{mb}}{\,d\nu}
\label{eq:nco}
\end{equation}
from \cite{co-observations-w43-carlhoff-2013} where the functions $f(\tex)$ and $Q(\tex)$ are defined as
% 

% \begin{eqnarray}
\begin{equation}
f(T_\mathrm{ex})=\frac{3h}{8\pi^3\mu^2}Q(T_\mathrm{ex})\left\{ \left[(J(T_\mathrm{ex})-J(T_\mathrm{bg})\right]\left[1-\exp\left(-\frac{T^{18}_\nu}{T_\mathrm{ex}}\right) \right] \right\}^{-1} \\
\end{equation}
\vspace{-0.4cm}
\begin{equation}
Q(T_\mathrm{ex})=\frac{k_B T_\mathrm{ex}}{J_u B_\mathrm{rot}h}\exp\left(\frac{B_\mathrm{rot}J_u(J_u+1)h}{k_B T_\mathrm{ex}}\right)
\end{equation}

% \end{eqnarray}
% 
with $J_u=2$ the upper level of the $J=2\rightarrow1$ transition, 
$\mu$ the molecule's dipole moment, and $B_\mathrm{rot}$ its rotational constant. 
The partition function $Q(\tex)$ is approximated for a linear molecule \citep{calculate-column-density-mangum&shirley-2015}. 
The factor $\taua/(1-\exp(-\taua))$ corrects the integrated intensity for possible
opacity broadening of the line \citep[see][]{co-opacity-correction-goldsmith-langer-1999}.
But this effect should be in any case small for the \cao{} line because \taua{} stays
below unity in most pixels. 

\begin{figure*}[!htb]
    \centering
%     \subfigure{
    \includegraphics[width=\textwidth]{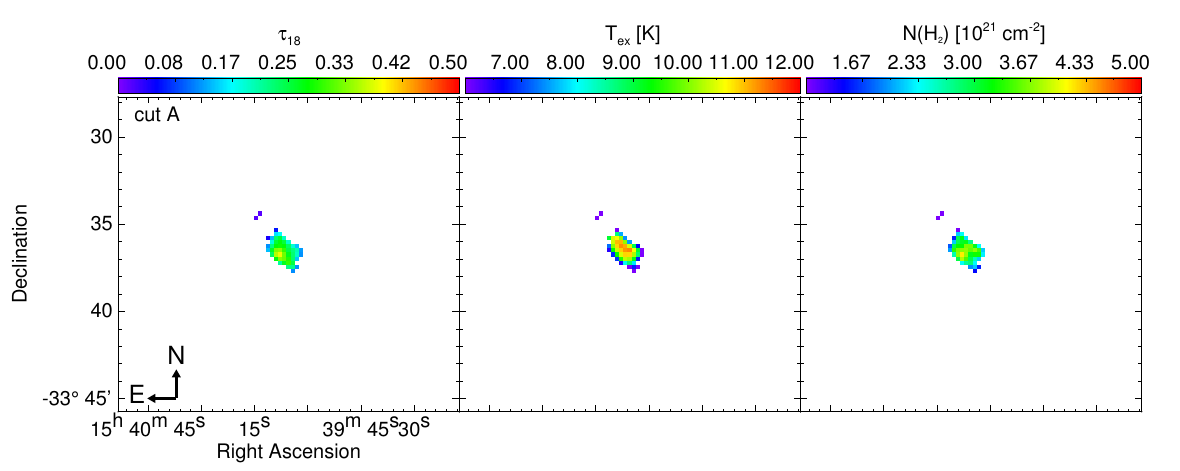}
%    \includegraphics[width=\textwidth]{Lupus_I_cutA_LTE.pdf}
%     }

%     \subfigure{
    \includegraphics[width=\textwidth]{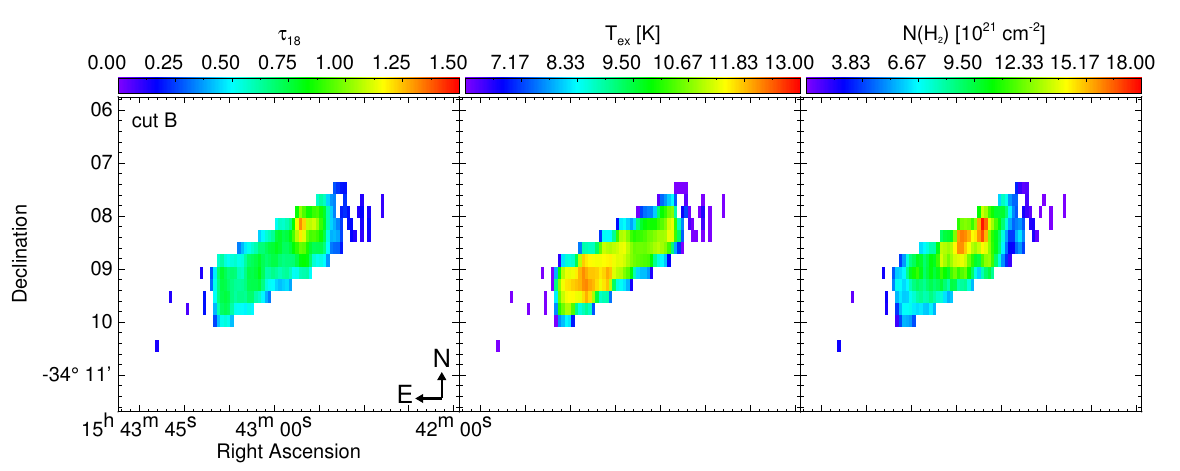}
  %  \includegraphics[width=\textwidth]{Lupus_I_cutB_LTE.pdf}
%     }

%     \subfigure{
    \includegraphics[width=\textwidth]{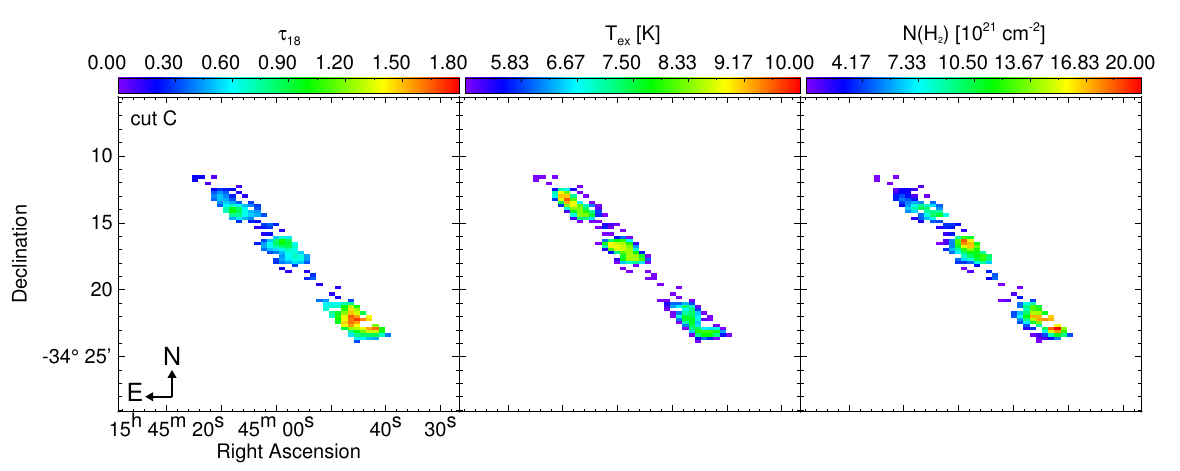}
%   \includegraphics[width=\textwidth]{Lupus_I_cutC_LTE.pdf}
%     }
\caption[Results of the LTE analysis for the three cuts]{Results of the LTE analysis for cut~A, B, and C (from top to bottom). The left panel shows 
the optical depth \taua, the middle panel the excitation temperature \tex, and the right panel the \hzw{}
column density. %\taud{} can easily be calculated by multiplying the colorbar values of \taua{} by $\Gamma=7.3$.
%The intensity scaling stays the same.
All maps were convolved with a Gaussian of two pixels which corresponds approximatively to the 
 beam size. 
%As an effect of that the edge pixels have seemingly lower values because of the adjacent undefined values.
Pixels within each cut for which the \cdot{} and the \caot{} spectrum had a $\mathrm{S/N}<4$ are left blank.}
\label{img:lte-results}
\end{figure*}
\begin{figure*}
    \centering
%     \subfigure{
% \resizebox{\hsize}{!}{
        \includegraphics[width=0.47\textwidth]{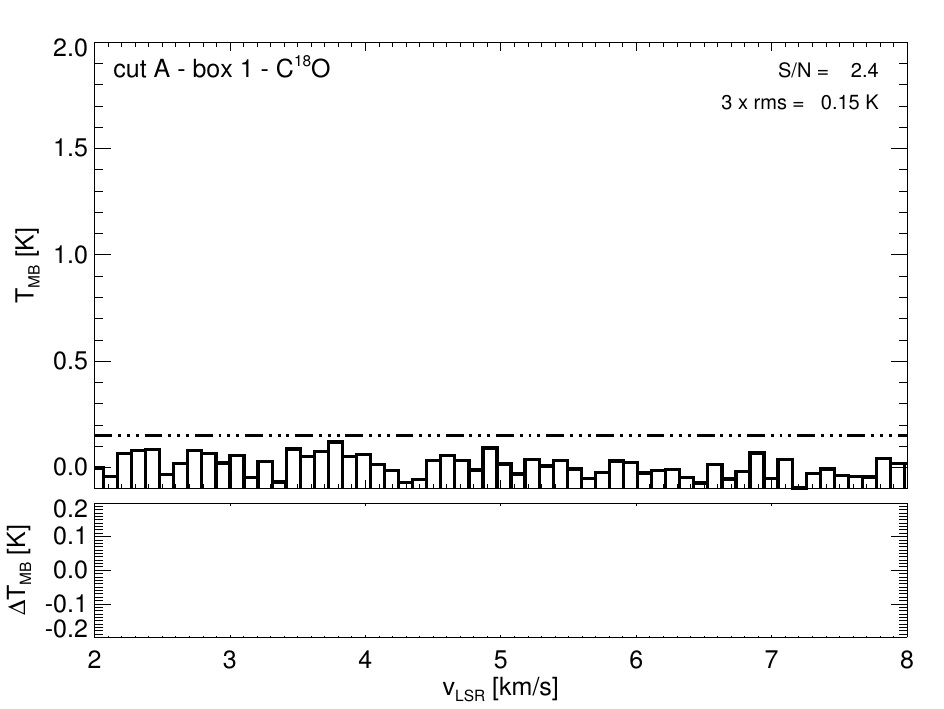}
%        \includegraphics[width=0.47\textwidth]{Lupus_I_cutA_average_box1_Gaussfit_C18O}
%         \includegraphics[width=\textwidth]{figure_cutA_C18O.png}
% 
% }
% 
%     }
%     ~ %add desired spacing between images, e. g. ~, \quad, \qquad, \hfill etc. 
%     \subfigure{
% 
        \includegraphics[width=0.47\textwidth]{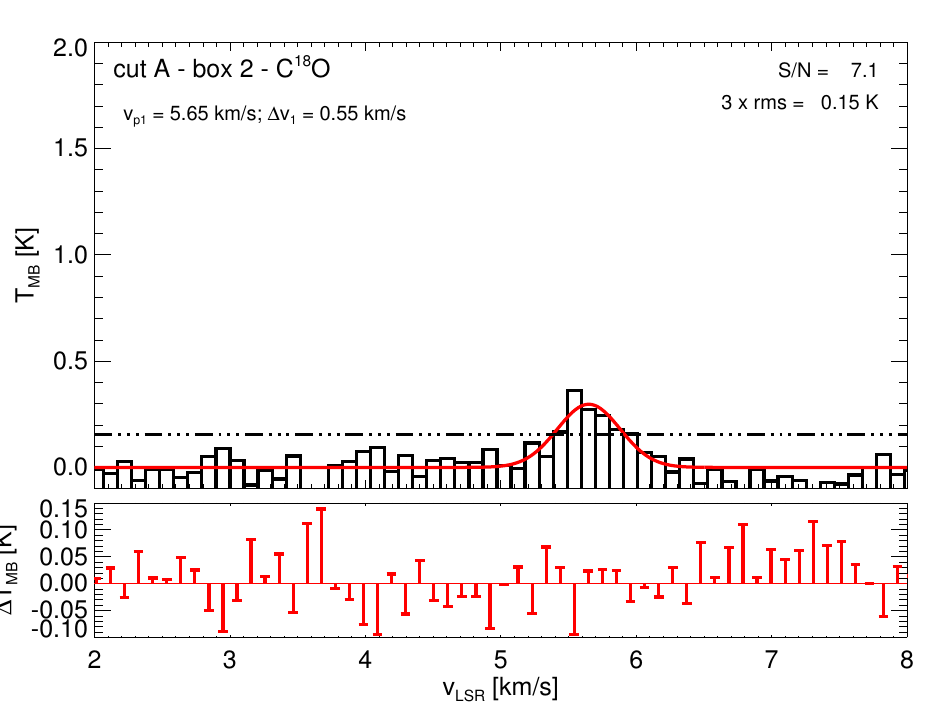}
%       \includegraphics[width=0.47\textwidth]{Lupus_I_cutA_average_box2_Gaussfit_C18O}
% %     }
% 
% %     ~ %add desired spacing between images, e. g. ~, \quad, \qquad, \hfill etc. 
%     %(or a blank line to force the subfigure onto a new line)
% %     \subfigure{
% % \vspace{7cm}

        \includegraphics[width=0.47\textwidth]{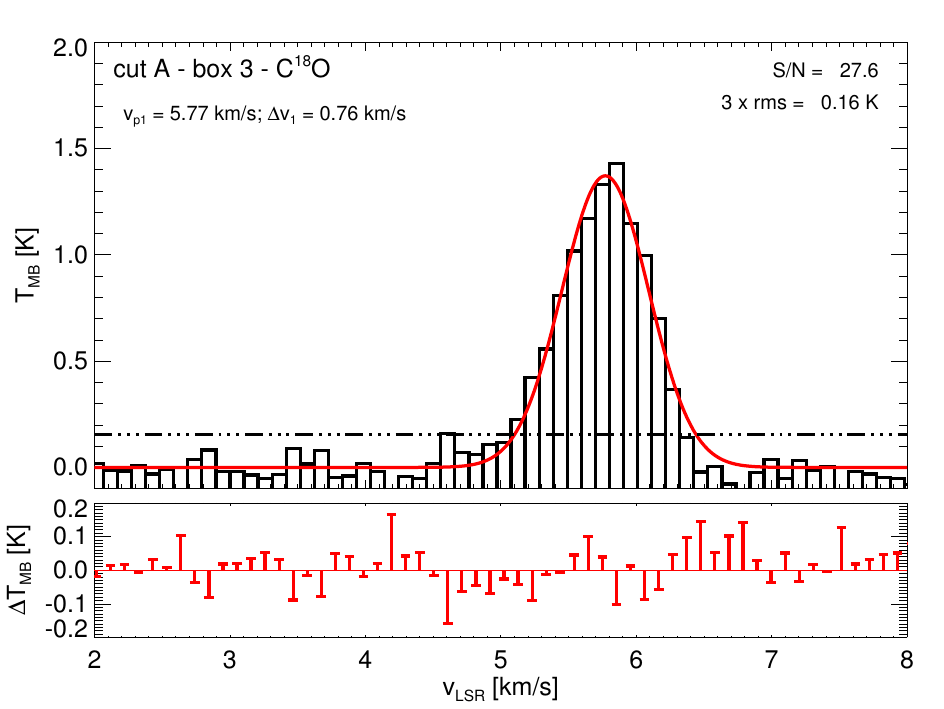}
%        \includegraphics[width=0.47\textwidth]{Lupus_I_cutA_average_box3_Gaussfit_C18O}
% %     }
% %     \sudfigure{
% 
        \includegraphics[width=0.47\textwidth]{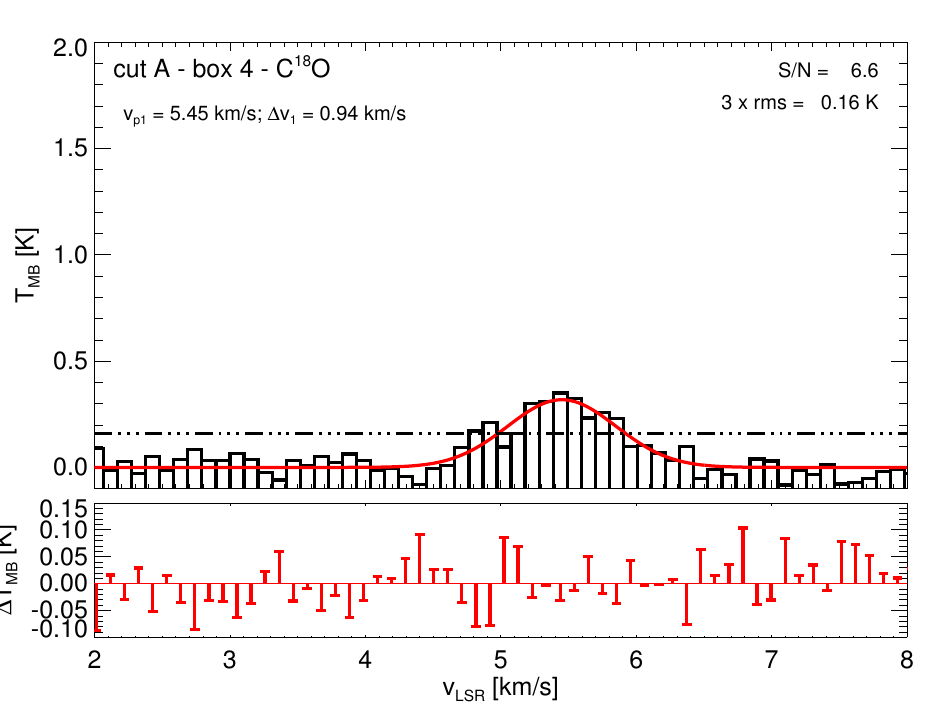}
%        \includegraphics[width=0.47\textwidth]{Lupus_I_cutA_average_box4_Gaussfit_C18O}
% %     }
% % 
% %     \subfigure{
% % \vspace{7cm}

        \includegraphics[width=0.47\textwidth]{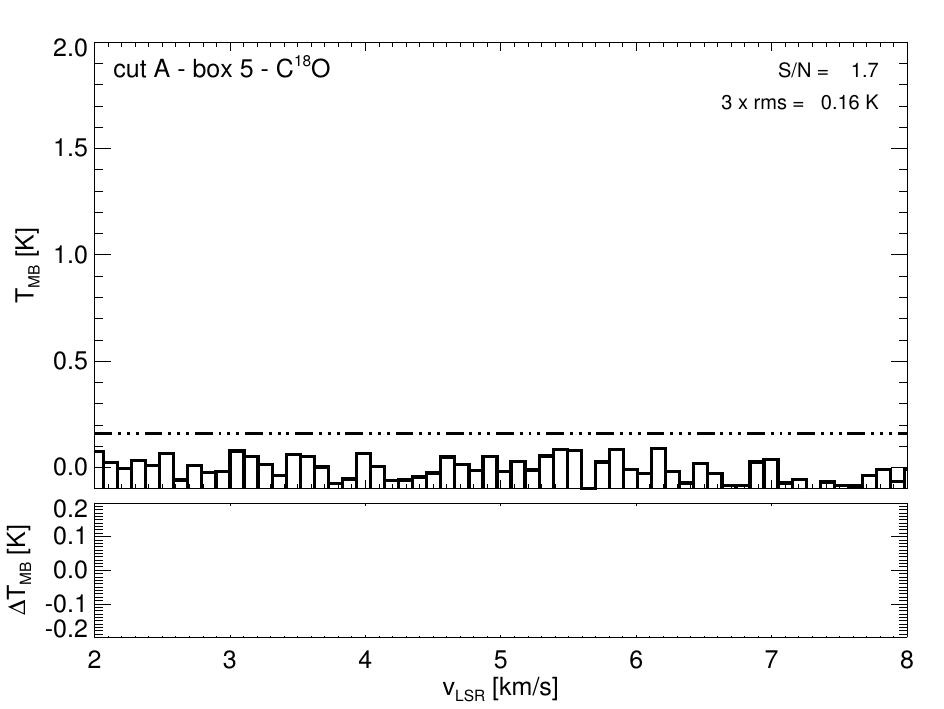}
%        \includegraphics[width=0.47\textwidth]{Lupus_I_cutA_average_box5_Gaussfit_C18O}
% %     }
% %     \subfigure{
% 
        \includegraphics[width=0.47\textwidth]{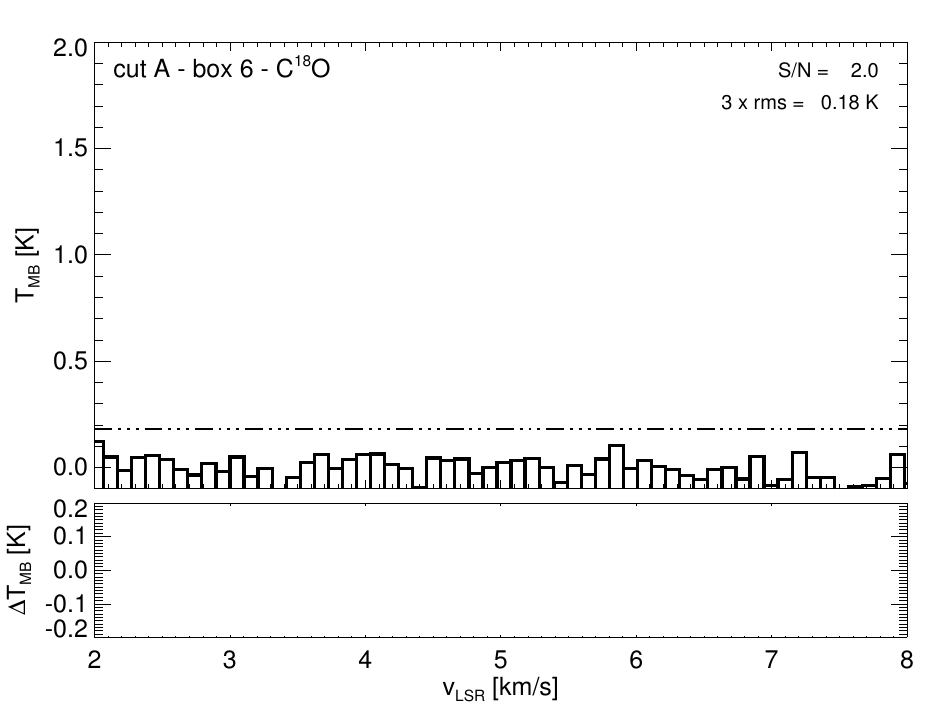}
%        \includegraphics[width=0.47\textwidth]{Lupus_I_cutA_average_box6_Gaussfit_C18O}
% %     }
    \caption[Gaussian fits to the average \cao{} spectra of the boxes in cut~A ]{Histograms of the average \caot{} spectra of the six boxes across cut~A with a 
    bin size of $0.1\kms$. 
    	 The black dash-dotted line marks the $3\times\mathrm{rms}$ limit.
    	 The values of $3\times\mathrm{rms}$ and the signal-to-noise ratio S/N are given in the upper right
    	 of each plot.
The red solid line shows the Gaussian fit to the spectrum. If more than one
component was fitted, each one of them is represented by a blue dashed
line and the red solid line is their sum.
   The residuals of the fit are given by the red bars in the small plot below each spectrum.
	 Peak positions $v_{\mathrm{p}_i}$ and FWHM $\Delta v_i$ of each component in one box 
	 are given in the upper left of each plot. A summary of all these values can be found
	 in Table~\ref{tbl:peaks_fwhm_gauss} where they are denoted by an upper index G.
	 %The numbering of the components 
	 %is done according to the grouping performed in the empirical analysis (see Table~\ref{tbl:peaks_fwhm}
	 %and Figure~\ref{img:peaks_fwhm})  and thus not always in increasing order. 
	 }
\label{img:cutA_C18O_Gauss}
\end{figure*}
\newpage
% \clearpage
\begin{figure*}
    \centering
%     \subfigure{
        \includegraphics[width=0.47\textwidth]{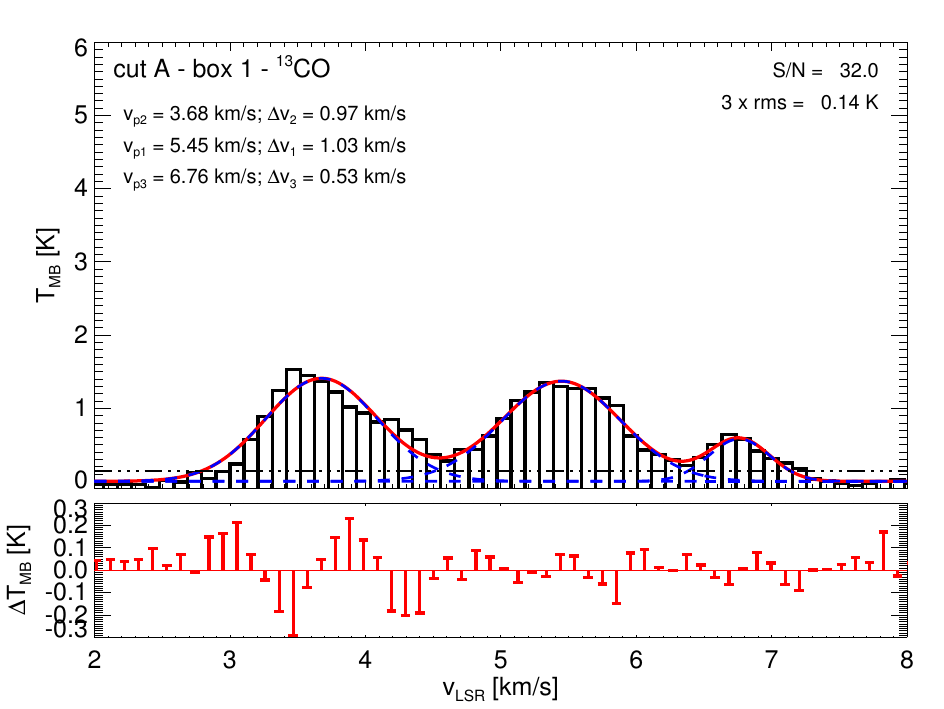}
%        \includegraphics[width=0.47\textwidth]{Lupus_I_cutA_average_box1_Gaussfit_13CO}
%     }
%     ~ %add desired spacing between images, e. g. ~, \quad, \qquad, \hfill etc. 
%     \subfigure{
        \includegraphics[width=0.47\textwidth]{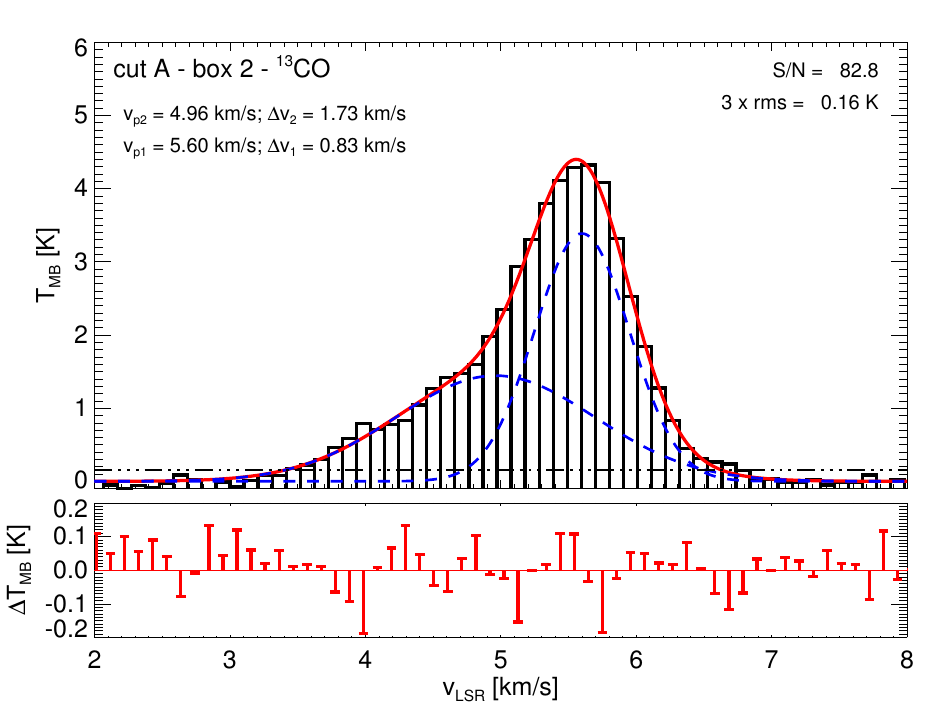}
%        \includegraphics[width=0.47\textwidth]{Lupus_I_cutA_average_box2_Gaussfit_13CO}
%     }
% \vspace{7cm}
%     ~ %add desired spacing between images, e. g. ~, \quad, \qquad, \hfill etc. 
%     \subfigure{
        \includegraphics[width=0.47\textwidth]{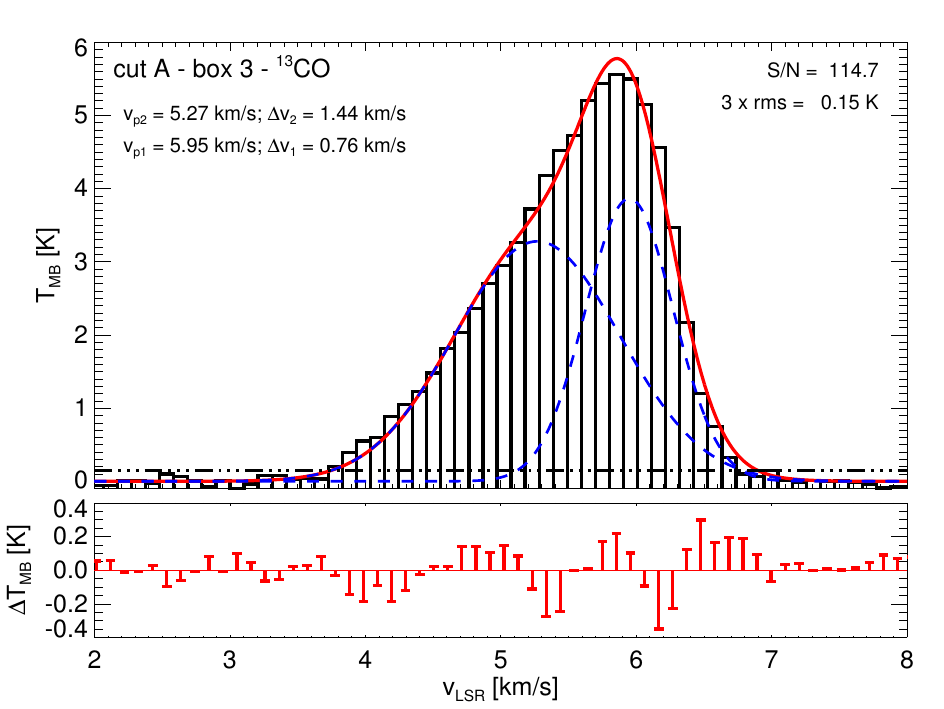}
 %       \includegraphics[width=0.47\textwidth]{Lupus_I_cutA_average_box3_Gaussfit_13CO}
%     }
%     \subfigure{
        \includegraphics[width=0.47\textwidth]{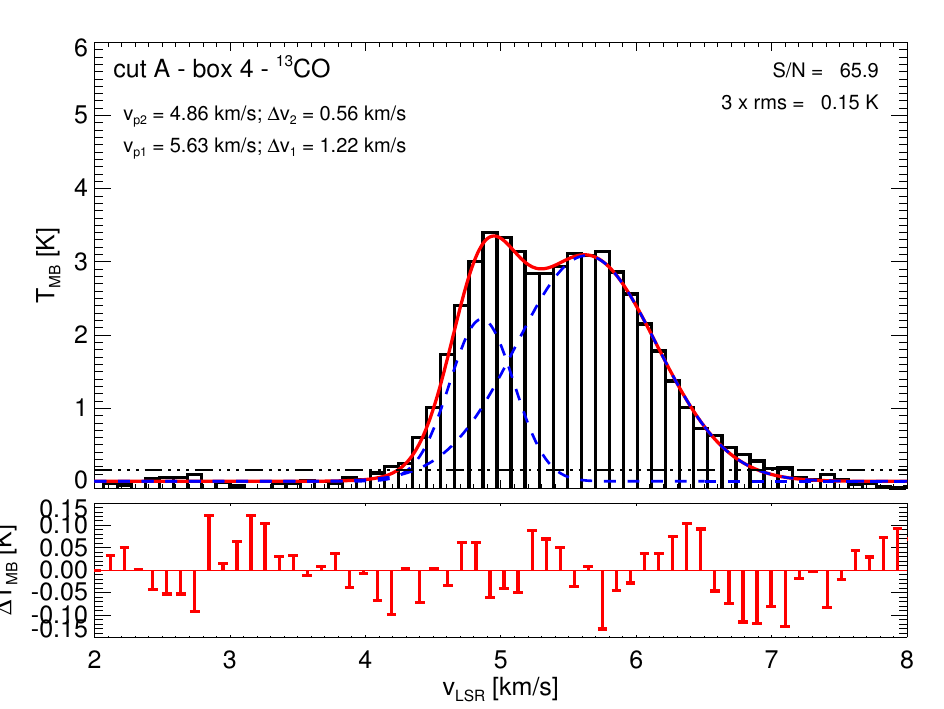}
 %       \includegraphics[width=0.47\textwidth]{Lupus_I_cutA_average_box4_Gaussfit_13CO}
%     }
% \vspace{7cm}
%     \subfigure{
        \includegraphics[width=0.47\textwidth]{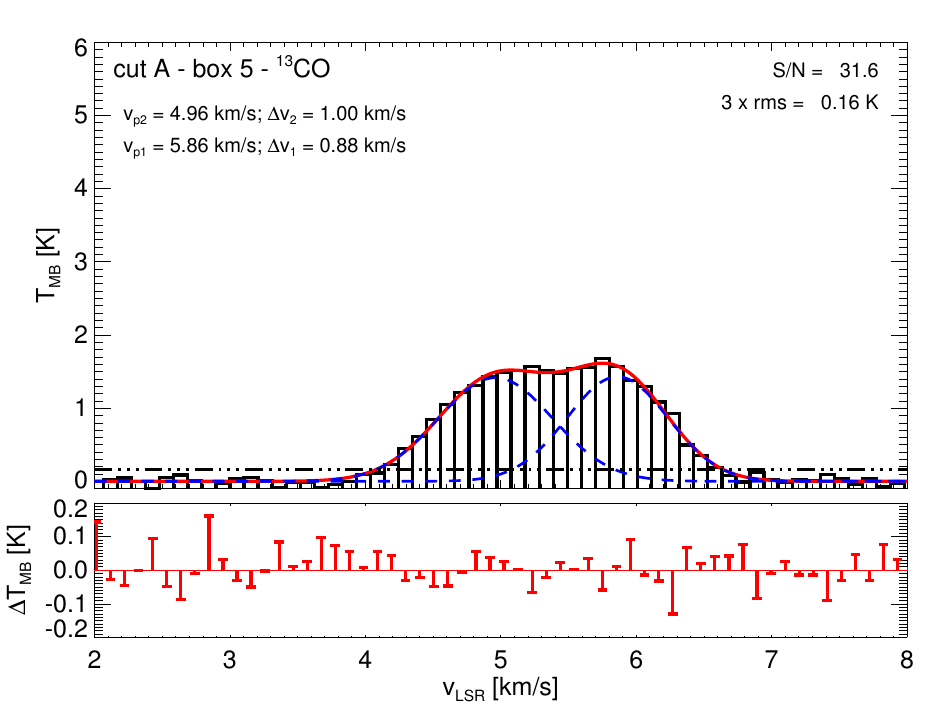}
 %      \includegraphics[width=0.47\textwidth]{Lupus_I_cutA_average_box5_Gaussfit_13CO}
%     }
%     \subfigure{
        \includegraphics[width=0.47\textwidth]{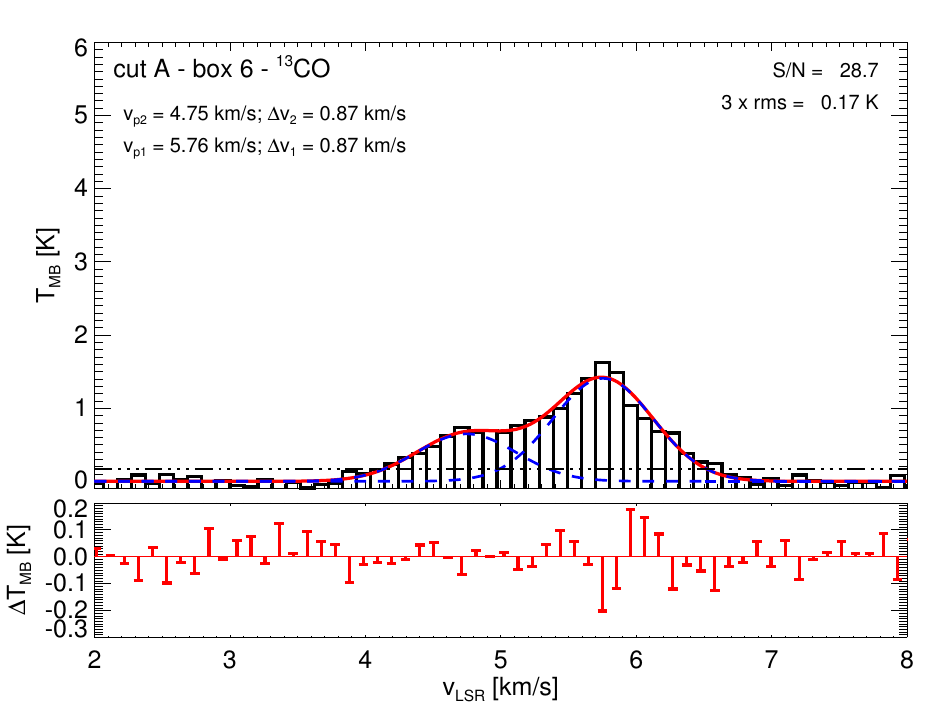}
 %       \includegraphics[width=0.47\textwidth]{Lupus_I_cutA_average_box6_Gaussfit_13CO}
%     }
    \caption[Gaussian fits to the average \cdo{} spectra of the boxes in cut~A ]{Histograms of the average \cdot{} spectra of the six boxes across cut~A with a 
    bin size of $0.1\kms$. 
    	 The black dash-dotted line marks the $3\times\mathrm{rms}$ limit.
    	 The values of $3\times\mathrm{rms}$ and the signal-to-noise ratio S/N are given in the upper right
    	 of each plot.
The red solid line shows the Gaussian fit to the spectrum. If more than one
component was fitted, each one of them is represented by a blue dashed
line and the red solid line is their sum.
   The residuals of the fit are given by the red bars in the small plot below each spectrum.
	 Peak positions $v_{\mathrm{p}_i}$ and FWHM $\Delta v_i$ of each component in one box 
	 are given in the upper left of each plot. A summary of all these values can be found
	 in Table~\ref{tbl:peaks_fwhm_gauss} where they are denoted by an upper index G.
	 %The numbering of the components 
	 %is done according to the grouping performed in the empirical analysis (see Table~\ref{tbl:peaks_fwhm}
	 %and Figure~\ref{img:peaks_fwhm})  and thus not always in increasing order. 
	 }
\label{img:cutA_13CO_Gauss}
\end{figure*}
% \newpage
% \clearpage
\begin{figure*}
    \centering
%     \subfigure{
        \includegraphics[width=0.47\textwidth]{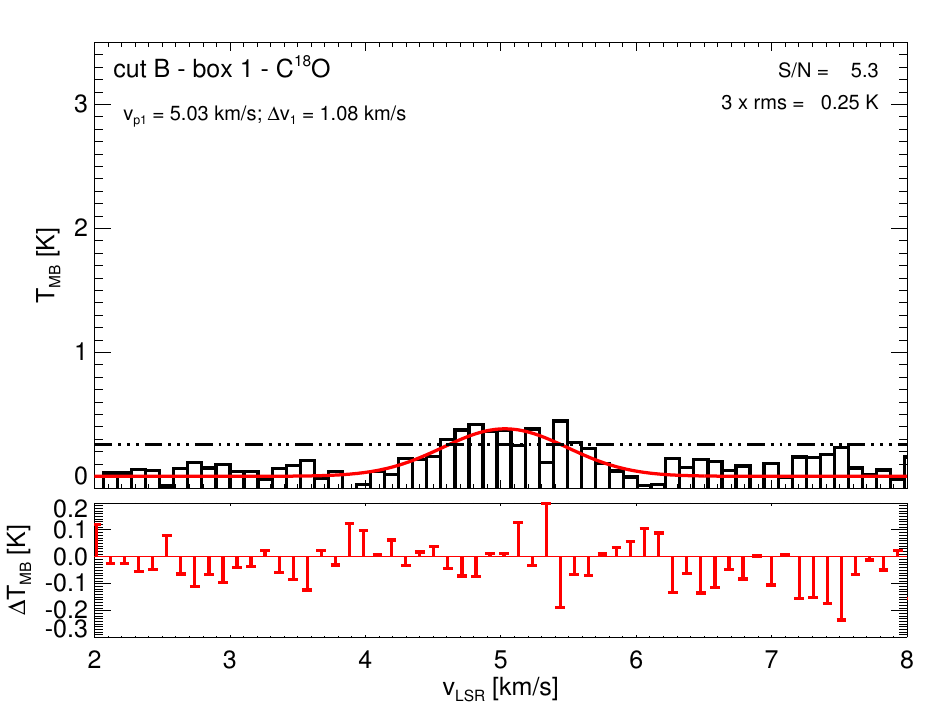}
%        \includegraphics[width=0.47\textwidth]{Lupus_I_cutB_average_box1_Gaussfit_C18O}
%     }
%     ~ %add desired spacing between images, e. g. ~, \quad, \qquad, \hfill etc. 
%     \subfigure{
        \includegraphics[width=0.47\textwidth]{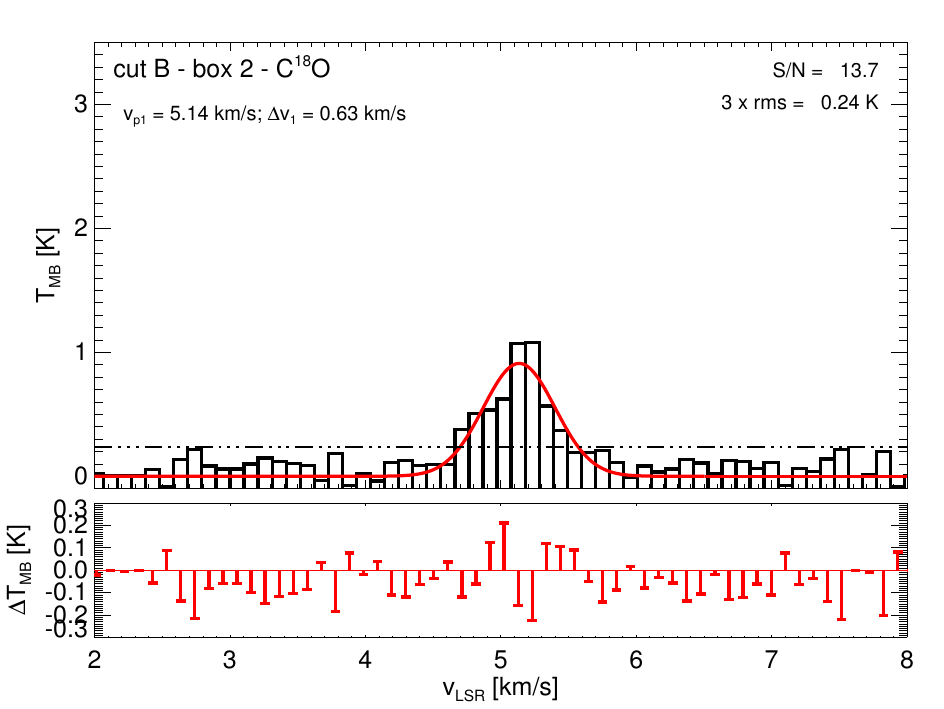}
%        \includegraphics[width=0.47\textwidth]{Lupus_I_cutB_average_box2_Gaussfit_C18O}
%     }
% \vspace{7cm}
%     ~ %add desired spacing between images, e. g. ~, \quad, \qquad, \hfill etc. 
%     \subfigure{
        \includegraphics[width=0.47\textwidth]{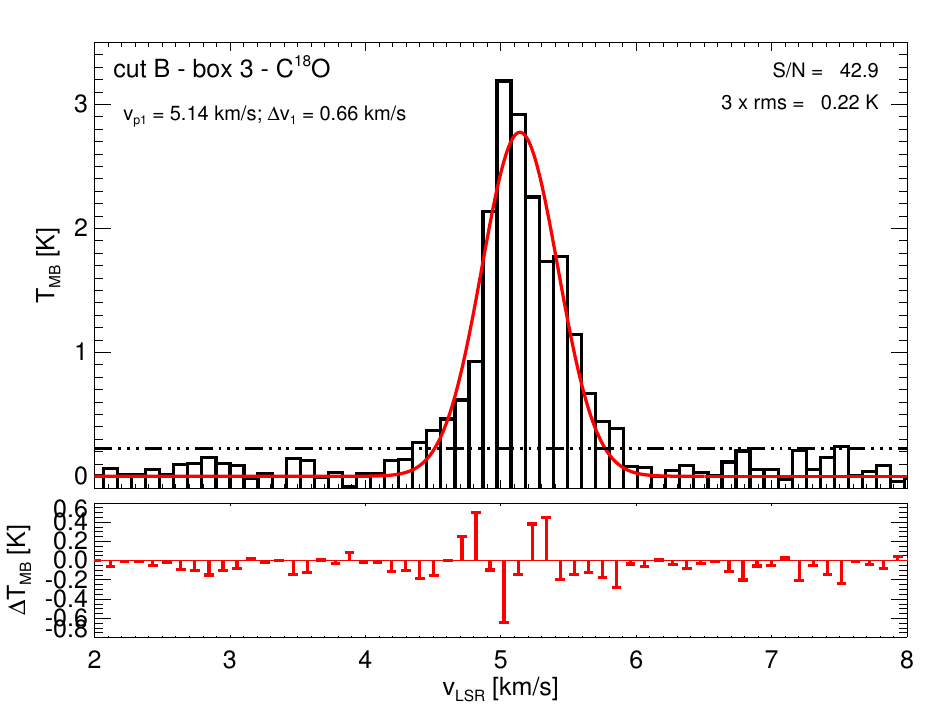}
%       \includegraphics[width=0.47\textwidth]{Lupus_I_cutB_average_box3_Gaussfit_C18O}
%     }
%     \subfigure{
        \includegraphics[width=0.47\textwidth]{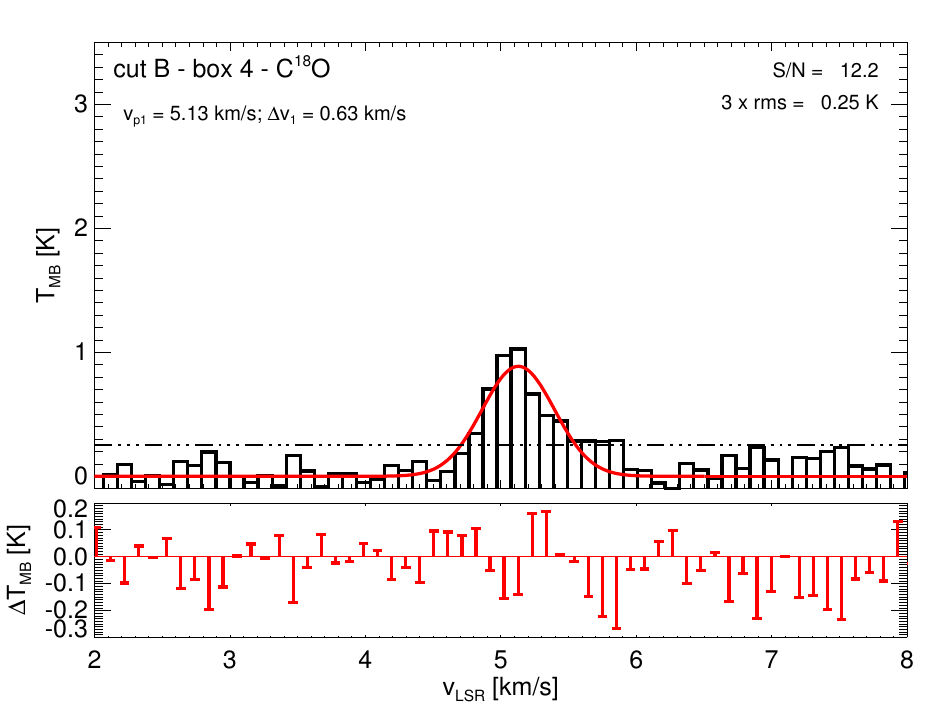}
%        \includegraphics[width=0.47\textwidth]{Lupus_I_cutB_average_box4_Gaussfit_C18O}
%     }
% \vspace{7cm}
%     \subfigure{
        \includegraphics[width=0.47\textwidth]{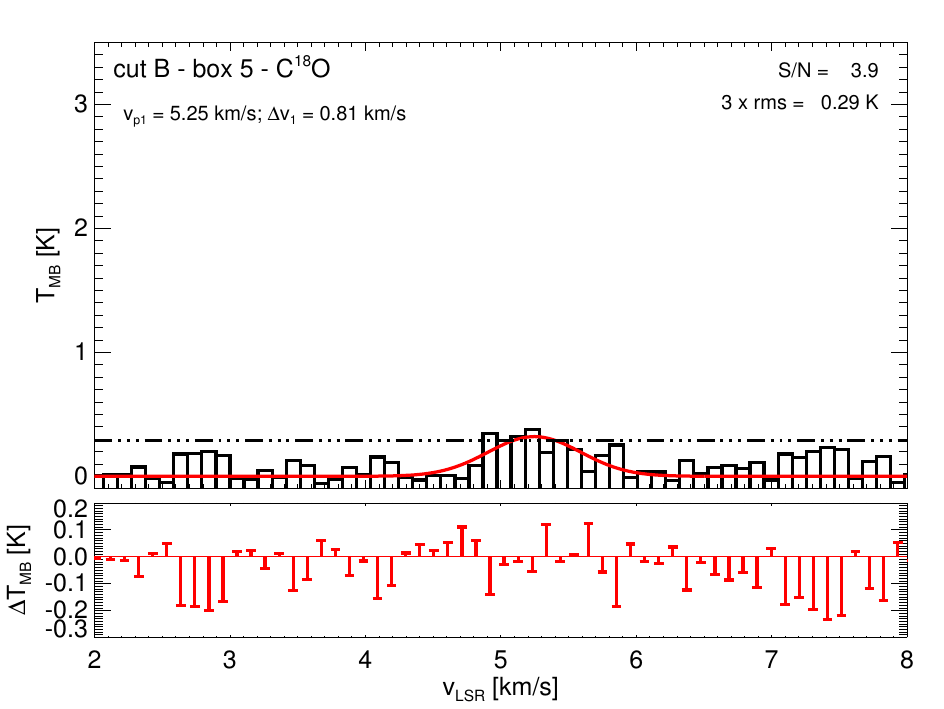}
 %       \includegraphics[width=0.47\textwidth]{Lupus_I_cutB_average_box5_Gaussfit_C18O}
%     }
%     \subfigure{
        \includegraphics[width=0.47\textwidth]{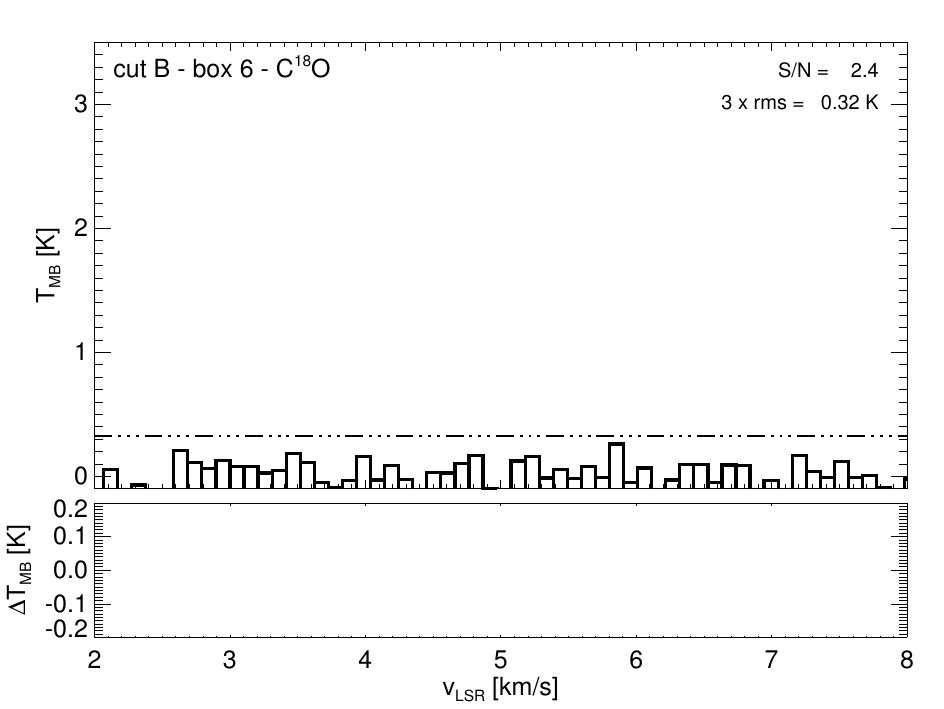}
 %       \includegraphics[width=0.47\textwidth]{Lupus_I_cutB_average_box6_Gaussfit_C18O}
%     }
    \caption[Gaussian fits to the average \cao{} spectra of the boxes in cut~B ]{Same as Figure~\ref{img:cutA_C18O_Gauss} but for cut~B.}
\label{img:cutB_C18O_Gauss}
\end{figure*}
% \newpage
% \clearpage
\begin{figure*}
    \centering
%     \subfigure{
        \includegraphics[width=0.47\textwidth]{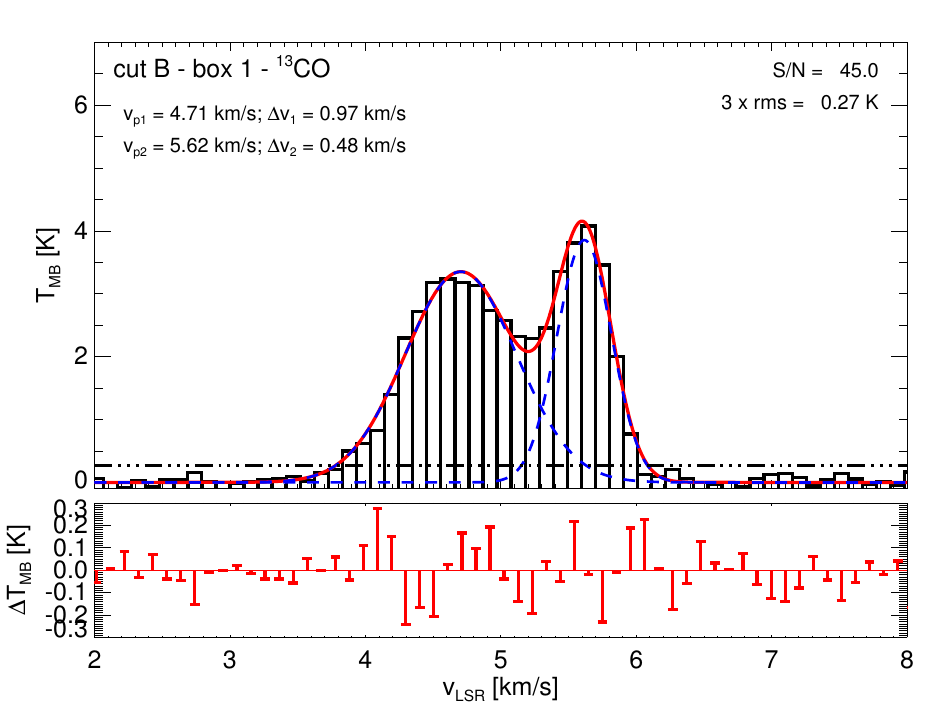}
%        \includegraphics[width=0.47\textwidth]{Lupus_I_cutB_average_box1_Gaussfit_13CO}
%     }
%     ~ %add desired spacing between images, e. g. ~, \quad, \qquad, \hfill etc. 
%     \subfigure{
        \includegraphics[width=0.47\textwidth]{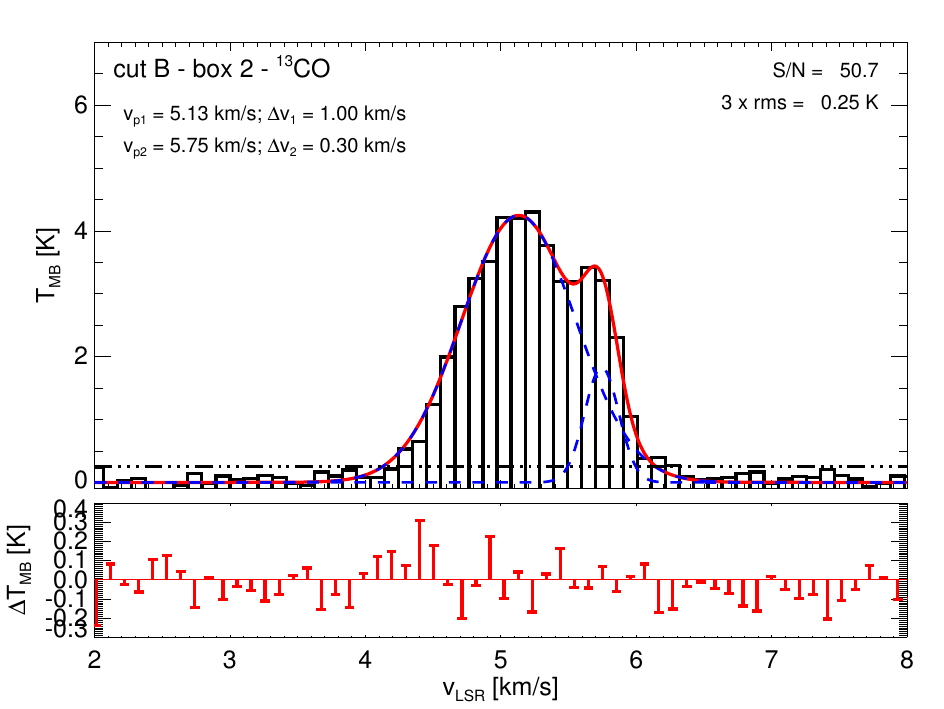}
%        \includegraphics[width=0.47\textwidth]{Lupus_I_cutB_average_box2_Gaussfit_13CO}
%     }
% \vspace{7cm}
%     ~ %add desired spacing between images, e. g. ~, \quad, \qquad, \hfill etc. 
%     \subfigure{
        \includegraphics[width=0.47\textwidth]{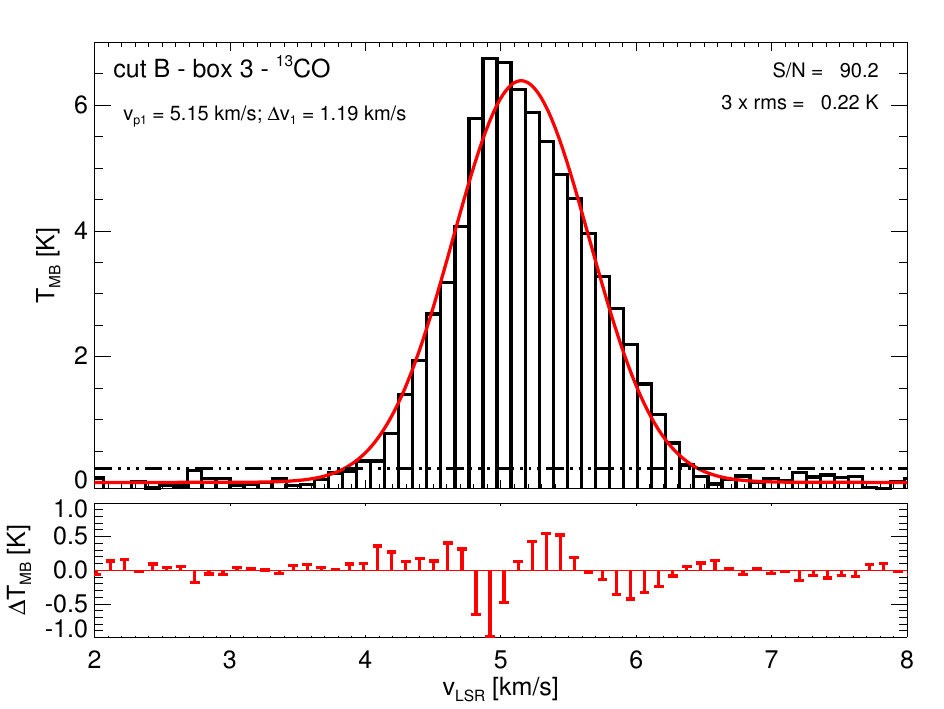}
 %       \includegraphics[width=0.47\textwidth]{Lupus_I_cutB_average_box3_Gaussfit_13CO}
%     }
%     \subfigure{
        \includegraphics[width=0.47\textwidth]{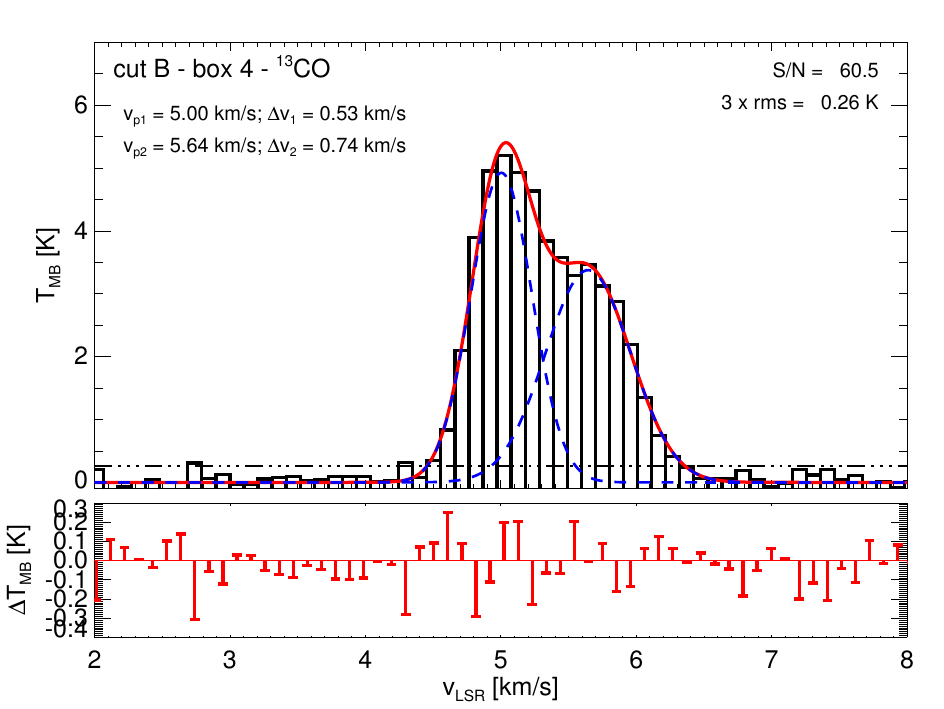}
%        \includegraphics[width=0.47\textwidth]{Lupus_I_cutB_average_box4_Gaussfit_13CO}
%     }
% \vspace{7cm}
%     \subfigure{
        \includegraphics[width=0.47\textwidth]{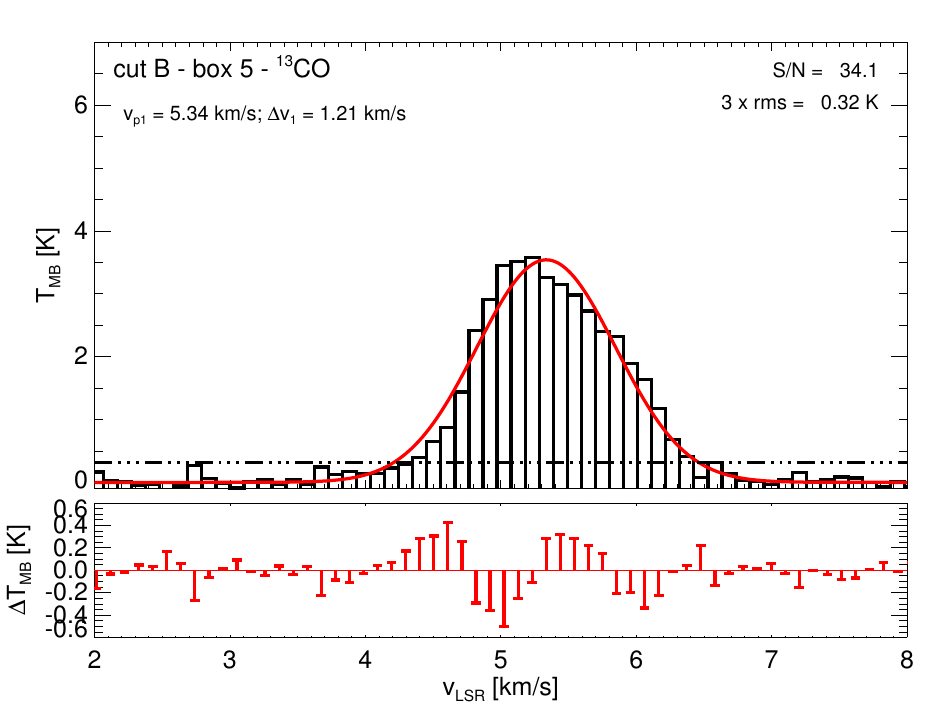}
%       \includegraphics[width=0.47\textwidth]{Lupus_I_cutB_average_box5_Gaussfit_13CO}
%     }
%     \subfigure{
        \includegraphics[width=0.47\textwidth]{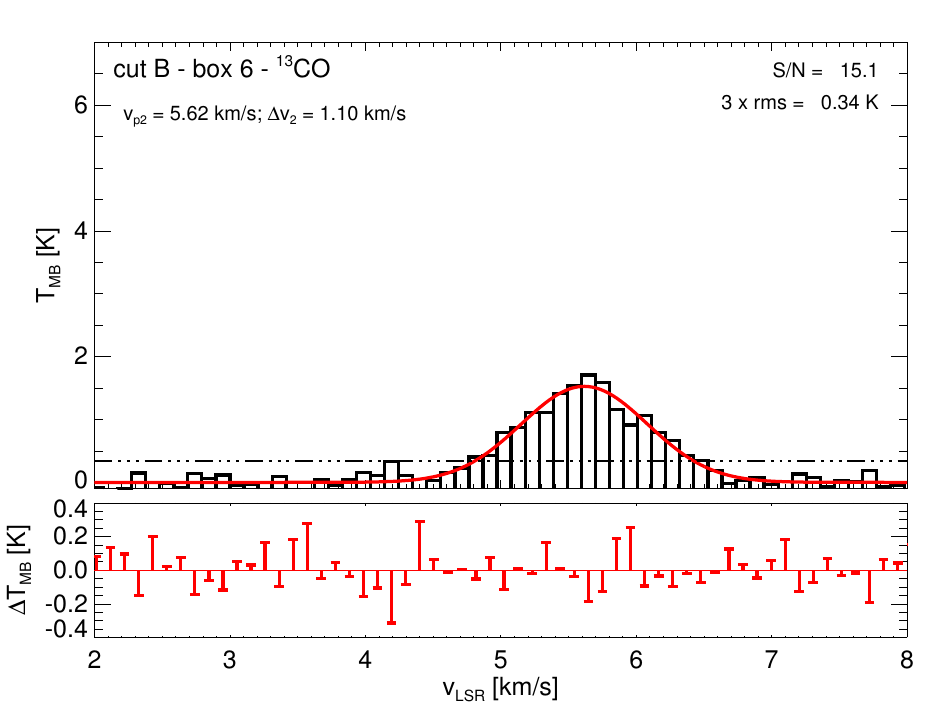}
%        \includegraphics[width=0.47\textwidth]{Lupus_I_cutB_average_box6_Gaussfit_13CO}
%     }
    \caption[Gaussian fits to the average \cdo{} spectra of the boxes in cut~B]{Same as Figure~\ref{img:cutA_13CO_Gauss} but for cut~B.}
\label{img:cutB_13CO_Gauss}
\end{figure*}
% \newpage
% \clearpage
\begin{figure*}
    \centering
%     \subfigure{
        \includegraphics[width=0.47\textwidth]{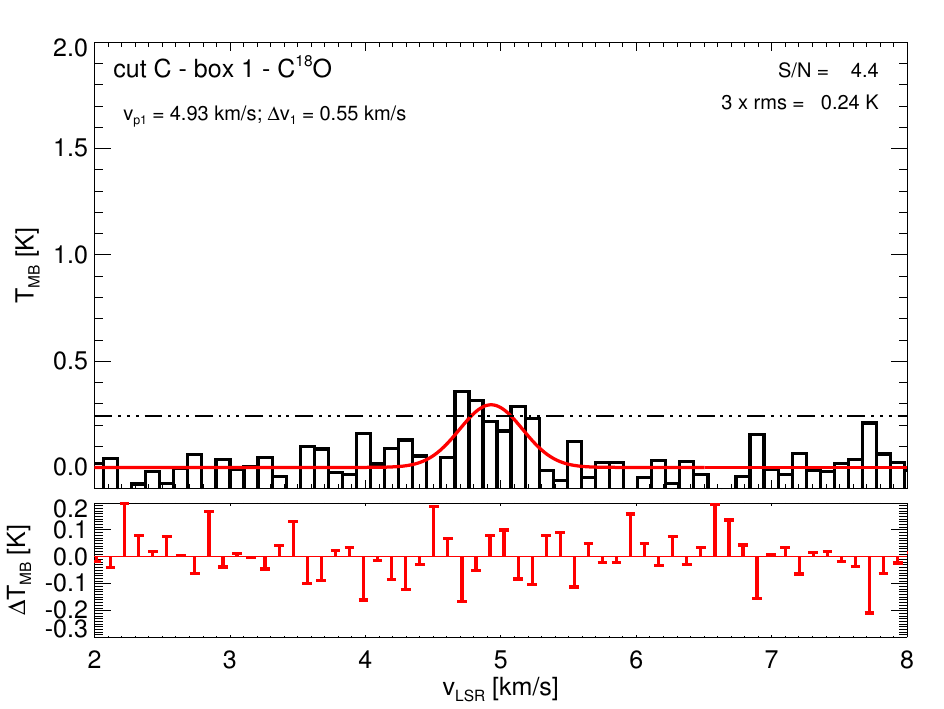}
%        \includegraphics[width=0.47\textwidth]{Lupus_I_cutC_average_box1_Gaussfit_C18O}
%     }
%     ~ %add desired spacing between images, e. g. ~, \quad, \qquad, \hfill etc. 
%     \subfigure{
        \includegraphics[width=0.47\textwidth]{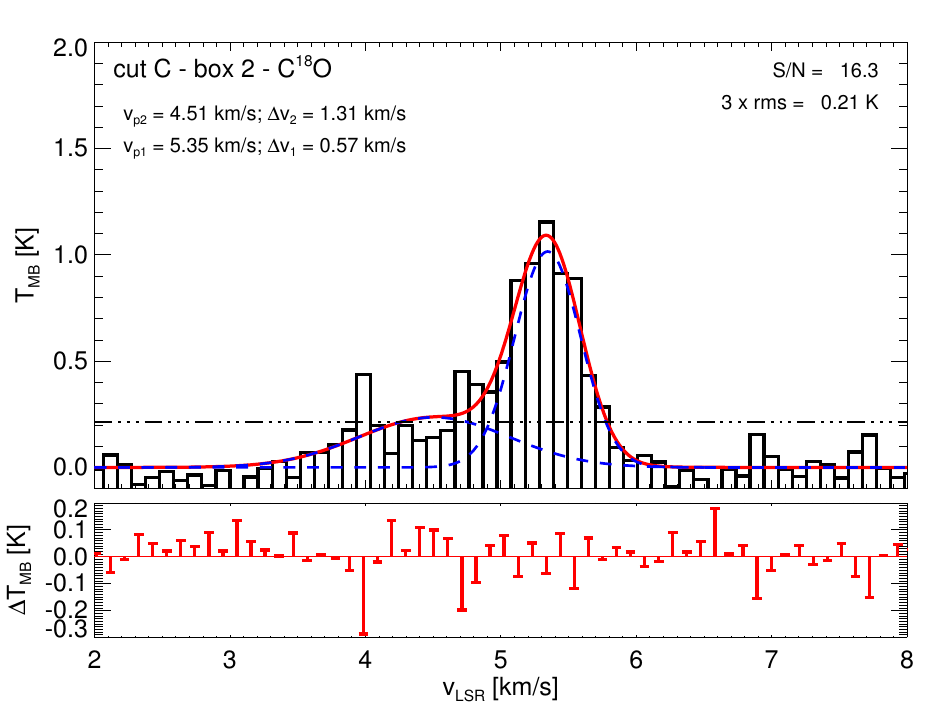}
%       \includegraphics[width=0.47\textwidth]{Lupus_I_cutC_average_box2_Gaussfit_C18O}
%     }
% \vspace{7cm}
%     ~ %add desired spacing between images, e. g. ~, \quad, \qquad, \hfill etc. 
%     \subfigure{
        \includegraphics[width=0.47\textwidth]{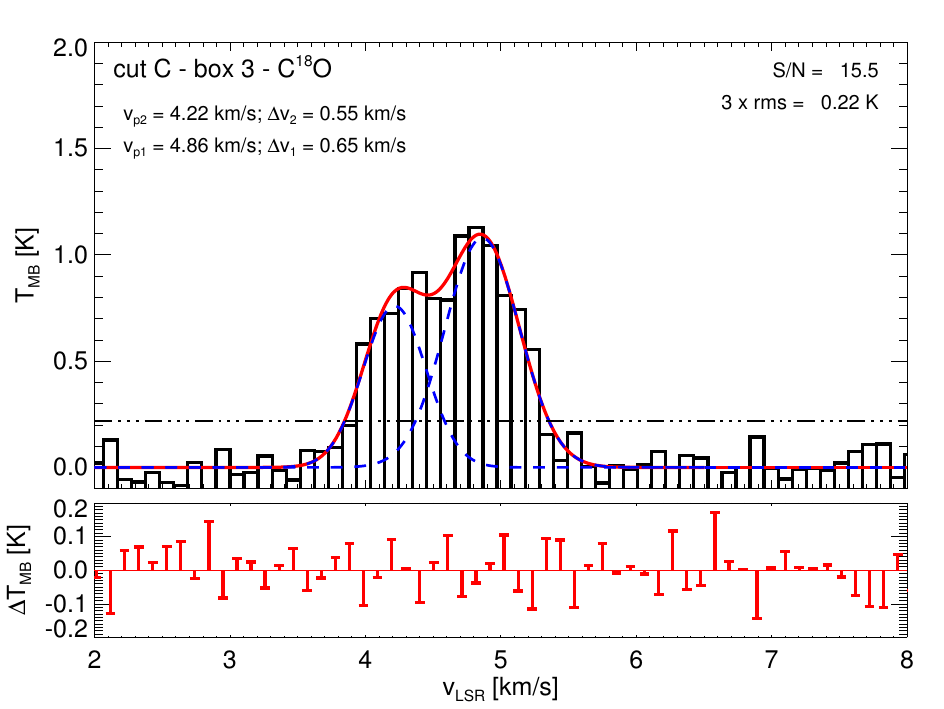}
%        \includegraphics[width=0.47\textwidth]{Lupus_I_cutC_average_box3_Gaussfit_C18O}
%     }
%     \subfigure{
        \includegraphics[width=0.47\textwidth]{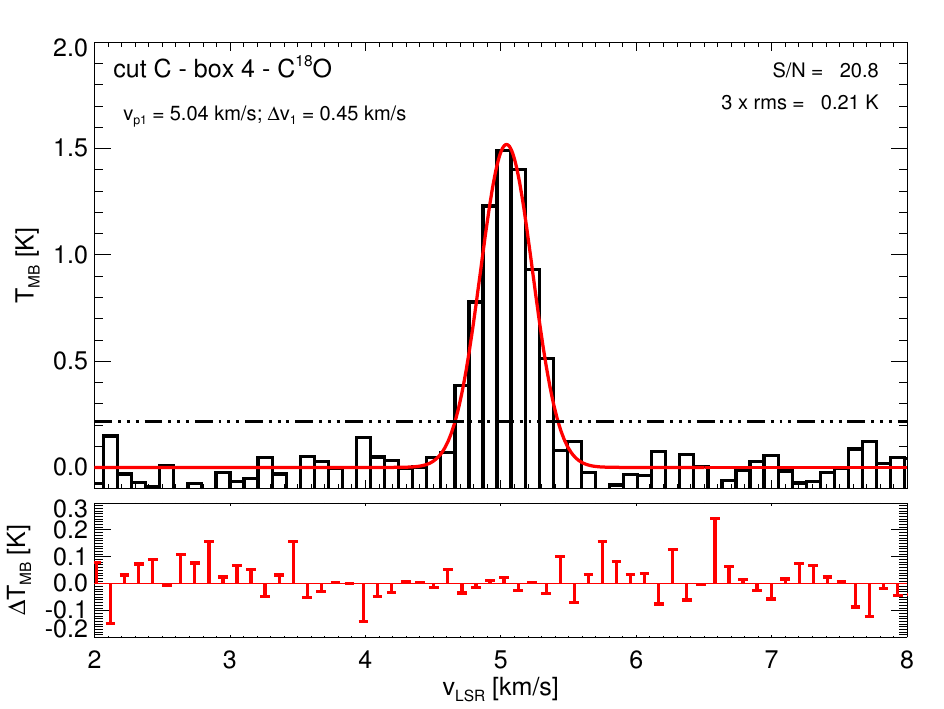}
  %      \includegraphics[width=0.47\textwidth]{Lupus_I_cutC_average_box4_Gaussfit_C18O}
%     }
% \vspace{7cm}
%     \subfigure{
        \includegraphics[width=0.47\textwidth]{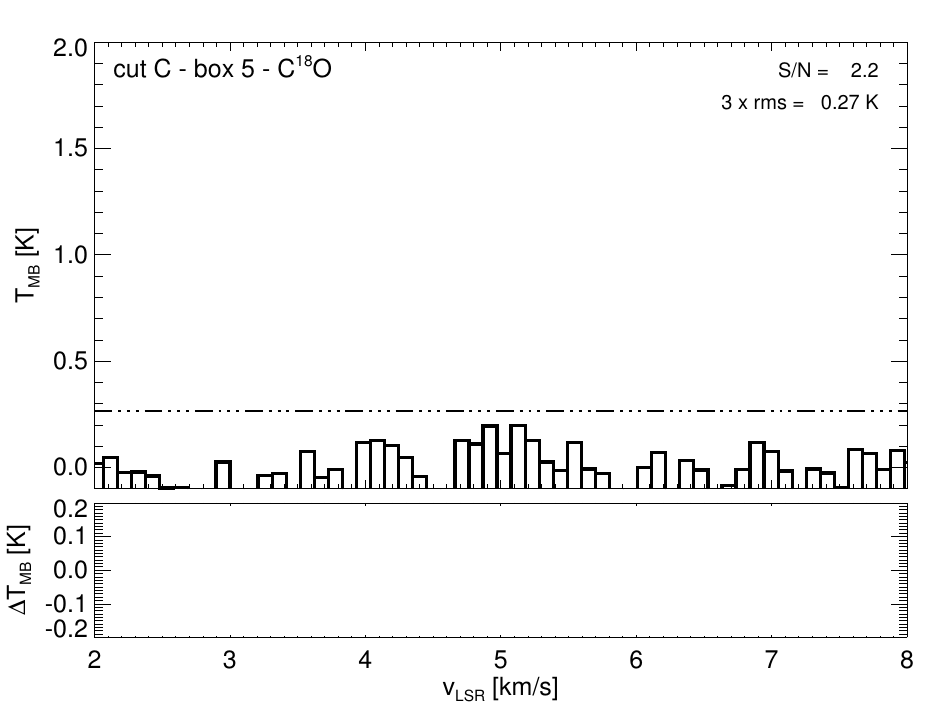}
%        \includegraphics[width=0.47\textwidth]{Lupus_I_cutC_average_box5_Gaussfit_C18O}
%     }
%     \subfigure{
        \includegraphics[width=0.47\textwidth]{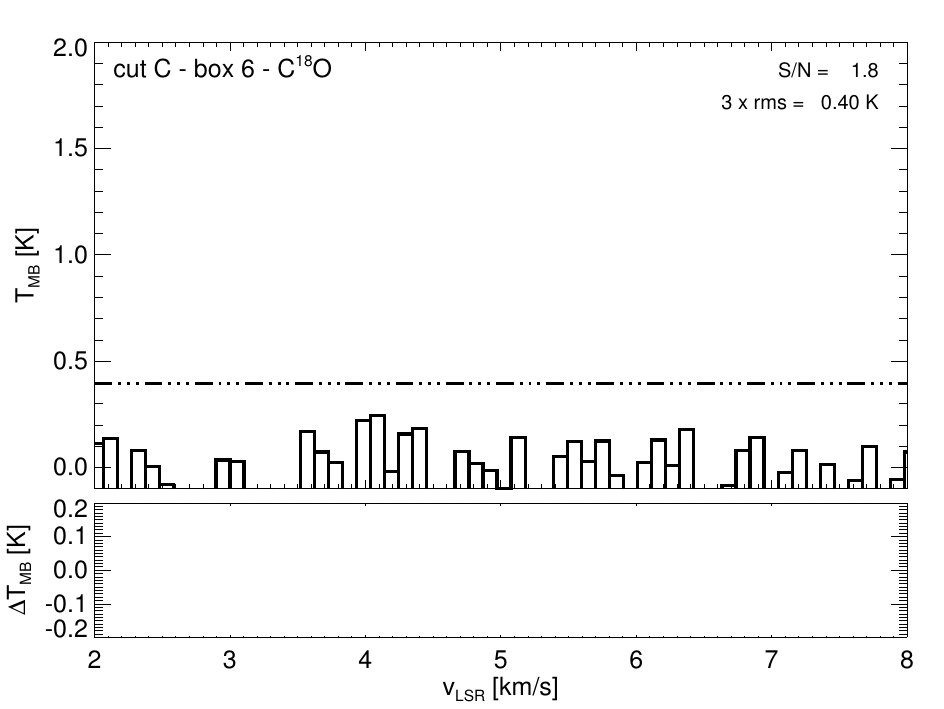}
%        \includegraphics[width=0.47\textwidth]{Lupus_I_cutC_average_box6_Gaussfit_C18O}
%     }
    \caption[Gaussian fits to the average \cao{} spectra of the boxes in cut~C]{Same as Figure~\ref{img:cutA_C18O_Gauss} but for cut~C.}
\label{img:cutC_C18O_Gauss}
\end{figure*}
% \newpage
% \clearpage
\begin{figure*}
    \centering
%     \subfigure{
        \includegraphics[width=0.47\textwidth]{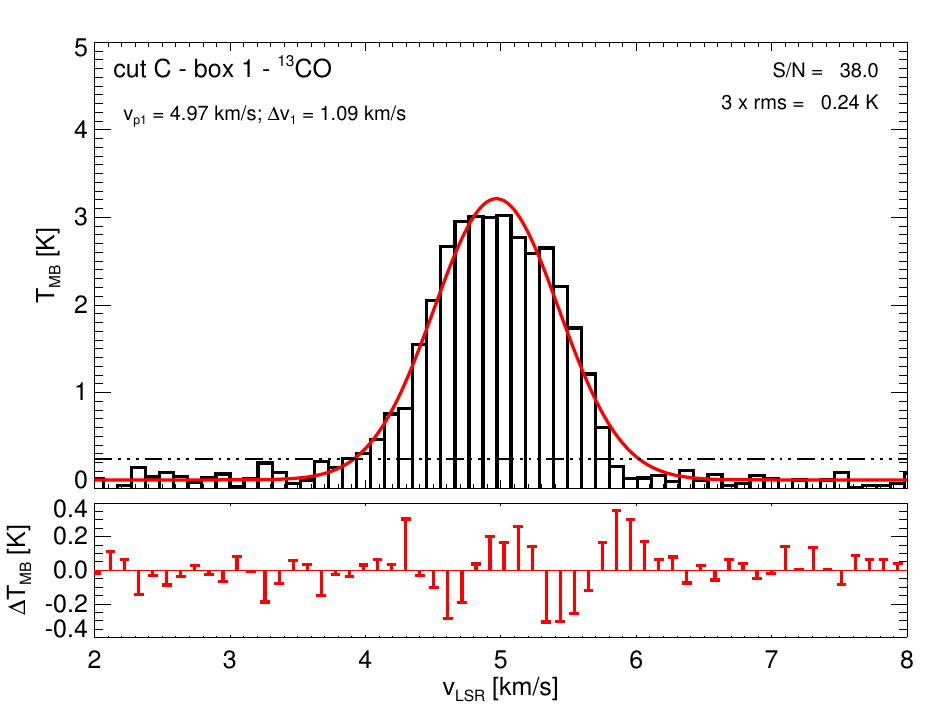}
%        \includegraphics[width=0.47\textwidth]{Lupus_I_cutC_average_box1_Gaussfit_13CO}
%     }
%     ~ %add desired spacing between images, e. g. ~, \quad, \qquad, \hfill etc. 
%     \subfigure{
        \includegraphics[width=0.47\textwidth]{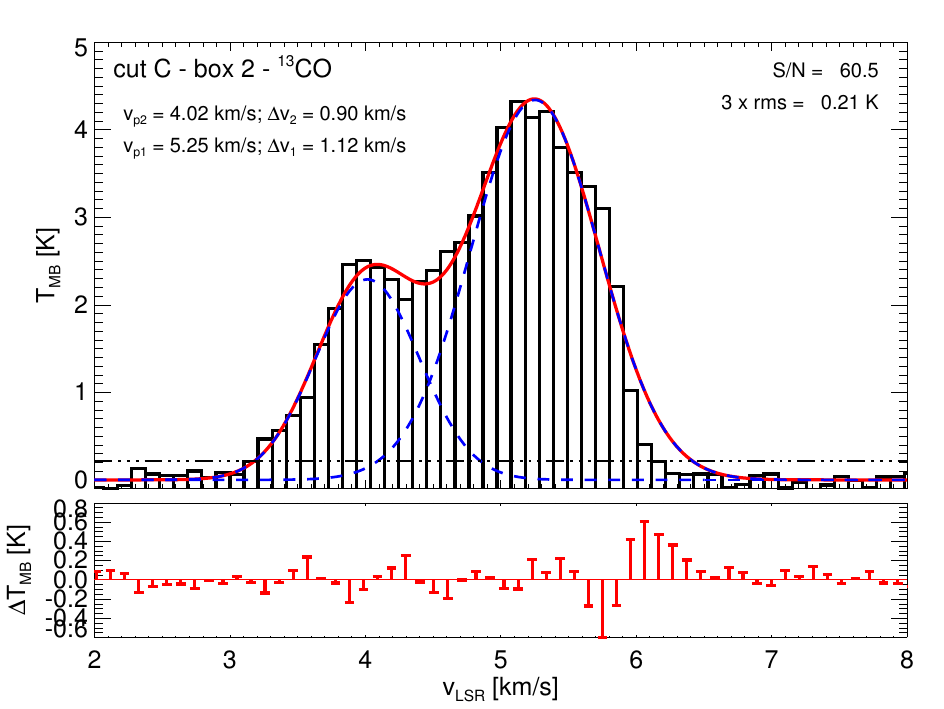}
%        \includegraphics[width=0.47\textwidth]{Lupus_I_cutC_average_box2_Gaussfit_13CO}
%     }
% \vspace{7cm}
%     ~ %add desired spacing between images, e. g. ~, \quad, \qquad, \hfill etc. 
%     \subfigure{
        \includegraphics[width=0.47\textwidth]{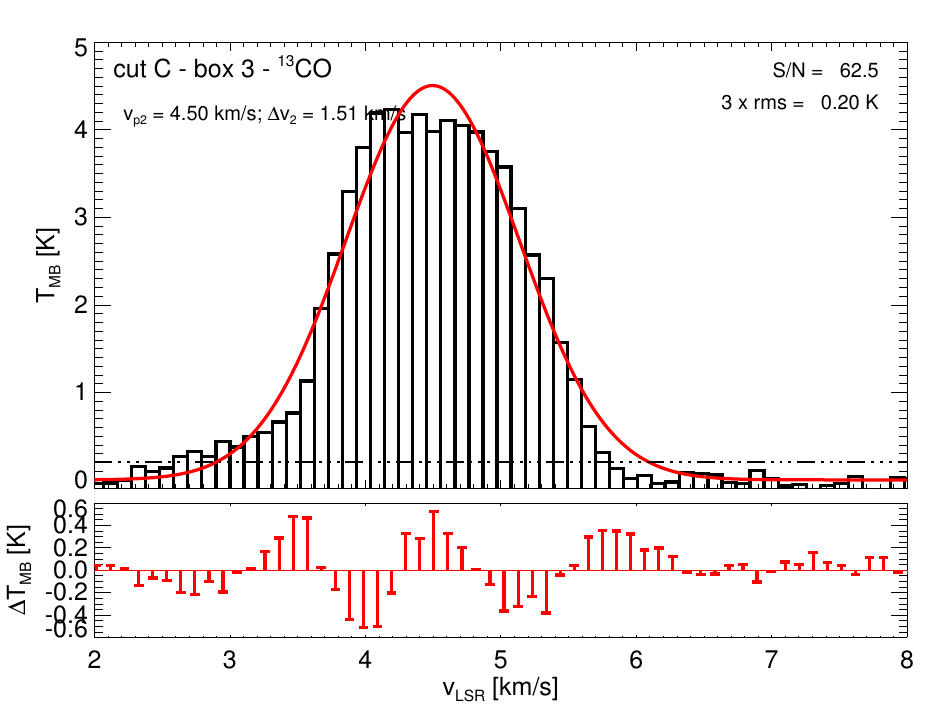}
%       \includegraphics[width=0.47\textwidth]{Lupus_I_cutC_average_box3_Gaussfit_13CO}
%     }
%     \subfigure{
        \includegraphics[width=0.47\textwidth]{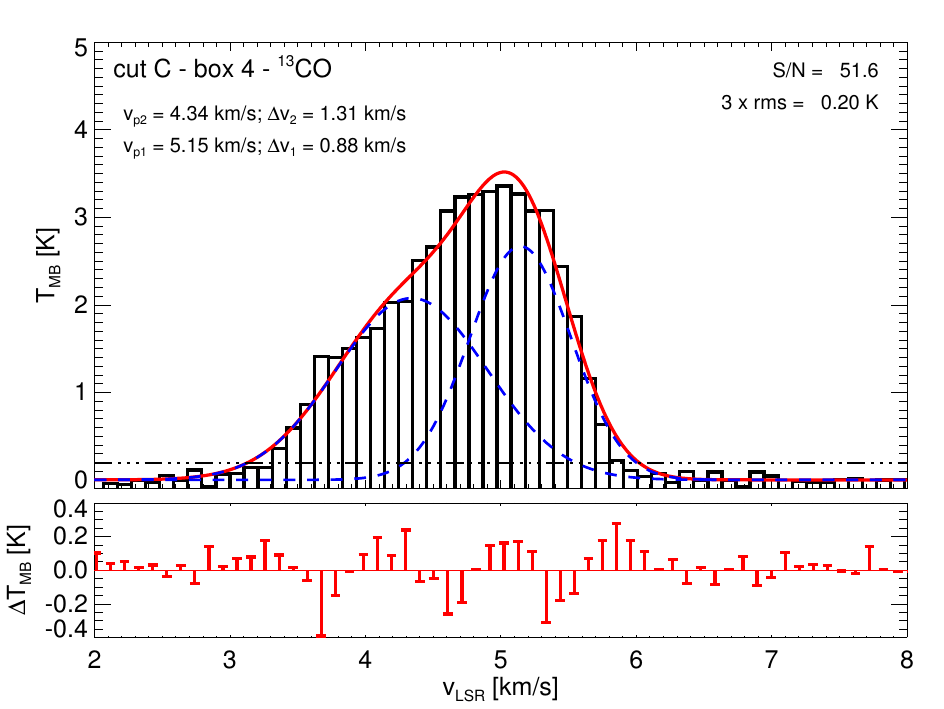}
%        \includegraphics[width=0.47\textwidth]{Lupus_I_cutC_average_box4_Gaussfit_13CO}
%     }
% \vspace{7cm}
%     \subfigure{
        \includegraphics[width=0.47\textwidth]{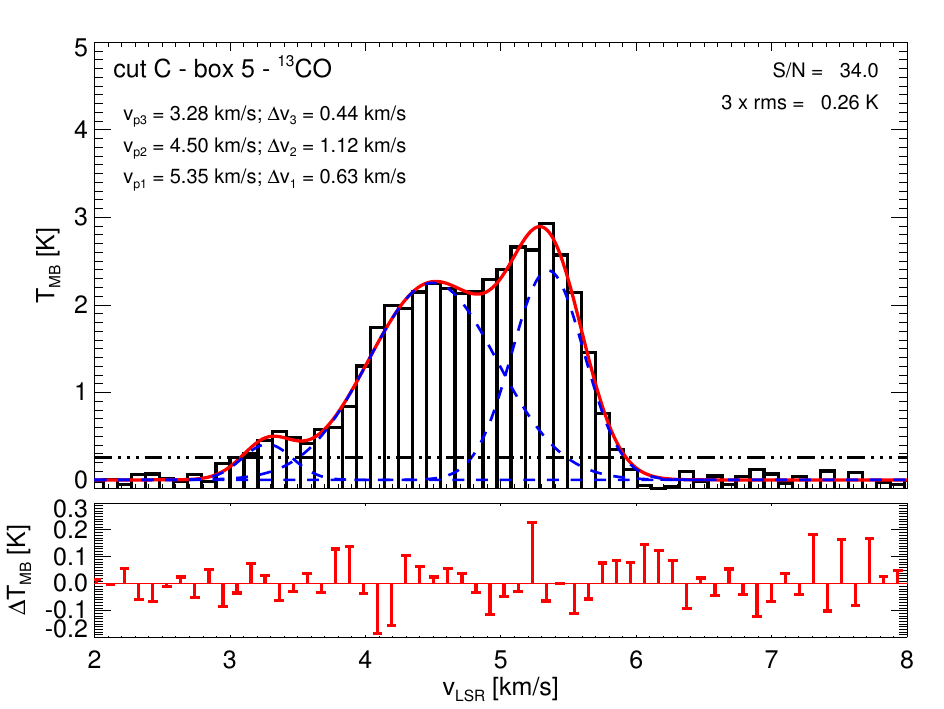}
%        \includegraphics[width=0.47\textwidth]{Lupus_I_cutC_average_box5_Gaussfit_13CO}
%     }
%     \subfigure{
        \includegraphics[width=0.47\textwidth]{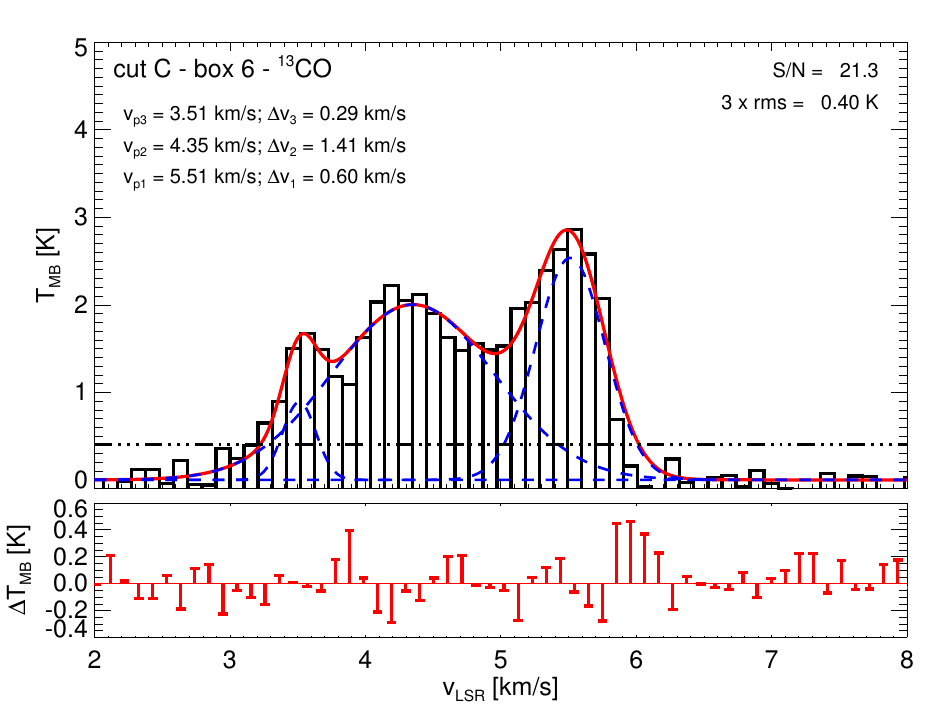}
%       \includegraphics[width=0.47\textwidth]{Lupus_I_cutC_average_box6_Gaussfit_13CO}
%     }
    \caption[Gaussian fits to the average \cdo{} spectra of the boxes in cut~C]{Same as Figure~\ref{img:cutA_13CO_Gauss} but for cut~C.}
\label{img:cutC_13CO_Gauss}
\end{figure*}
\end{appendix}

\bibliographystyle{aa} % style aa.bst
\bibliography{literatur} % your references Yourfile.bib

% \begin{thebibliography}{}
% 
% \end{thebibliography}

\Online

\end{document}